\documentclass{aa}  

\usepackage{graphicx}

\usepackage{txfonts}
\usepackage{multirow}
\usepackage{array}
\usepackage{enumitem}
\usepackage[pdftex, colorlinks=true, linkcolor=blue, citecolor=blue, filecolor=blue, urlcolor=blue]{hyperref}
\graphicspath{{./}{figures/}}							

\newcolumntype{H}{>{\setbox0=\hbox\bgroup}c<{\egroup}@{}}  

\newcommand{\hc}{\ensuremath{H_\mathrm{0}}}

\begin{document}

\title{Accelerating lensed quasar discovery and modeling with physics-informed variational autoencoders}

\author{
	Irham T. Andika \inst{\ref{affil:tum}, \ref{affil:mpa}}
}

\author{
	Irham T. Andika \inst{\ref{affil:tum}, \ref{affil:mpa}}
	\and
	Stefan Schuldt \inst{\ref{affil:unimi}, \ref{affil:inafmi}}    
    \and
    Sherry H. Suyu \inst{\ref{affil:tum}, \ref{affil:mpa}, \ref{affil:asiaa}}
    \and
    Satadru Bag \inst{\ref{affil:tum}, \ref{affil:mpa}}
	\and
	Raoul Ca\~nameras \inst{\ref{affil:lam}}
    \and
    Alejandra Melo \inst{\ref{affil:mpa}, \ref{affil:tum}}
    \and
	Claudio Grillo \inst{\ref{affil:unimi}, \ref{affil:inafmi}}    
    \and
	James H. H. Chan \inst{\ref{affil:amnh}, \ref{affil:cuny}}    
}

\institute{
	Technical University of Munich, TUM School of Natural Sciences, Department of Physics, James-Franck-Str. 1, D-85748 Garching, Germany \\
	\email{irham.andika@tum.de}
	\label{affil:tum}
	\and
	Max-Planck-Institut f\"{u}r Astrophysik, Karl-Schwarzschild-Str. 1, D-85748 Garching, Germany
	\label{affil:mpa}
	\and
	Dipartimento di Fisica, Universit\`a degli Studi di Milano, via Celoria 16, I-20133 Milano, Italy
	\label{affil:unimi}
    \and
    INAF -- IASF Milano, via A. Corti 12, I-20133 Milano, Italy
    \label{affil:inafmi}
	\and
	Academia Sinica Institute of Astronomy and Astrophysics (ASIAA), 11F of ASMAB, No.1, Section 4, Roosevelt Road, Taipei 10617, Taiwan
	\label{affil:asiaa}
    \and
    Aix-Marseille Université, CNRS, CNES, LAM, Marseille, France
    \label{affil:lam}
    \and Department of Astrophysics, American Museum of Natural History, Central Park West and 79th Street, NY 10024-5192, USA
    \label{affil:amnh}
    \and Department of Physics and Astronomy, Lehman College of the CUNY, Bronx, NY 10468, USA
    \label{affil:cuny}
}


\date{}

 
\abstract{
Strongly lensed quasars provide valuable insights into the rate of cosmic expansion, the distribution of dark matter in foreground deflectors, and the characteristics of quasar hosts.
However, detecting them in astronomical images is difficult due to the prevalence of non-lensing objects.
To address this challenge, we developed a generative deep learning model called VariLens, built upon a physics-informed variational autoencoder. 
This model seamlessly integrates three essential modules: image reconstruction, object classification, and lens modeling, offering a fast and comprehensive approach to strong lens analysis.
VariLens is capable of rapidly determining both (1) the probability that an object is a lens system and (2) key parameters of a singular isothermal ellipsoid (SIE) mass model -- including the Einstein radius ($\theta_\mathrm{E}$), lens center, and ellipticity -- in just milliseconds using a single CPU.
A direct comparison of VariLens estimates with traditional lens modeling for 20 known lensed quasars within the Subaru Hyper Suprime-Cam (HSC) footprint shows good agreement, with both results consistent within $2\sigma$ for systems with $\theta_\mathrm{E}<3\arcsec$.
To identify new lensed quasar candidates, we began with an initial sample of approximately 80 million sources, combining HSC data with multiwavelength information from Gaia, UKIRT, VISTA, WISE, eROSITA, and VLA.
After applying a photometric preselection aimed at locating $z>1.5$ sources, the number of candidates was reduced to 710\,966. 
Subsequently, VariLens highlights 13\,831 sources, each showing a high likelihood of being a lens.
A visual assessment of these objects results in 42 promising candidates that await spectroscopic confirmation. 
These results underscore the potential of automated deep learning pipelines to efficiently detect and model strong lenses in large datasets, substantially reducing the need for manual inspection.
}


\titlerunning{Accelerating lensed quasar discovery and modeling with physics-informed variational autoencoders}
\authorrunning{Andika et al.} 

\maketitle

%

\nolinenumbers 

\section{Introduction} \label{sec:intro}

Quasars serve as excellent tracers for investigating the assembly of supermassive black holes (SMBHs) and their host galaxies in the distant Universe \citep{2020ARA&A..58...27I,2023ARA&A..61..373F}.
When coupled with strong gravitational lensing, these systems offer unique perspectives on various astronomical phenomena. 
For instance, lensing can amplify the apparent flux and enhance the spatial resolution of target sources, allowing for the accurate characterization of quasar hosts that might otherwise be very faint or compact to observe \citep[e.g.,][]{2021MNRAS.501..269D,2021ApJ...917...99Y,2022MNRAS.517.3377S,2023ApJ...943...25G}.
Lensing also allows us to investigate the foreground mass profile, offering valuable clues as to the dark matter characteristics and the mechanisms influencing the evolution of large-scale structures \citep{2022arXiv221010790S}.
Furthermore, due to their variability, lensed quasars are exceptional tools for measuring the rate of cosmic expansion, quantified by the Hubble constant, \hc, through time-delay cosmography \citep{1964MNRAS.128..307R,2022A&ARv..30....8T}.

Previous studies have discovered hundreds of lensed quasars \citep[e.g.,][and the references within]{2022A&A...659A.140C,2023MNRAS.520.3305L,2024A&A...682A..47D}, but only a few tens of sources have been harnessed for time-delay analysis \citep{2020MNRAS.494.6072S,2020MNRAS.498.1420W}. 
This is partly because only systems with a large enough separation ($\Delta \theta \gtrsim 1 \arcsec$), an adequate variability, and time delays ranging from weeks to years offer enhanced precision in \hc\ measurements \citep{2015MNRAS.450.1042R,2020A&A...640A.105M,2023A&A...672A...2E,2023arXiv231209311Q}.
Given the evident tension between \hc\ values calculated from the cosmic microwave background (CMB) radiation and those inferred from closer sources \citep[e.g.,][]{2019NatAs...3..891V}, expanding the number of cosmography-grade lensed quasars will help address this issue \citep{2020A&A...641A...6P,2022ApJ...934L...7R}.


With the emergence of next-generation facilities such as the \textit{Vera C. Rubin} Observatory's Legacy Survey of Space and Time \citep[LSST;][]{2019ApJ...873..111I}, scheduled to observe around 20\,000 $\deg^2$ of the southern sky across six bandpasses, we foresee a significant rise in the number of lens candidates. 
To be precise, LSST is anticipated to unveil approximately 2000 lensed quasars, among which about 200 are expected to be quadruply lensed sources \citep{2020MNRAS.498.1420W,2022AJ....163..139Y}.
However, locating them is challenging given that the fraction of strong lenses in all galaxies per unit area of the sky is as low as $\lesssim 10^{-3}$ \citep{2010MNRAS.405.2579O}.
This motivates us to develop a method capable of accurately identifying cosmography-grade systems amidst more numerous contaminants.

In our previous work, we introduced novel strategies for detecting galaxy-quasar lenses in multiband imaging surveys \citep{2023ApJ...943..150A, 2023A&A...678A.103A}.
The method combines multiple convolutional neural networks (CNNs) and vision transformers (ViTs), resulting in a streamlined approach for locating lens candidates in the Hyper Suprime-Cam Subaru Strategic Program \citep[hereafter HSC;][]{2022PASJ...74..247A} dataset with a low number of false positives.
Considering the similarity between the HSC data and the anticipated quality of LSST images, we expect our model to perform at a comparable level with upcoming LSST data.
Building upon our previous approach, we investigated the application of deep generative networks to refine our lens-finding strategy, specifically through the use of a physics-informed variational autoencoder.
Additionally, we introduced two critical advancements in our method: (1) expanding the photometric preselection criteria to include more lensed quasar candidates, and (2) implementing fast, automated lens modeling to efficiently identify the most promising sources suitable for time-delay analysis.
The latter point is crucial, given the anticipated rapid growth of photometric data in the coming years, contrasted with the limited availability of spectroscopic resources.

The structure of this paper is as follows. 
In Section~\ref{sec:data}, we describe the data collection process and detail the simulation used to generate mock galaxy-quasar lens systems, which serve as training data for our integrated lens-finding and modeling algorithm.
Section~\ref{sec:networks} provides an in-depth look at the neural network architecture employed in this study.
In Section~\ref{sec:result}, we evaluate the performance of our model and present the lensed quasar candidates identified through our analysis.
Finally, Section~\ref{sec:conclusion} summarizes our findings and presents the conclusions drawn from this work.

All cosmological-related calculations were conducted using the $\Lambda$CDM model, where $\hc = 72~\text{km~s}^{-1}~\text{Mpc}^{-1}$ and $\Omega_\mathrm{m} = 1 - \Omega_\Lambda = 0.3$ \citep{2017MNRAS.465.4914B,2020A&A...641A...6P}. 
It is essential to mention that the mock lensed quasar images generated for the training dataset are independent of the exact value of \hc.
Additionally, the magnitudes are expressed using the AB convention.
The reddening caused by Galactic extinction was corrected using the map from \cite{1998ApJ...500..525S} and filter adjustments from \citet{1999PASP..111...63F} and \cite{2011ApJ...737..103S}. 
This correction was performed using the \texttt{dustmaps}\footnote{\href{https://dustmaps.readthedocs.io/en/latest/}{https://dustmaps.readthedocs.io/en/latest/}} library developed by \cite{2018JOSS....3..695G}.

\section{Data acquisition and training set construction} \label{sec:data}

Our search for lensed quasars involves two stages.
First, we will conduct data mining through a comprehensive analysis of multiband catalogs in this section.
Next, we will assess the likelihood of candidates being lensed quasars as opposed to other types of sources using a deep learning classifier in Section~\ref{sec:networks}.

In the first stage, our goal is to improve candidate purity by focusing on objects that, according to their photometric catalog data, have a higher chance of being lenses.
This practice provides an efficient strategy to distinguish candidates from the majority of contaminants while minimizing computational resources. 
This section provides a detailed explanation of the initial phase of our search method. 
Additionally, we will discuss how the data is utilized to create the training dataset for the deep learning classifier through strong lens simulations.

\subsection{Primary photometric catalog} \label{sec:hsc_data}

To construct the main catalog, we utilize the wide-depth data from the third release of the public HSC survey dataset \citep{2022PASJ...74..247A}. 
These data are acquired with the Hyper Suprime-Cam on the Subaru 8.2m telescope \citep{2018PASJ...70S...8A,2019PASJ...71..114A}, which captures 670 deg$^2$ of the sky across five bands, reaching a depth of up to 26~mag ($5\sigma$ for point sources), with a 0\farcs168 pixel scale and 0\farcs6–0\farcs8 seeing. 
Including partially observed areas, the latest dataset extends to approximately 1300 deg$^2$. 
This HSC data will be used to select the parent sample for our lensed quasar candidates.

We choose all sources with 5$\sigma$ detection in all bands, which corresponds to magnitude limits of  $[gri]<26.0$, $z<25.0$, and $y<24.0$.
We then use the flux and magnitude values from the \texttt{CModel} photometry reported in the HSC \texttt{pdr3\_wide.forced} table, which explicitly accounts for the point spread function (PSF) in the measurements.
The following flags are applied to ensure reliable photometry: 
\begin{enumerate}[noitemsep]
	\item \texttt{[grizy]\_sdsscentroid\_flag} is False
	\item \texttt{[grizy]\_pixelflags\_edge} is False
	\item \texttt{[grizy]\_pixelflags\_interpolatedcenter} is False
	\item \texttt{[grizy]\_pixelflags\_saturatedcenter} is False
	\item \texttt{[grizy]\_pixelflags\_crcenter} is False
	\item \texttt{[grizy]\_pixelflags\_bad} is False
	\item \texttt{[grizy]\_cmodel\_flag} is False
\end{enumerate}
After that, science images, each with 70 pixels on a side, along with their PSF models, are retrieved using the HSC data access tools\footnote{
\href{https://hsc-gitlab.mtk.nao.ac.jp/ssp-software/data-access-tools}{https://hsc-gitlab.mtk.nao.ac.jp/ssp-software/data-access-tools}
}. 
At this stage, 86\,792\,741 unique sources, which constitute our parent sample, have passed our preliminary magnitude thresholds and flag criteria. 
To determine the spectroscopic classifications of these sources where available, we crossmatch this parent sample with databases of previously discovered quasars \citep{2021arXiv210512985F, 2023ARA&A..61..373F}, stars/galaxies \citep{2023arXiv230107688A}, strong lenses\footnote{
The previously identified galaxy-scale strong lens systems are gathered from two primary sources: the Master Lens Database (MLD; available online at \href{https://test.masterlens.org/}{https://test.masterlens.org/}) and the Gravitationally Lensed Quasar Database (GLQD; accessible at \href{https://research.ast.cam.ac.uk/lensedquasars/}{https://research.ast.cam.ac.uk/lensedquasars/}).
}, and brown dwarfs\footnote{
The catalog of brown dwarfs include late-M stars, as well as L and T dwarfs.
} \citep{2018ApJS..234....1B, 2019MNRAS.489.5301C}.

\begin{table}[htb!]
	\caption{Overview of the selection criteria used to identify lensed quasar candidates.}
	\label{tab:preselection}
	\centering
	\begin{tabular}{clc}
		\hline\hline
		Step & Selection & Candidates \\
		\hline
		1 & Parent samples from Section~\ref{sec:hsc_data} &  86\,792\,741 \\
		2 & Sources with a neighbor within 2\arcsec & 7\,193\,353 \\
		   & or available spectroscopic data \\
		3 & S/N ($g,\ r,\ i,\ z,\  y$) > 5 & 6\,568\,040\\
		4 & Detection in X-ray, IR, or radio & 1\,929\,190  \\
		5 & $i$-band Kron radius > 0\farcs5 & 710\,966\\
		6 & Network classification & 18\,300 \\
		7 & Astrometric information & 13\,831  \\
			& ~\textbullet~ Astrometric excess noise $<10$~mas \\
			& ~\textbullet~ Proper motion significance $<10\sigma$ \\
		8 & Grade A and B classifications from  & 42 \\
			& the visual check detailed in Section~\ref{sec:lenscand} &  \\
		\hline
	\end{tabular}
\end{table}

\subsection{Auxiliary data}

Near-infrared (NIR) data are gathered from multiple surveys, including the VISTA Kilo-degree Infrared Galaxy Survey \citep[VIKING;][]{2013Msngr.154...32E},  the VISTA Hemisphere Survey \citep[VHS;][]{2013Msngr.154...35M}, the United Kingdom Infrared Telescope (UKIRT) Infrared Deep Sky Survey \citep[UKIDSS;][]{2007MNRAS.379.1599L}, and the UKIRT Hemisphere Survey \citep[UHS;][]{2018MNRAS.473.5113D}.
For this study, we utilize available photometry in the $J$, $H$, and $K$ (or $K_\mathrm{s}$) filters. 
Notably, VHS and VIKING cover a significant portion of the southern hemisphere, while UKIDSS and UHS cover vast regions of the northern sky.
At the time of our search, data collection for both UKIDSS and VIKING had been completed.
By comparison, UHS had only made $J$-band photometry available, whereas VHS had provided data in both the $J$ and $K_\mathrm{s}$ filters across the majority of sky regions.
Thus, the exact photometric data available for each object depends on its sky position.

In addition to the NIR data, the unWISE mid-infrared (MIR) measurements are used  \citep{2019ApJS..240...30S}, which catalogs nearly two billion sources detected by the Wide-field Infrared Survey Explorer \citep[WISE;][]{2010AJ....140.1868W}. 
With its $\sim$0.7 magnitude deeper imaging and enhanced source extraction in crowded regions, the unWISE catalog provides superior data quality compared to the original WISE survey.
To merge the HSC data with the infrared (IR) catalogs mentioned above, we apply a 2$\arcsec$ crossmatching radius between corresponding targets.

Given that some quasars are X-ray bright, we incorporated data from the eROSITA telescope mounted on the Spektrum-Roentgen-Gamma (SRG) satellite. 
Specifically, we employed the first data release of the eROSITA All-Sky Survey \citep[eRASS;][]{2024A&A...682A..34M}, which provides soft X-ray measurements in the energy band 0.2–2.3 keV.
The crossmatching process is then performed against our optical data using a radius of 15\arcsec.

Next, to identify radio-bright sources, we also included data from the Karl G. Jansky Very Large Array Sky Survey \cite[VLASS;][]{2020PASP..132c5001L}.
VLASS is a radio survey designed to map the sky at 2--4 GHz (S-band). 
For this analysis, we looked for the radio counterparts of our optical sources within a radius of 2\farcs5, using the second epoch of the VLASS Quick Look catalog \citep{2020RNAAS...4..175G,2021ApJS..255...30G,2023ApJS..267...37G}.

With the complete multiwavelength dataset in hand, we prioritized sources showing signal-to-noise ratio (S/N) of at least 3$\sigma$ in the X-ray, IR, or radio.
Ultimately, a total of 1\,929\,190 sources met this criterion.
It is worth noting that the combination of multiwavelength information plays a crucial role in distinguishing between quasars, stars, and brown dwarfs \citep[e.g.,][]{2020ApJ...903...34A,2022AJ....163..251A}.
Additionally, this crossmatching method is effective for filtering out artifacts like cosmic rays or transient sources that appear in only one survey, helping to ensure cleaner data \citep[e.g.,][]{2022PhDT.........1A}.
A summary of our selection criteria is provided in Table~\ref{tab:preselection} for further details.

\subsection{Real galaxy and simulated quasar samples} \label{sec:galaxy_quasar}

The training dataset used in this study is adapted from \cite{2023A&A...678A.103A}. 
The mock lenses, are created by overlaying simulated point sources onto real galaxy images. 
The deflectors in this dataset are sourced from galaxy samples observed in the Sloan Digital Sky Survey Data Release 18 catalog \citep[SDSS;][]{2023arXiv230107688A}.
Specifically, we select objects classified as ``GALAXY'' by the SDSS spectroscopic pipeline having velocity dispersions of $50 \leq \sigma_\mathrm{v} \leq 500$~km~s$^{-1}$ and S/N of $\sigma_\mathrm{v}/\sigma_\mathrm{v, err} > 5$.
This specific $\sigma_\mathrm{v}$ range is selected to exclude systems with measurements that are either too small, too large,  or inaccurate.
Furthermore, we restrict our selection to deflectors within redshift of $z < 4$ since the prevalence of the lensing optical depth for high-$z$ quasars originates from ellipticals around $z \sim 1$ \citep{2015ApJ...805...79M,2019ApJ...870L..12P}.
These selected deflectors are then cross-matched to our parent data with a search radius of 1\arcsec\ to obtain their photometric measurements and  image cutouts when available. 
As a result, we obtain a sample of 138\,556 deflectors, predominantly made up of luminous red galaxies (LRGs) with typical velocity dispersions around $\sigma_\mathrm{v} \sim 250$~km~s$^{-1}$, extending to $z \sim 1.5$ (see Figure~\ref{fig:lensgal_dist}).

To simulate the point-source emissions of background quasars, we generate a thousand quasar spectra, uniformly distributed across rest-frame 1450\,$\AA$ absolute magnitudes ranging from $M_{1450} = -30$ to $-20$ and redshifts between $z = 1.5$ and $7.2$.
The simulation is performed using the \texttt{SIMQSO}\footnote{\url{https://simqso.readthedocs.io/en/latest/}} module \citep{2013ApJ...768..105M}, following the methodology outlined by \citet[][see their Sect. 2.2 for more information]{2023ApJ...943..150A}.
Our quasar model is composed of a continuum emission characterized by a broken power-law shape. 
The continuum slopes ($\alpha_\nu$) follow a Gaussian function with means of $-1.5$ for rest wavelengths of $\lambda_\mathrm{rest} \leq 1215~\AA$ and $-0.5$ at $\lambda_\mathrm{rest}> 1215~\AA$, assuming a fixed dispersion of 0.3. 
The model further incorporates pseudo-continuum iron emissions using templates from \cite{1992ApJS...80..109B} and \cite{2001ApJS..134....1V}. 
Additionally, broad and narrow emission lines are included, adhering to the strengths, ratios, and widths observed in SDSS quasars \citep[e.g.,][]{2013AJ....145...10D,2016AJ....151...44D,2017AJ....154...28B}.

Additionally, the generated spectra account for intergalactic medium (IGM) absorption through the Ly$\alpha$ forest along the line of sight \citep{2010ApJ...721.1448S,2011ApJ...728...23W}). 
For quasars with $z \gtrsim 5.5$, we apply the Ly$\alpha$ damping wing effect using the formula provided by \cite{1998ApJ...501...15M}, with a proximity zone size of 3~Mpc and neutral hydrogen fractions varying randomly between 0\% and 10\% \citep[e.g.,][]{2019A&A...631A..85E,2020ApJ...903...34A}.
Furthermore, internal reddening due to dust is applied to the spectra using the model from \cite{2000ApJ...533..682C}, with $E(B-V)$ values randomly chosen between $-0.02$ and 0.14. 
Negative reddening values are used to generate quasar models with bluer continua than the original templates provide. 
Photometry is subsequently derived from these mock data, and its uncertainty is computed using the magnitude--error relation specific to each survey \citep[e.g.,][]{2016ApJ...829...33Y}.

\subsection{Synthetic multiband lensed quasar images with strong lensing simulation} \label{sec:lens_simulation}

In the following stage, we use a singular isothermal ellipsoid (SIE) approximation to represent the profile of the lens mass \citep[e.g.,][]{1998ApJ...502..531B,2022AandA...668A..73R,2023A&A...671A.147S}. 
This model is characterized by 5 parameters: the Einstein radius ($\theta_\mathrm{E}$), the axis ratio ($q$), the position angle ($\phi$), and the image centroid coordinates ($x$, $y$). 
The Einstein radius can be directly estimated from the observed galaxy velocity dispersion using the following equation:
\begin{equation}
	\label{eq:sie_lens}
	\theta_\mathrm{E} = 4\pi \frac{\sigma_v^2}{c^2} \frac{D_\mathrm{ds}}{D_\mathrm{s}},
\end{equation}
where $c$ is the speed of light, and $D_\mathrm{ds}$ is the angular diameter distance from the lens to the source, while $D_\mathrm{s}$ is the distance from the observer to the source. 
As shown by the distance ratio in Equation~\ref{eq:sie_lens}, the Einstein radius $\theta_\mathrm{E}$ does not depend on $H_0$.

The dimensionless surface mass density, or convergence, of the SIE profile used here can be described as
\begin{equation}
	\kappa(r) = \frac{\theta_\text{E}}{(1+q)\ r},
\end{equation}
where \( r \) denotes the elliptical radius given by
\begin{equation}
	r = \sqrt{x^2 + \frac{y^2}{q^2}},
\end{equation}
and \( q \) represents the axis ratio, defined as
\begin{equation}
	q = \sqrt{\frac{1 - \sqrt{e_x^2 + e_y^2}}{1 + \sqrt{e_x^2 + e_y^2}}},
\end{equation}
where \( e_x \) and \( e_y \) are the components of the complex ellipticity in the coordinate plane of $x$--$y$ of the lens system.
These variables can also be expressed as
\begin{align}
	e_x &= \frac{1 - q^2}{1 + q^2} \cos(2\phi),\\
	e_y &= \frac{1 - q^2}{1 + q^2} \sin(2\phi).
\end{align}

To account for extra mass beyond the observed image region, we include external shear parameters ($\gamma_\mathrm{ext, 1}$, $\gamma_\mathrm{ext, 2}$). 
These parameters are related to the total shear strength ($\gamma_\text{ext}$), which is given by
\begin{equation}
	\gamma_\text{ext} = \sqrt{\gamma_\text{ext,1}^2 + \gamma_\text{ext,2}^2},
\end{equation}
and the shear is oriented by
\begin{equation}
	\phi_\text{ext} = \left\{
	\begin{array}{ll}
		\frac{s}{2} & \text{if } \gamma_\text{ext,1} \geq 0 \text{ and } \gamma_\text{ext,2} \geq 0, \\
		\frac{\pi - s}{2} & \text{if } \gamma_\text{ext,1} < 0 \text{ and } \gamma_\text{ext,2} \geq 0, \\
		\frac{\pi + s}{2} & \text{if } \gamma_\text{ext,1} < 0 \text{ and } \gamma_\text{ext,2} < 0, \\
		\frac{2\pi - s}{2} & \text{if } \gamma_\text{ext,1} \geq 0 \text{ and } \gamma_\text{ext,2} < 0,
	\end{array}
	\right.
\end{equation}
with \( s \) defined as
\begin{equation}
	s = \arcsin \left( \frac{|\gamma_\text{ext,2}|}{\gamma_\text{ext}} \right).
\end{equation}

We determine the SIE parameters \(x\), \(y\), \(e_x\), and \(e_y\) by modeling the light profile of each deflector using its HSC $i$-band data. 
This approach assumes that the mass profile follows the light distribution. 
Our fitting method utilizes a mixture of elliptical S\'{e}rsic and exponential light profiles, implemented in the \texttt{PyAutoGalaxy}\footnote{\url{https://pyautogalaxy.readthedocs.io/en/latest/}}, a publicly available tool for studying galaxy morphologies \citep{2018MNRAS.478.4738N,2023JOSS....8.4475N}. 
Afterward, we apply random external shears, with position angles selected randomly between 0 and 180~$\deg$ and strengths drawn from a Gaussian distribution with a mean of 0 and a standard deviation of 0.058 \citep[e.g.,][]{2022AandA...662A...4S}.

Next, images of simulated lenses are generated by combining each real galaxy with a randomly chosen mock quasar. 
The quasar is placed behind the lens at a random angular distance ($\beta$) within the range $0\farcs01 \leq \beta \leq \theta_\mathrm{E}$. 
The parameter $\beta$ can be decomposed into its components ($x_\mathrm{s}$, $y_\mathrm{s}$) as follows:
\begin{equation}
	\beta = \sqrt{x^2_\mathrm{s} + y^2_\mathrm{s}}.
\end{equation}
The background quasar image is then traced onto the lens plane, and the deflection angles and magnifications are calculated based on the lensing equation using \texttt{PyAutoLens}\footnote{\url{https://pyautolens.readthedocs.io/en/latest/}} \citep[version 2024.1.27.4;][]{2021JOSS....6.2825N}. 
The resulting multiplied quasar images are then convolved with the HSC PSF model and overlaid onto the original HSC galaxy images.

The process of pairing and positioning quasars can be performed up to 500 times to determine a viable lens configuration that meets the following conditions: (1) the simulated image must exhibit a significant lensing effect with a magnification factor of $>5$, (2) the maximum $y$-band flux of the lensed quasar must be detected at $> 5\sigma$ above the median background noise, and (3) the $y$-band magnitude of the lensed quasar must be $>15$~mag to prevent the inclusion of excessively bright objects or saturated images.
If these conditions are not met, the current deflector is discarded, and the process continues with the next one. 
Additionally, systems with $\theta_\mathrm{E} \geq 5\arcsec$ are excluded from consideration, as the largest Einstein radius observed in galaxy-scale lensing cases are not exceeding this limit \citep{2007ApJ...671L...9B,2019A&A...631A..40S,2021A&A...646A.126S}. 
Ultimately, 137\,552 mock lenses that meet our criteria are obtained from the preliminary pool of 138\,556 deflectors and 1000 mock quasars.

\begin{figure*}[htb!]
	\centering
	\resizebox{\hsize}{!}{\includegraphics{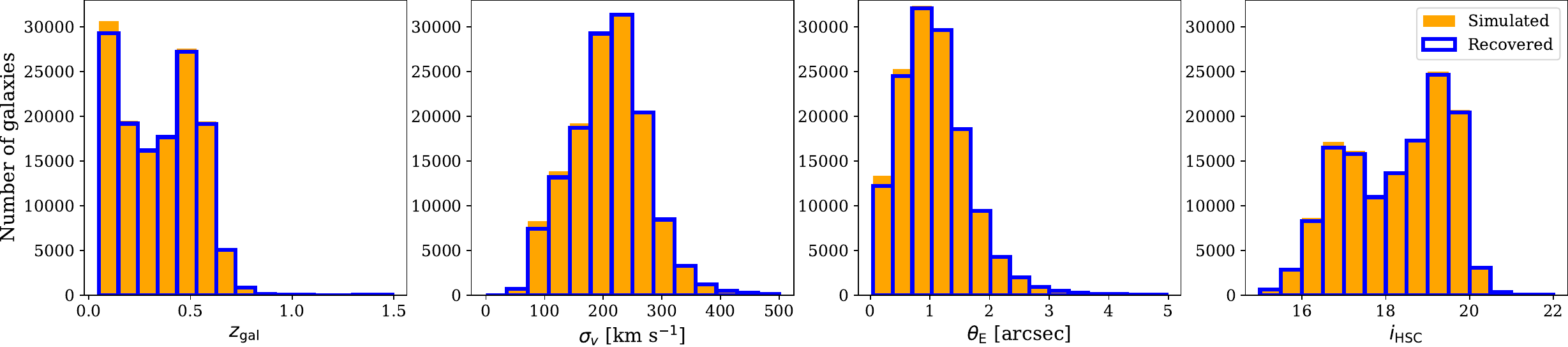}}
	\caption{
		Distribution of redshifts ($z_\mathrm{gal}$), stellar velocity dispersions ($\sigma_v$), Einstein radii ($\theta_\mathrm{E}$), and HSC $i$-band magnitudes ($i_\mathrm{HSC}$) for the galaxies used to simulate the lens systems. 
		The orange histograms represent the mock lens configurations in all training, validation, and test datasets, while the mock lenses correctly identified by our classifier are highlighted with blue lines.
	}
	\label{fig:lensgal_dist}
\end{figure*}

\begin{figure}[htb!]
	\centering
	\resizebox{\hsize}{!}{\includegraphics{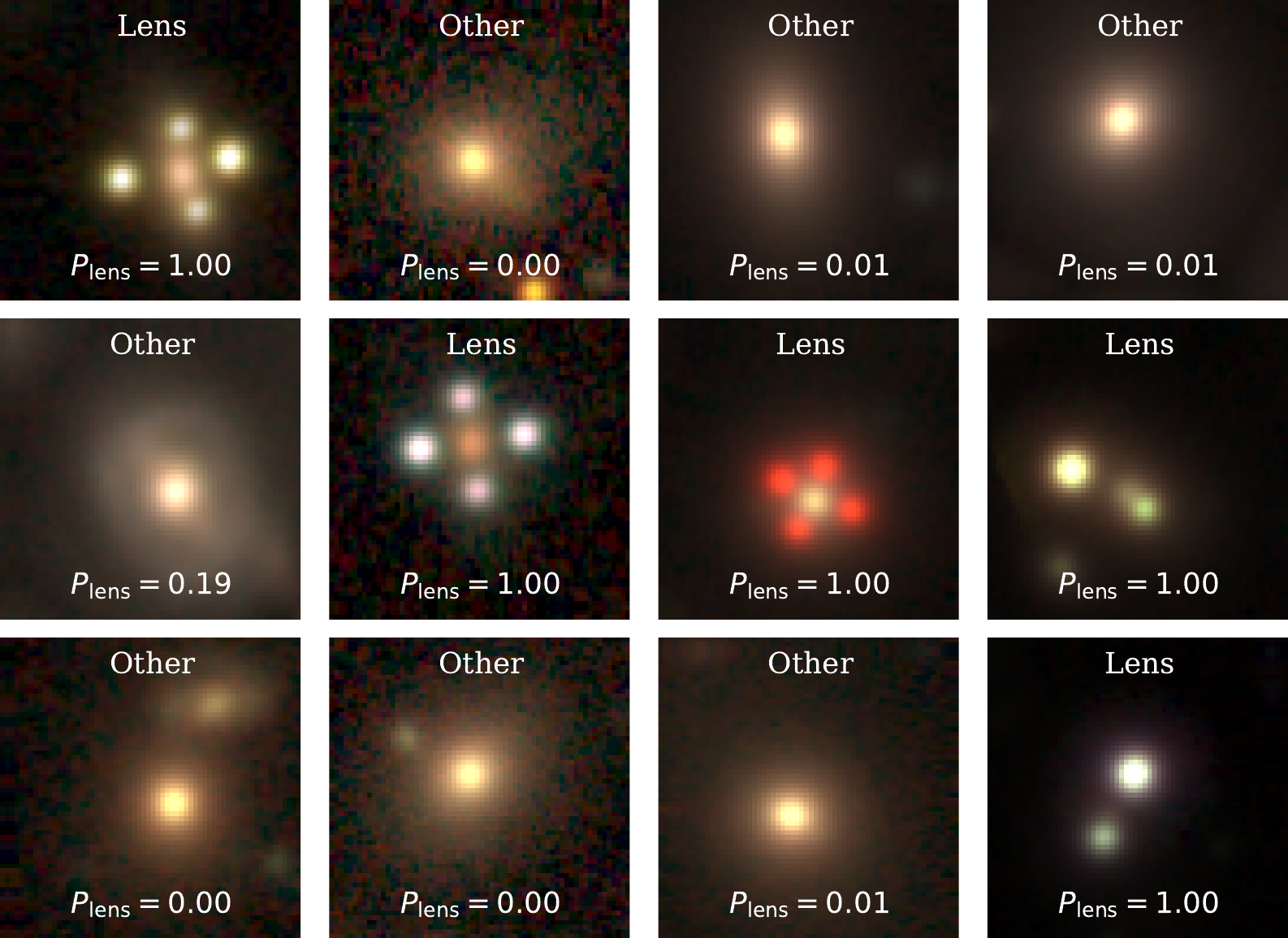}}
	\caption{
		Examples of mock lenses and other contaminants used for training the networks, with the inferred lens probability for each image indicated.
		By overlaying the multiply imaged source's light following an SIE+$\gamma_\text{ext}$ lens configuration onto a real galaxy image, we are able to construct realistic galaxy-quasar lens mocks.
	}
	\label{fig:mock_lens}
\end{figure}

Figure~\ref{fig:lensgal_dist} illustrates the distribution of lens galaxy redshifts, velocity dispersions, Einstein radii, and $i$-band magnitudes used in our simulation. 
For the color images of the simulated lens systems, see Figure~\ref{fig:mock_lens}. 
Our deflector redshifts are primarily concentrated at $z \approx 0.5$ and extend up to $z \sim 1.5$, as noted earlier.
There is a notable increase in the number of deflector galaxies up to $i \approx 19.5$, followed by a sharp decline at fainter magnitudes. 
Consequently, our training dataset is biased to larger and brighter lens galaxies.
This distribution pattern primarily reflects the selection criteria of SDSS for its spectroscopic targets, adhering to the selection set by \cite{2013AJ....145...10D,2016AJ....151...44D} and \cite{2016ApJS..224...34P} for studying large-scale structures in the Universe. 
Most of these targets are LRGs at $z \lesssim 1$, which are particularly useful for examining baryon acoustic oscillations and, thus, understanding the Universe's expansion \citep[e.g.,][]{2022MNRAS.516.4307W,2022MNRAS.511.5492Z}. 
The magnitude limits for galaxies included in the spectroscopic surveys are $i = 19.9$ and 21.8 for SDSS III and IV, respectively.

\section{Lens detection and parameter estimation with neural networks} \label{sec:networks}

\subsection{Input data preparation and augmentation}

Our training datasets are divided into two groups to align with the two distinct modules in our network architecture: regression and classification. 
The first group, hereafter DS1, is used to train the regressor and consists of all the simulated lensed quasars generated in the preceding section.
The second dataset, named DS2, includes half of the previously created mock lenses and half of the real galaxies without lensed quasars, which will be used to train our classifier.
Both DS1 and DS2 contain 137\,552 objects each. 
Again, while DS1 is entirely composed of mock lenses, DS2 is a balanced mix of simulated lensed quasars and real galaxies.

Data augmentation is applied to both DS1 and DS2 but with different setups. 
For DS1, only translations of up to 3 pixels are used.
Transformations such as rotations and flips are avoided, as they would modify the lensing configuration, requiring updates to the ground truth lens parameters and thereby increasing computational complexity.
In contrast, for DS2, we apply random rotations of $\pm\pi/2$, translations of up to 10 pixels, and vertical or horizontal flips on the fly during the training loop.
In the case of the classification task, this approach boosts the number of training data and helps the network learn to correctly categorize different perspectives of the same image.

The training images are initially constructed from five HSC bands, each measuring 70 pixels per side, corresponding to an angular size of approximately 11\farcs8. 
After undergoing the augmentation process, we crop the images to $64\times 64$ pixels, that is, equivalent to 10\farcs8, to focus on the sources at the center.
These image cutouts are then adjusted using min-max scaling to normalize fluxes between zero and one, followed by square-root stretching to enhance low-flux features and improve visual appearance \citep[e.g.,][]{2021AandA...653L...6C,2023A&A...678A.103A}.
This scaling is performed on each $grizy$ image cube for every source, ensuring that the relative pixel brightness across different bandpasses is preserved, thereby retaining the colors of the sources.

\subsection{Physics-informed network architecture}

Autoencoders (AEs) are neural networks designed to capture a compressed, or latent, representation of input data and then reconstruct the input from this representation \citep[see, for example,][and the references therein]{2020arXiv200305991B}. 
Variational autoencoders (VAEs) extend the traditional AE by introducing a probabilistic approach to the latent space \citep{2019arXiv190602691K}. 
Instead of mapping inputs to fixed points in the latent space, VAEs encode input data to a distribution over the latent space, typically modeled as a Gaussian. 
The encoder outputs parameters of this distribution (mean and variance), and the latent representation is sampled from this distribution during training. 
This probabilistic framework allows VAEs to create new data points by sampling from the distribution learned in the latent space and subsequently decoding these samples.

\begin{figure}[htb!]
	\centering
	\resizebox{\hsize}{!}{\includegraphics{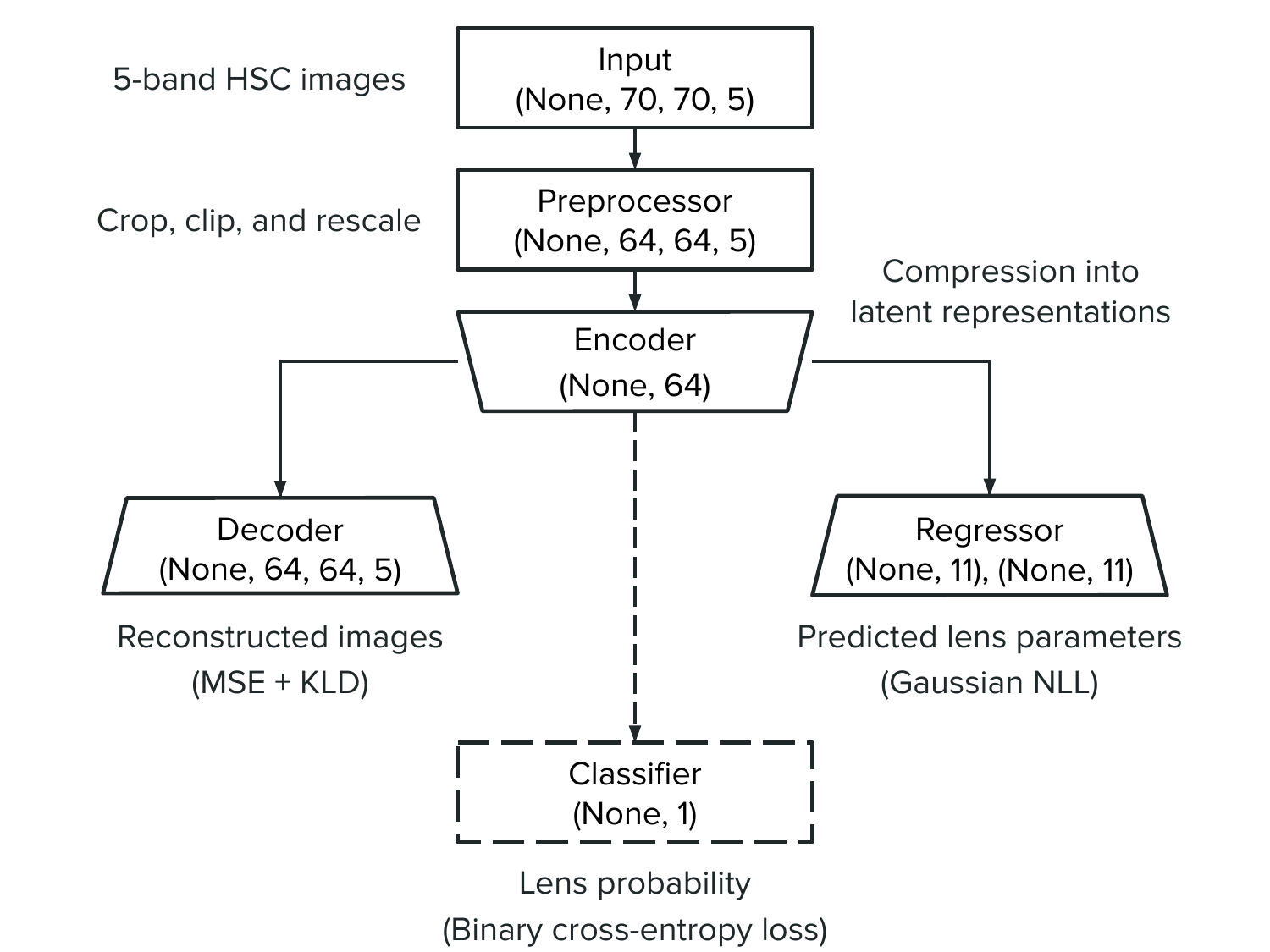}}
	\caption{
		Simplified overview of the VariLens architecture.
		The networks include three main components: the encoder, decoder, and regressor.
		The input consists of a batch of 5-band HSC images, initially sized at $70\times70$ pixels.
		The batch size is flexible, with ``None'' indicating it can vary based on user specifications.
		These images are cropped, clipped, and rescaled to standardize pixel values before reaching the encoder. 
		The encoder subsequently takes the preprocessed $64\times64$ pixel images and compresses them into a 64-dimensional latent representation. 
		The decoder then reconstructs this latent representation back into the original cropped $64\times64$ pixel images.
		At the same time, the regressor estimates lens and source parameters, guiding the latent distribution to ensure it is physics-informed.
		Once the networks are fully trained, the decoder and regressor are removed and replaced by a single dense layer serving as the classification head. 
		Transfer learning is then applied to fine-tune the classifier, optimizing it to effectively distinguish between lensing and non-lensing systems.
	}
	\label{fig:arch_compact}
\end{figure}

In this paper, we present a specialized version of VAE, which we name VariLens. 
This architecture integrates a physical model into the loss function to detect strongly lensed quasars and estimate the associated physical properties such as SIE+$\gamma_\text{ext}$ mass parameters, source positions, and redshifts.
VariLens features an encoder with a branching structure designed for two distinct tasks, as illustrated in Figure~\ref{fig:arch_compact}.
The primary branch connects directly to the decoder, following the conventional VAE framework where the encoded latent representations are utilized to reconstruct the input data. 
Meanwhile, the secondary branch, referred to as the regressor, diverges after the latent space representation, leading to a dense layer dedicated to predicting the physical parameters and their associated errors. 
More details of the VariLens networks can be found in Appendix~\ref{sec:full_architecture}.

The encoder in VariLens consists of four convolutional layers with 16, 32, 64, and 128 filters, each paired with batch normalization (BN) and leaky ReLU activation layers.
These convolutional layers employ strides of 2 and `same' padding. 
Then, the compressed data is flattened, and a dense layer with 256 units is added, which also includes a BN layer and leaky ReLU activation.
Following that, two dense layers, each with 64 units, are added to output two separate vectors: one for the mean and one for the variance of the latent variable distribution.
During the training, the reparameterization trick is employed to sample from this distribution, allowing the gradient to be backpropagated through the network by treating the randomness as a separate step from the network’s parameters, thus enabling the network to learn both the mean and variance effectively \citep{2019arXiv190602691K}.

The decoder in VariLens mirrors the encoder in its design but in reverse. 
It begins with a dense layer with 128 units, which is then batch normalized and leaky ReLU activated.
After that, we reshape it into a tensor that matches the dimensions required for the subsequent transposed convolutional layers\footnote{
	Transposed convolutional layers are also known as deconvolutional layers.
}.
The tensor above is passed through four transposed convolutional layers with filter sizes of 128, 64, 32, and 16, each followed by a BN layer and a leaky ReLU activation. 
These layers are configured with strides of 2 and `same' padding to progressively increase the spatial dimensions of the feature maps, ultimately reconstructing the original input size of $64 \times 64$ pixels.
The final layer of the decoder is a convolutional layer with five filters and a sigmoid activation function, which outputs the reconstructed image. 
This layer ensures that the output image matches the original input format, with pixel values scaled to the appropriate range. 

The optimization of the encoder and decoder is partly achieved using two loss functions.
The first one is the reconstruction loss, parameterized using mean squared error (MSE), defined as: 
\begin{equation} 
	\mathcal{L}_{\text{MSE}} = \frac{1}{N} \sum_{i=1}^{N} \left( x_i - \hat{x}_i \right)^2, 
\end{equation} 
where \( x_i \) and \( \hat{x}_i \) represent the \(i\)-th observed and reconstructed images, respectively, and \( N \) is the count of data points.
Specifically, $x_i$ is a vector representing 5-bands image with dimensions of $64\times64$ pixels, resulting in $N=20\,480$ data points.
This loss measures how well the decoder can reconstruct the input data from its latent space representation. 
Minimizing this term improves the accuracy of the reconstruction. 
The second component is the Kullback-Leibler divergence \citep[KLD;][]{2019arXiv190602691K}, given by: 
\begin{equation} 
	\mathcal{L}_{\text{KLD}} = \frac{1}{2} \sum_{j=1}^{D} \left( 1 + \log(\sigma_{z, j}^2) - \mu_{z, j}^2 - \sigma_{z, j}^2 \right), 
\end{equation} 
where \( \mu_{z, j} \) and \( \sigma_{z, j} \) are the mean and standard deviation of the \( j \)-th latent variable, respectively, and \( D \) denotes dimensionality of the latent space. 
This term quantifies the divergence between the learned distribution of latent variables and a standard normal distribution.
It regularizes the latent space by encouraging the latent variables to conform to the Gaussian prior distribution. 

We then design the regressor to predict two values for each physical parameter using a Gaussian distribution \citep[e.g.,][]{2017ApJ...850L...7P,2023A&A...671A.147S}. 
This method yields not just a point estimate but also a mean value along with an uncertainty quantified at the $1\sigma$ level. 
To assess the regressor's performance, we utilize the Gaussian negative log-likelihood (NLL) as the loss function, expressed as:
\begin{equation}
    \mathcal{L}_{\text{NLL}} = \frac{1}{M} \sum_{k=1}^{M} \left[ \frac{(\eta_k^\text{pred} - \eta_k^\text{true})^2}{2 \sigma_{\eta, k}^2} + \frac{1}{2} \log(2 \pi \sigma_{\eta, k}^2) \right],
\end{equation}
where \( \eta_k^\text{pred} \) represents the vector of predicted parameters, \( \eta_k^\text{true} \) denotes the vector of true parameters, \( \sigma_{\eta, k} \) is the standard deviation for the \( k \)-th parameter, and \( M \) is the total number of parameters. 
The loss for a particular lens system is calculated by summing over the \(M\) parameters \(\eta\) and their corresponding uncertainties \(\sigma_{\eta}\).
These parameters encompass the properties of the foreground lens, including the mass center coordinates (\(x_\text{gal}\), \(y_\text{gal}\)), complex ellipticities (\(e_\text{x}\), \(e_\text{y}\)), Einstein radius (\(\theta_\text{E}\)), external shear components (\(\gamma_\mathrm{ext, 1}\), \(\gamma_\mathrm{ext, 2}\)), and deflector's redshift (\(z_\text{gal}\)). 
Additionally, \(\eta\) includes the redshift (\(z_\text{qso}\)) of the background quasar and its position (\(x_\text{qso}\), \(y_\text{qso}\)) in the source plane.
In total, \(\eta\) comprises 11 physical parameters with its corresponding uncertainties \(\sigma_{\eta}\) also comprising 11 parameters.

The total loss for optimizing VariLens, denoted as \( \mathcal{L} \), integrates three components:
\begin{equation}
	\mathcal{L} = \mathcal{L}_{\text{MSE}} + \mathcal{L}_{\text{KLD}} + \mathcal{L}_{\text{NLL}},
 	\label{eq:loss}
\end{equation}
which ensure accurate data reconstruction, proper latent space regularization, and precise fitting of the regression output to the true parameters.

\subsection{Normalization of parameters}

Due to the differing ranges of different physical parameters, we apply min-max scaling to consistently map them to the range of $[0, 1]$ using:
\begin{equation}
	\eta^\text{scaled} = \frac{\eta - \eta_\text{min}}{\eta_\text{max} - \eta_\text{min}}.
\end{equation}
Here, $\eta$ represents the original parameter value while $\eta_\text{min}$ ($\eta_\text{max}$) is the minimum (maximum) value of the related parameter within the training dataset. 
This normalization ensures that each parameter contributes equally to the loss function, leading to improved optimization and model performance.
Scaling back the network output back to the original ranges can be done by applying:
\begin{equation}
	\eta = \eta^\text{scaled} \cdot (\eta_\text{max} - \eta_\text{min}) + \eta_\text{min},
\end{equation}
To revert the scaled uncertainties $\sigma_{\eta}^\text{scaled}$ back to the original $\sigma_{\eta}$, we use:
\begin{equation}
	\sigma = \sigma^\text{scaled} \cdot (\eta_\text{max} - \eta_\text{min}),
\end{equation}
where the deviations are not shifted by \(\eta_\text{min}\) since they are relative to the predicted median \(\eta\).
To ensure the regressor predicts values between 0 and 1, we include a sigmoid layer before splitting the outputs into 11 values for the median \(\eta\) and 11 values for the uncertainty \(\sigma_{\eta}\).

Furthermore, to calibrate our model's predicted uncertainties to match the expected Gaussian confidence intervals, we derive scaling factors for each parameter in $\sigma_\eta$ by comparing the empirical percentiles of the absolute prediction errors with theoretical Gaussian percentiles \citep[see also][]{2023A&A...671A.147S}. 
Specifically, we first compute the absolute errors between our model's predictions and the true values, then estimate the 68.3\%, 95.4\%, and 99.7\% percentiles of these errors. 
The scaling factor is then defined as the average ratio of these error percentiles to the predicted uncertainty levels, $\sigma_{\eta}$, output by VariLens. 
This method effectively adjusts our model's uncertainty estimates to better reflect the actual error distribution, ensuring that, on average, 68.3\%, 95.4\%, and 99.7\% of predictions fall within 1, 2, and 3 standard deviations, respectively.

\subsection{Model optimization}

The VariLens training process is done in two phases: (1) simultaneous image reconstruction and physical parameter regression, and (2) classification.
In the first phase, the physics-informed VAE network -- which includes the encoder, decoder, and regressor -- is optimized using DS1 with an initial learning rate of \(10^{-4}\). 
The objective is to minimize the loss function specified in Equation~\ref{eq:loss}. 
The training process consists of several critical steps, including dataset partitioning, batch processing, and iterative optimization. 
The dataset is divided into training, validation, and test sets in a 70:20:10 ratio to ensure comprehensive model evaluation and avoid overfitting. 
Further, the data is segmented into smaller batches of 128 samples, enabling efficient computational handling. 
During training, the network undergoes iterative forward and backward propagation. 
In the forward pass, input data is processed through the network to produce predictions, which are then evaluated against ground-truth labels using a predefined loss function. 
Backpropagation follows, where the loss function's gradients are computed and utilized to adjust the model's weights and biases through a stochastic gradient descent method like the adaptive moment estimation algorithm \citep{2014arXiv1412.6980K}. 
This iterative refinement process systematically improves the model's predictive accuracy and overall performance\footnote{
We utilized computers equipped with two Intel\textsuperscript{\textregistered} Xeon\textsuperscript{\textregistered} Processor E5 v4 Family CPUs, providing a total of 20 cores (hyperthreaded to 40) and 512 GB of RAM to optimize VariLens. 
The training process took approximately 8 hours to complete without utilizing GPU acceleration. 
During the inference phase, each source required only a few milliseconds to compute, yielding the predicted lens probability and SIE+$\gamma_\text{ext}$ parameters.
}.

In the second phase, we utilize the pretrained VariLens encoder as the base model for a downstream classification task.
This is accomplished by adding a linear classifier, consisting of a single dense layer with one neuron and a sigmoid activation function, on top of the encoder. 
This dense layer serves as the classification head, tasked with differentiating between lens and non-lens categories through the use of a binary cross-entropy loss function.
The process outlined above, known as transfer learning, comprises two main steps. 
First, all weights and biases in the pretrained encoder are frozen, and only the classification head is trained starting with a learning rate of \(10^{-4}\). 
After optimizing the classification head, we proceed with fine-tuning the entire model by unfreezing all trainable parameters in both the encoder and the classification head, using a reduced initial learning rate of \(10^{-5}\).

Throughout the training process, a learning rate scheduler is used to decrease the learning rate by a factor of 0.2 if the model's performance does not improve over five consecutive epochs \citep[e.g.,][]{2023ApJ...943..150A}. 
Additionally, early stopping is implemented to halt training if the loss does not decrease by at least \(10^{-4}\) over 10 consecutive epochs. 
Usually, the training process requires between 100 and 200 epochs to converge, after which the optimized parameters are saved. 
Ultimately, the optimal model is determined by the set of weights and bias parameters that produce the minimum loss on the validation dataset.

\begin{figure*}[htb!]
	\centering
	\resizebox{0.45\hsize}{!}{\includegraphics{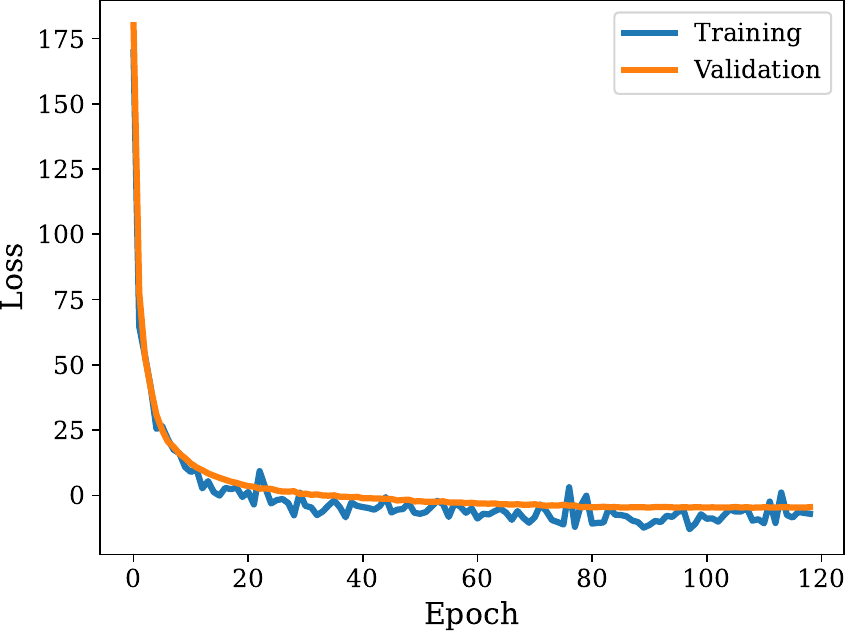}}
	\hspace{0.05\hsize}
	\resizebox{0.45\hsize}{!}{\includegraphics{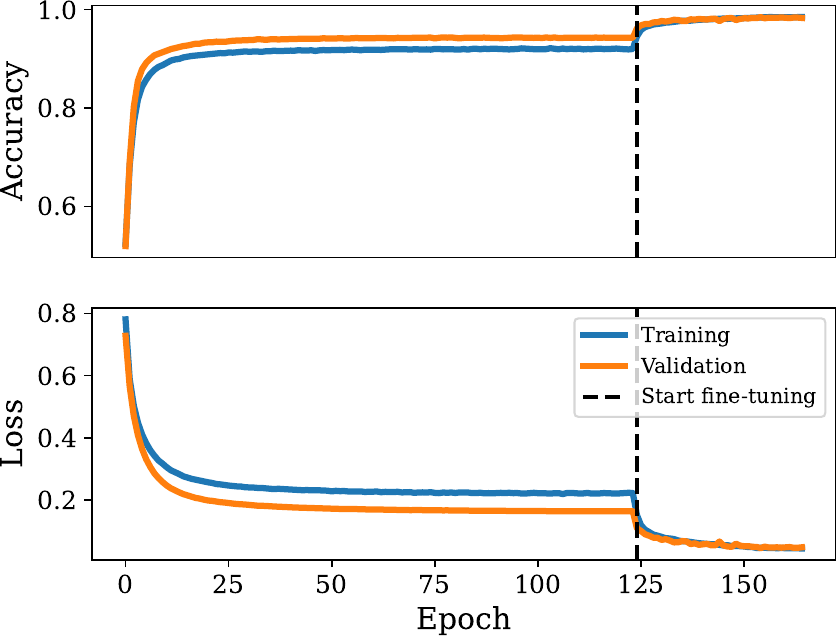}}
	\caption{
		Loss and accuracy curves over training epochs for the VariLens model.
		The left panel shows optimizations for the physics-informed VAE, while the right panel displays the one for the classifier module.
		The metrics, evaluated on the training and validation datasets, are depicted by the blue and orange lines, respectively.
		For the classifier, the dashed black line indicates the epoch when network fine-tuning begins.
		Prior to fine-tuning, the VariLens classifier appeared to perform worse on the training dataset compared to the validation samples. 
		This could be due to the model not having fully converged or found the optimal weights at that stage.
	}
	\label{fig:accuracy_loss}
\end{figure*}

A typical approach to assess the network's learning capability is by analyzing the loss and accuracy curves over training epochs, as shown in Figure~\ref{fig:accuracy_loss}. 
In the case of VariLens optimization, both training and validation losses decrease and stabilize, following similar trends after several epochs. 
The absence of overfitting signatures -- for example, the lack of a continued decline in training loss with a concurrent rise in validation loss -- demonstrates that our networks generalize effectively.
We note that the metrics discussed are not the only indicators of model performance. 
In the section that follows, we will present additional metrics that provide a more detailed evaluation of our model's effectiveness in regression and classification tasks.

\section{Results and discussion} \label{sec:result}

\subsection{Image reconstruction and latent space exploration}

\begin{figure}[htb!]
	\centering
	\resizebox{\hsize}{!}{\includegraphics{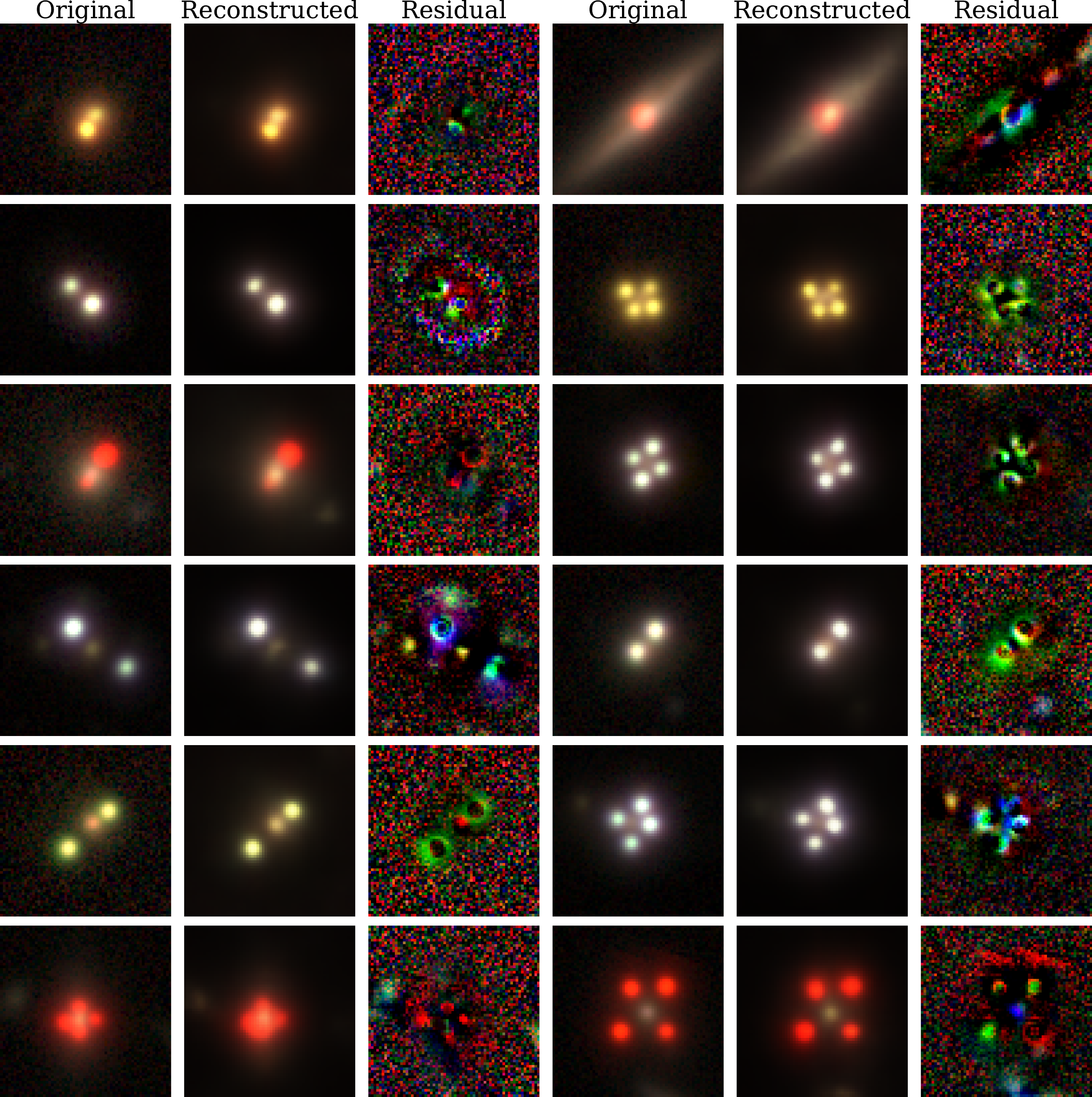}}
	\caption{
		Image reconstruction examples. 
		The original data containing mock lensed quasars, the reconstructed images predicted by VariLens, and the difference between them (i.e., residuals) are shown. 
		These HSC $grz$ images originate from sources within the test dataset.
	}
	\label{fig:reconstruction}
\end{figure}

As mentioned earlier, a key application of VariLens is its ability to perform image reconstruction. 
This process takes 5-band HSC images as inputs and effectively reconstructs them, reducing unnecessary details while preserving essential information.
An illustration of this process is presented in Figure~\ref{fig:reconstruction}, where the decoder showcases good performance in reconstructing the input mock lensed quasars.
A potential direction for the future development of VariLens is to upgrade the decoder for tasks beyond image reconstruction, including deblending foreground lenses and background sources, as well as noise reduction \citep[e.g.,][]{2019arXiv191103867M,2021MNRAS.500..531A}. 

We then conduct unsupervised explorations of the latent space produced by VariLens to assess its potential for discriminating lensed quasars from non-lensed sources. 
This is achieved by employing two robust dimensionality reduction techniques.
The first one is principal component analysis (PCA), which reduces the high-dimensional data onto a lower-dimensional space by identifying principal components -- that is, the directions of maximum variance. 
Each principal component's importance is quantified by its eigenvalue, which indicates the proportion of the total variance captured along that component. 
In other words, by analyzing these eigenvalues, we can assess how much of the overall data variance is explained by the selected components. 
This method effectively reveals underlying patterns and main axes of variation in the data. 
However, it is important to note that PCA relies on linear assumptions and may not effectively capture more complex, nonlinear relationships within the data.

\begin{figure}[htb!]
	\centering
	\resizebox{\hsize}{!}{\includegraphics{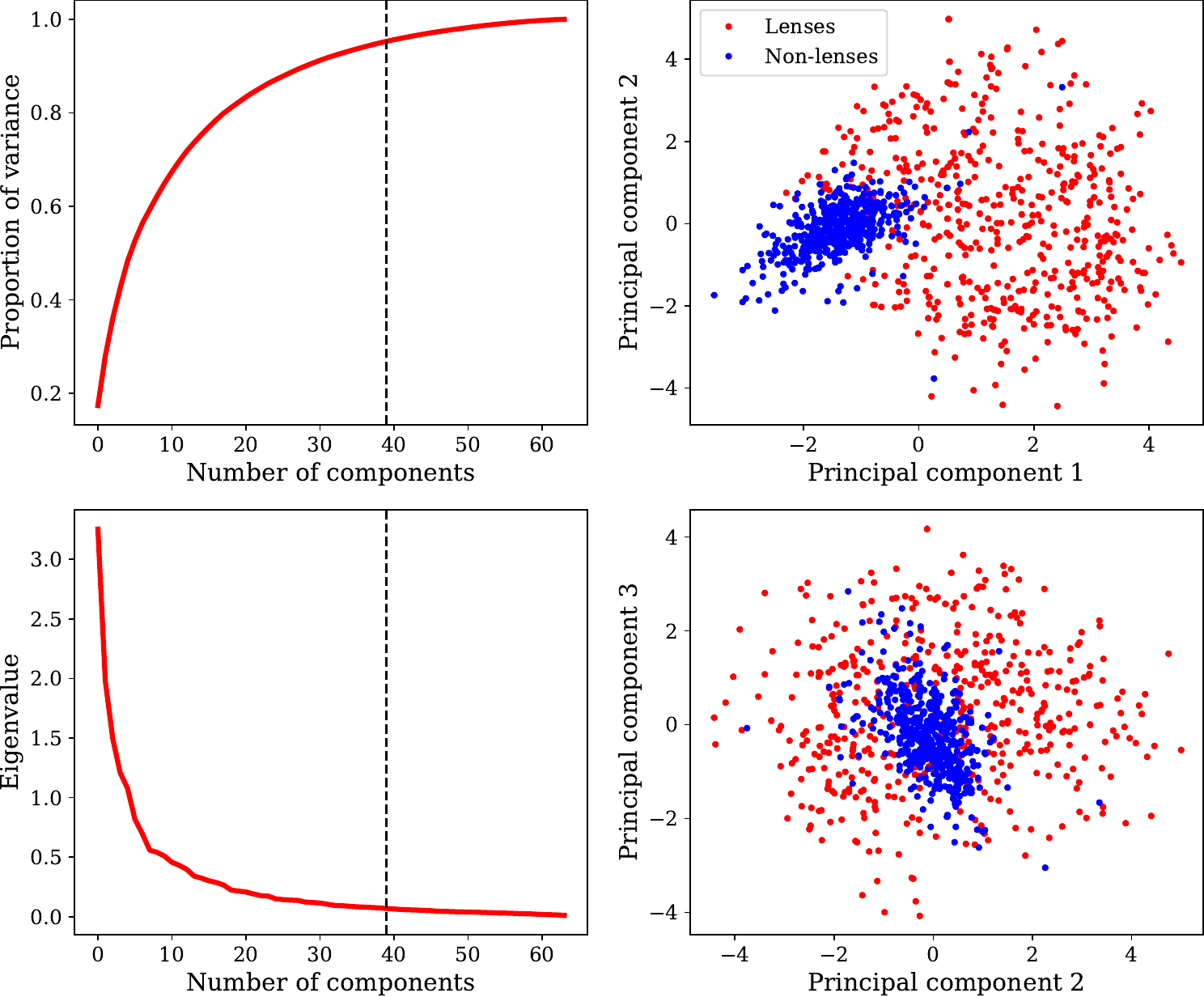}}
	\caption{
		Visualization of a subset of the data, including mock lensed quasars and contaminants, after dimensionality reduction using principal component analysis (PCA). 
		The vertical dashed black lines in the left panels mark the number of principal components required to explain 95\% of the variance and the corresponding eigenvalues. 
		The right panels show the first three PCA components, which reveal minimal separability between lenses (red) and non-lenses (blue).
	}
	\label{fig:pca}
\end{figure}

\begin{figure}[htb!]
	\centering
	\resizebox{\hsize}{!}{\includegraphics{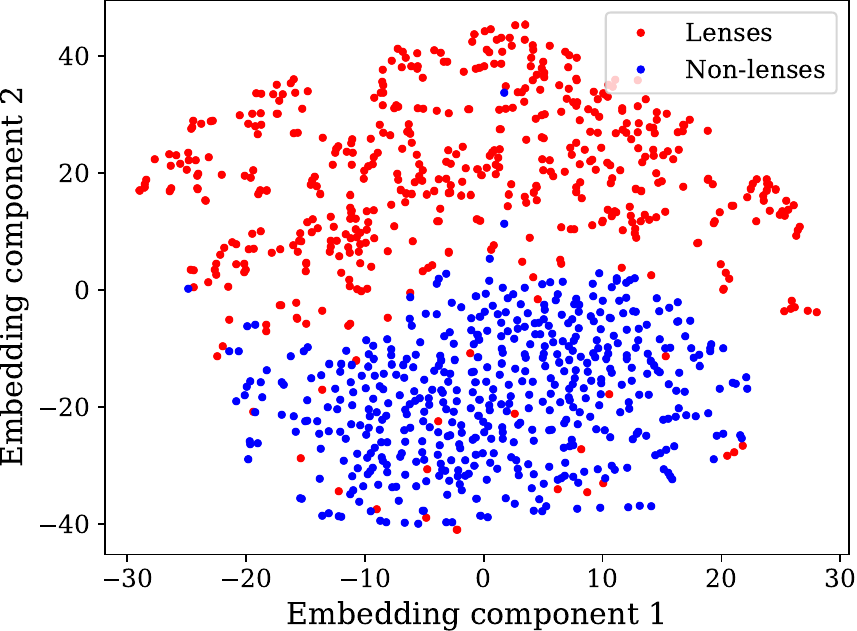}}
	\caption{
		Dimensionality reduction using two-dimensional t-distributed Stochastic Neighbor Embedding (t-SNE) on the encoded data. 
		Projecting the high-dimensional data into two embedding components allows t-SNE to effectively visualize distinct clusters, with lensed quasars (red) and non-lensed sources (blue) showing clear separation.
	}
	\label{fig:tsne}
\end{figure}

The second technique we employ is t-distributed Stochastic Neighbor Embedding (t-SNE), a nonlinear dimensionality reduction method designed to maintain local structures within the data.
This approach focuses on maintaining pairwise distances between data points, ensuring that similar points stay close together and dissimilar points are separated. 
Hence, t-SNE is especially effective for visualizing complex clusters and nonlinearly separable relationships. 
The results of the PCA and t-SNE visualizations are presented in Figures~\ref{fig:pca}~and~\ref{fig:tsne}, respectively.
While PCA provides little evidence of separability between lensed and non-lensed sources, the two-dimensional t-SNE indicates that these classes are reasonably distinguishable.
This difference can be attributed to the linear nature of PCA, which may overlook nonlinear relations. 
In contrast, the nonlinear approach of t-SNE effectively captures and visualizes complex relationships, providing a more accurate description of class separation.
This indicates that, although the VariLens encoder was initially trained on a dataset of only mock lenses, it effectively learns to organize the latent space in a way that distinguishes between lensed and non-lensed quasars.

\subsection{Fine-tuned classifier} \label{sec:classifier}
The classifier module of VariLens provides probability estimates given the input data. 
A probability \( P_\mathrm{lens} = 1 \) signifies a high likelihood that the image features a lensed quasar. 
Conversely, \( P_\mathrm{lens} = 0 \) suggests that the image is unlikely to contain a lens and is more likely to include other types of sources, such as a ordinary galaxy, quasar, star, or even a photometric artifact.

To evaluate the overall performance of the trained model, we use the receiver operating characteristic (ROC) curve. 
The area under the ROC curve (AUROC) indicates the model's ability to distinguish between two classes as the decision threshold is adjusted. 
In this assessment, lensed quasars are designated as positives (P), while non-lenses and other contaminating sources are defined as negatives (N). 
True positives (TP) represent instances where the model correctly identifies lensed quasars, whereas true negatives (TN) are cases where non-lenses are accurately recognized. 
False positives (FP) occur when the model mistakenly classifies contaminants as lensed quasars, and false negatives (FN) are instances where the model fails to detect lensed quasars. 
The ROC curve visualizes the trade-off between the false-positive rate (FPR) and the true-positive rate (TPR) across different thresholds for the test dataset, where:
\begin{equation}
	\mathrm{
		TPR = \frac{TP}{P} = \frac{TP}{TP + FN}
	},\quad \text{and} \quad
	\mathrm{
		FPR = \frac{FP}{N} = \frac{FP}{FP + TN}
	}.
\end{equation}

\begin{figure}[htb!]
	\centering
	\resizebox{\hsize}{!}{\includegraphics{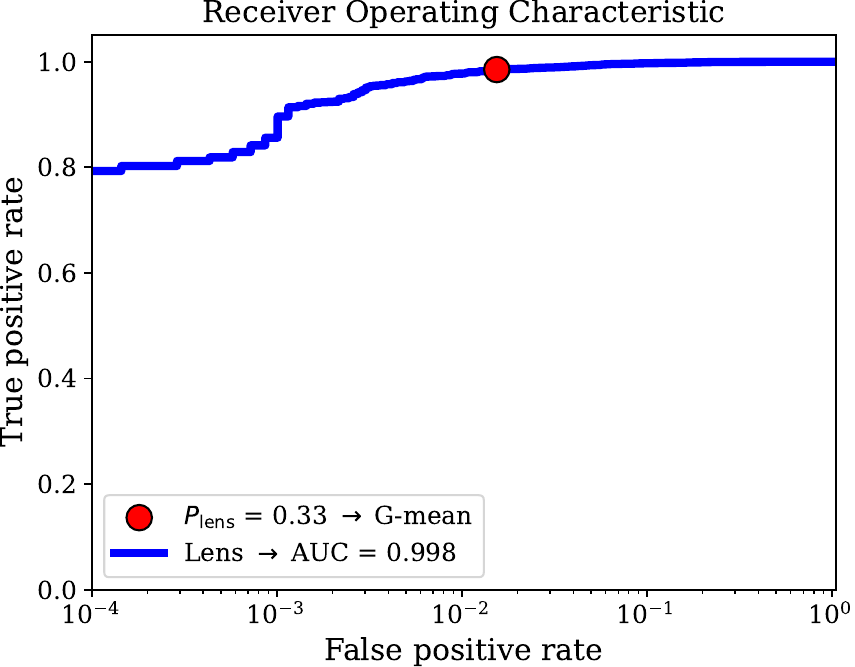}}
	\caption{
		Receiver operating characteristic (ROC) curve, depicted with a blue line, along with the corresponding area under the curve (AUC).
		The false positive rate and true positive rate at the chosen $P_\mathrm{lens}$ threshold are marked with a red circle.
	}
	\label{fig:roc_curve}
\end{figure}

The ROC curve is generated by progressively increasing the probability threshold. 
A perfect classifier achieves an AUROC of 1, indicating flawless performance, while a random classifier yields an AUROC of 0.5. 
Our findings reveal an ROC curve with high AUROC values, demonstrating the classifier's outstanding accuracy.
To balance the TPR and FPR as depicted by the ROC curve, we use the geometric mean (G-mean)\footnote{
	The G-mean can be expressed as $\text{G-mean} = \sqrt{\text{TPR} \times (1 - \text{FPR})}$.
}. 
The optimal G-mean score determines the best $P_\mathrm{lens}$ threshold that balances TPR and FPR, maximizing true positives while minimizing false positives. 
As illustrated in Figure~\ref{fig:roc_curve}, the optimal threshold for $P_\mathrm{lens}$ in our model is 0.33, which corresponds to FPR and TPR values of 2\% and 98\%, respectively. 
It is crucial to note that lowering the $P_\mathrm{lens}$ threshold below the optimal value significantly increases the number of candidates but at the cost of reduced quality, making visual inspection more challenging and less efficient. 
Therefore, it is essential to strike a balance between completeness and purity to ensure a manageable number of candidates for further observations.

\subsection{Selection completeness}

\begin{figure}[htb!]
	\centering
	\centering
	\resizebox{\hsize}{!}{\includegraphics{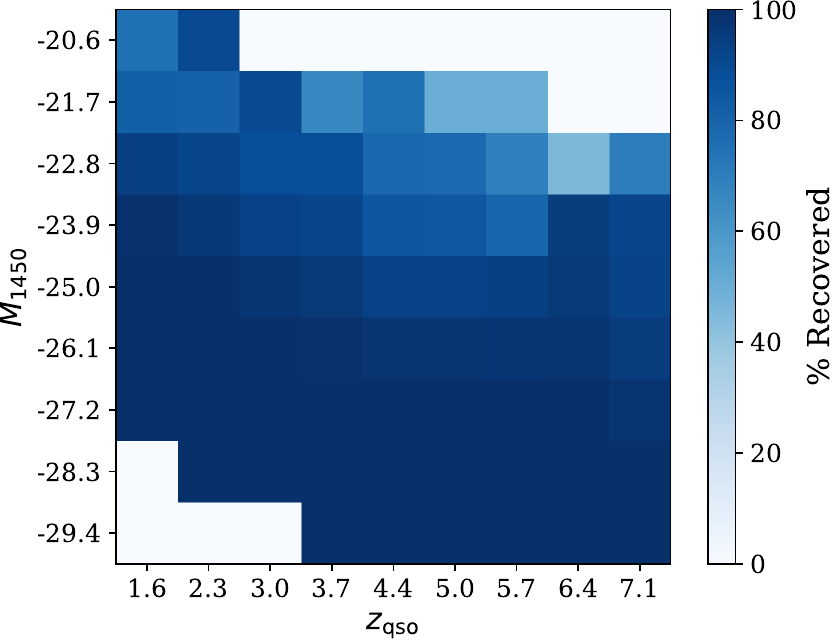}}
	\caption{
		Quasar selection function adopted in this work. 
		The recovery rate percentage indicates the fraction of mock quasars within each ($M_{1450}$, $z$) bin that our classifier successfully recognizes.
	}
	\label{fig:completeness}
\end{figure}

As previously discussed, the quasar samples in our simulation are uniformly distributed over a redshift range of \(1.5 \leq z \leq 7.2\) and an absolute magnitude range of \(-30 \leq M_{1450} \leq -20\). 
Without the amplification provided by strong lensing, our classifier is capable of detecting only those quasars with \(M_{1450} \lesssim -22\) at redshifts \(z \gtrsim 6\).
However, lensing can push this detection threshold to lower luminosities, depending on the magnification factor.	
To explore this further, we quantify our selection function (or completeness) as the fraction of mock quasars with specified values of $M_{1450}$, $z$, and spectral energy distributions (SEDs) that meet our selection criteria. 
The corresponding results are presented in Figure~\ref{fig:completeness}. Quasars with lower intrinsic brightness are only detectable if they undergo substantial magnification, which is relatively uncommon. 
Consequently, our completeness rate decreases for quasars with lower luminosity at any given redshift. 
Moreover, as discussed in Section~\ref{sec:lens_simulation}, we exclude sources with $y \leq 15$~mag to avoid unusually bright objects or saturated images. 
As a result, we do not recover extremely luminous quasars with $M_{1450} \lesssim -27$ at lower redshifts ($z \lesssim 3$), although such quasars may not even exist in reality.

As illustrated in Figure~\ref{fig:lensgal_dist}, our classifier successfully recovers the majority of lensing configurations with minimal bias.
However, a slight reduction in the recovery rate is noticeable for systems where the deflector redshifts are $z_\mathrm{gal} \gtrsim 1$ and magnitudes are $i \gtrsim 21$. 
This decline is likely due to the fainter apparent fluxes of more distant lens galaxies, which makes them more challenging to detect. 
While these sources can still be identified when the lensed quasars are well separated, the detection becomes more challenging for compact lens systems with smaller masses. 
For example, Figure~\ref{fig:lensgal_dist} indicates a modest decrease in the recovery rate for lenses with $\sigma_v \lesssim 100$~km~s$^{-1}$ and $\theta_\mathrm{E} \lesssim 0\farcs5$.
This is expected, as the HSC data is constrained by seeing limits of $\approx0\farcs6$ \citep{2022PASJ...74..247A}. 
Consequently, any lensing configuration with $\theta_\mathrm{E}$ smaller than this threshold would be difficult to resolve, complicating the inference of any lensing parameters.

\subsection{Inference of lens parameters} \label{sec:regression_inference}

\begin{figure*}[htb!]
	\centering
	\resizebox{\hsize}{!}{\includegraphics{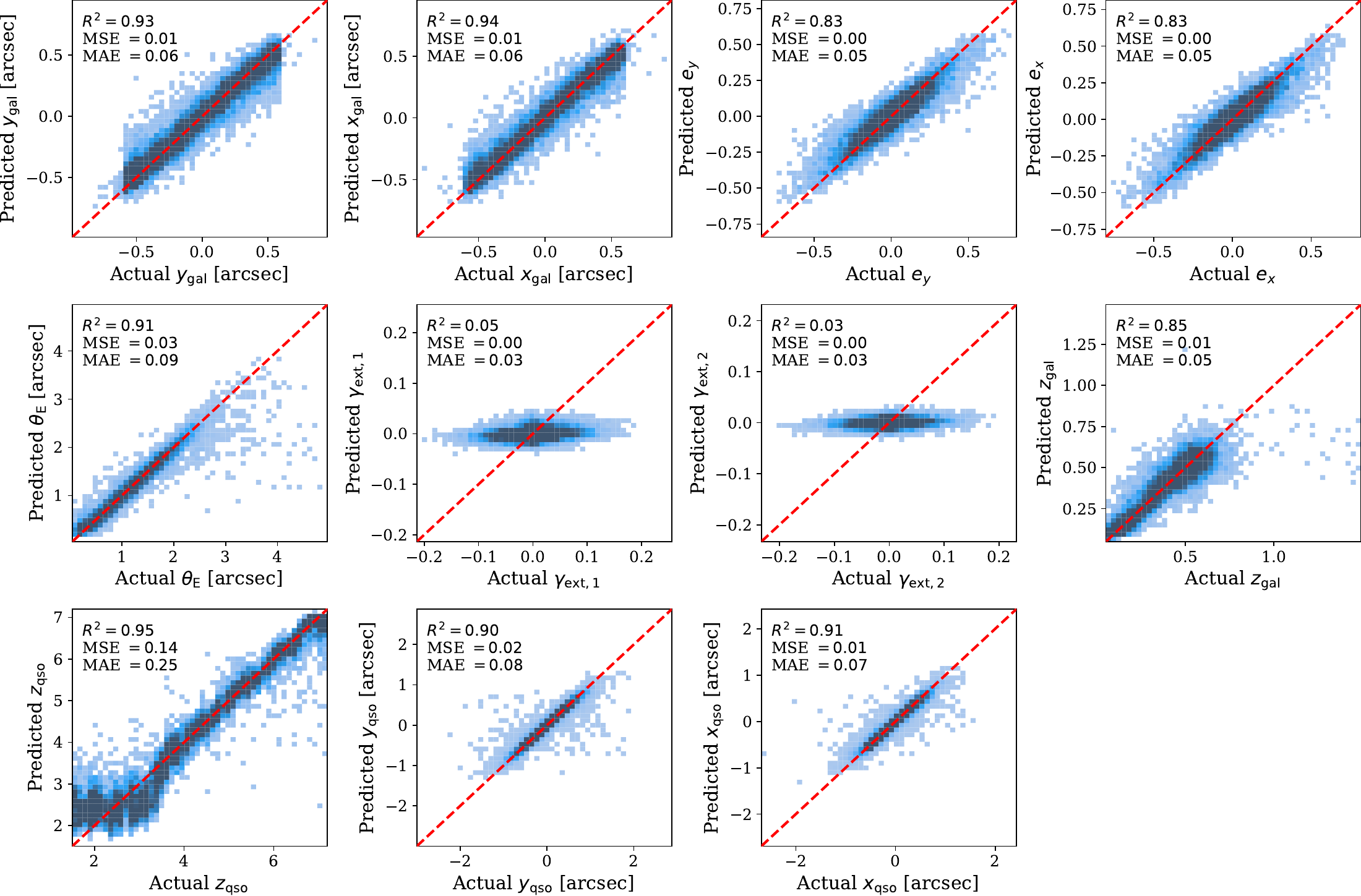}}
	\caption{
		Comparison of true and predicted physical parameters.
		The metrics $R^2$, MSE, and MAE for sources in the test dataset are displayed in each panel.
		Darker regions indicate higher concentrations of sources within the 2D histogram bins. 
		The red dashed line represents the 1:1 relationship between the predicted and actual values.
	}
	\label{fig:regression}
\end{figure*}

We evaluate here the performance of VariLens in determining the SIE+$\gamma_\text{ext}$ lens parameters, redshifts, and positions of the quasars.
Figure~\ref{fig:regression} illustrates the comparison between predicted and actual values on the test dataset, featuring images that the network has not seen during training.
For each parameter, we present the distribution of values using 2D colored histograms. Additionally, we assess the model's quality using three key metrics: the coefficient of determination ($R^2$), mean squared error (MSE), and mean absolute error (MAE). 
The definitions and calculations of these metrics can be found in Appendix~\ref{sec:regression_metrics}. 
Specifically, $R^2$ indicates the proportion of variance in the dependent variable that can be explained by the independent variables, with values approaching 1 reflecting exceptional model accuracy. 
MSE measures the average of the squared differences between predicted and actual values, highlighting larger discrepancies. 
In contrast, MAE represents the average of the absolute differences, offering a more straightforward measure of prediction accuracy.
Both MSE and MAE are lower for more accurate models, with MAE being less sensitive to outliers compared to MSE.

As illustrated in Figure~\ref{fig:regression}, the network performs well in predicting most parameters, with the exception of the external shear. 
Predictions for the lens centers and complex ellipticities closely match the true values, showing no significant bias to either underestimation or overestimation.
While the $R^2$ scores for ellipticities are slightly lower than those for lens centers, this difference, although not substantial, indicates that measuring galaxy shapes may be more challenging than determining their positions within ground-based images.
Nonetheless, the high $R^2$ scores (\(\gtrsim 0.8\)), along with MSE and MAE values near zero, demonstrate that VariLens accurately models the observed galaxies.

Regarding the Einstein radii, VariLens provides accurate estimates up to $\theta_\mathrm{E} \approx 2$\arcsec.
However, its performance declines beyond this range, likely due to the low amount of systems in the training dataset, leading to the underrepresentation of lenses with larger $\theta_\mathrm{E}$.
As larger Einstein radii correspond to greater lens mass, this population of massive lensing galaxies is rare in the Universe.
A similar issue is observed in the redshift predictions for deflector galaxies. 
The network performs well up to $z_\text{gal} \approx 0.7$ but struggles with higher redshifts. 
This decline in performance is due to the SDSS instrument's limiting magnitude, which allows spectroscopic observations only for brighter galaxies at high redshifts, resulting in the exclusion of fainter deflectors (see also Section~\ref{sec:galaxy_quasar}). 
Consequently, the number of samples at $z_\text{gal} \gtrsim 0.7$ is significantly reduced, leading to the underrepresentation of high-$z$ deflectors in the training dataset.
One potential solution to address these issues is to resample the lenses to achieve a uniform distribution of $\theta_\mathrm{E}$ and $z_\mathrm{gal}$, rather than relying on their natural distribution, as suggested in other studies \citep[e.g.,][]{2021A&A...646A.126S}. 
This approach would allow the network to be exposed to the entire spectrum of parameter values, thereby mitigating the effects of selection bias. 
However, we did not implement this correction in our current study and will consider it for future work.

A key innovation of VariLens is the incorporation of both source (quasar) redshift and positional measurements. 
While the network recovers source positions with high accuracy, there is notable scatter in redshift estimations for sources at \( z \lesssim 3 \).
The accuracy of redshift estimates improves significantly at higher redshifts. 
Notably, at \( z \approx 3 \), the Ly\(\alpha\) emission is redshifted to an observed wavelength of \( \approx 4864 \) Å, while the mid-wavelength of the HSC \( g \)-band is \( \approx 4798 \) Å. 
As the quasar's Ly\(\alpha\) emission moves out of the \( g \)-band and Ly$\alpha$ forest absorption becomes more pronounced, a phenomenon known as \( g \)-band dropout occurs -- that is, the quasar appears very faint in the \( g \)-band while only bright enough in the redder bands. 
At even higher redshifts, this trend extends to producing dropouts in the \( r \), \( i \), and \( z \) filters, a phenomenon known as the Lyman break.
The ability of VariLens to infer such a Lyman-break effect enhances its redshift estimation accuracy, demonstrating the network's effectiveness in learning that complex phenomenon.

VariLens currently struggles to correctly estimate the external shear components (\(\gamma_\text{ext,1}\), \(\gamma_\text{ext,2}\)), where it always predicts values close to zero. 
It seems that minor distortions caused by the external shear are not well captured in the HSC ground-based images, likely due to PSF variations across different lens systems being comparable to the shear effect. 
This challenge is also noted by \cite{2023A&A...671A.147S}, who faced similar difficulties in predicting shear in HSC data. 
Accurate shear measurement appears to be feasible only in more idealized datasets, where higher image resolution, stable PSFs, and prior lens light subtraction improve results \citep[e.g.,][]{2018arXiv180800011M}. 
It is worth noting that the need for accurate external shear estimation depends on the specific scientific goals, as it has a negligible impact on the statistical studies of lens mass and quasar emissions conducted here.

Comparing the performance of VariLens to other modeling networks is challenging due to differences in underlying assumptions. 
While most previous studies, as discussed below, primarily focus on galaxy-galaxy lenses, our work complements them by specifically targeting lensed quasar cases.
Nevertheless, we briefly summarize here the approaches taken in previous studies.
The concept of harnessing CNNs to estimate SIE parameters such as \( e_\text{x} \), \( e_\text{y} \), and \( \theta_\text{E} \) was initially raised by \citet{2017Natur.548..555H}, who focused on mock datasets resembling those captured by the Hubble Space Telescope (HST). 
Their methodology involved subtracting the lens light prior to applying CNNs for lens parameter estimation.
Further advancements on this idea were made by \citet{2017ApJ...850L...7P}, who presented estimations of uncertainty and external shear components. 
Later, \cite{2021MNRAS.505.4362P} designed a model to deduce the lensing parameters, which they applied to simulated Euclid data with a resolution of 0\farcs1. 
Building on the work of \citet{2017Natur.548..555H}, they incorporated error estimation techniques and introduced a hybrid method that combines neural networks with traditional, non-machine learning-based models to refine parameter inference.

Variations in image resolution, type of bandpasses, quality of training datasets, and differing number of predicted parameters coupled with their intrinsic degeneracies make direct comparisons between the previous models with VariLens's performance challenging.
\cite{2023A&A...671A.147S} presents the most similar work in terms of image properties, which employed a residual network (ResNet) to predict SIE+$\gamma_\text{ext}$ parameters for $griz$-bands of HSC data.
However, their work primarily addresses galaxy-galaxy lensing, which exploits the detailed shape of the lensed arc rather than relying solely on the source position, as is common in lensed quasar studies. 
We evaluated VariLens on their test dataset of 31 lensed galaxies, but the results were suboptimal, with significant uncertainties in our predictions. 
This outcome underscores the distinct challenges between modeling lensed galaxies and quasars, particularly the added complexity and information provided by lensed arcs in galaxy-galaxy lensing.

Additionally, considerable progress has been made in the domain of automated modeling methods that do not rely on machine learning \citep[e.g.,][]{2018MNRAS.478.4738N, 2022MNRAS.517.3275E, 2022A&A...668A..73R, 2022A&A...666A...1S, 2023A&A...672A...2E, 2022ApJ...935...49G, 2023MNRAS.518.1260S,
2023A&A...673A..33S}. 
These approaches generally surpass neural networks in accuracy and precision but are substantially more time-intensive, often requiring processing durations ranging from several hours to weeks.
In contrast, the inference time of VariLens, along with other previously mentioned neural network-based methods, can be reduced to a fraction of a second, making them significantly more efficient than traditional lens modeling.

\subsection{Evaluation with known lensed quasars} \label{sec:known_lenses_eval}

\subsubsection{Classifier performance}

\begin{figure*}[htb!]
	\centering
	\centering
	\resizebox{\hsize}{!}{\includegraphics{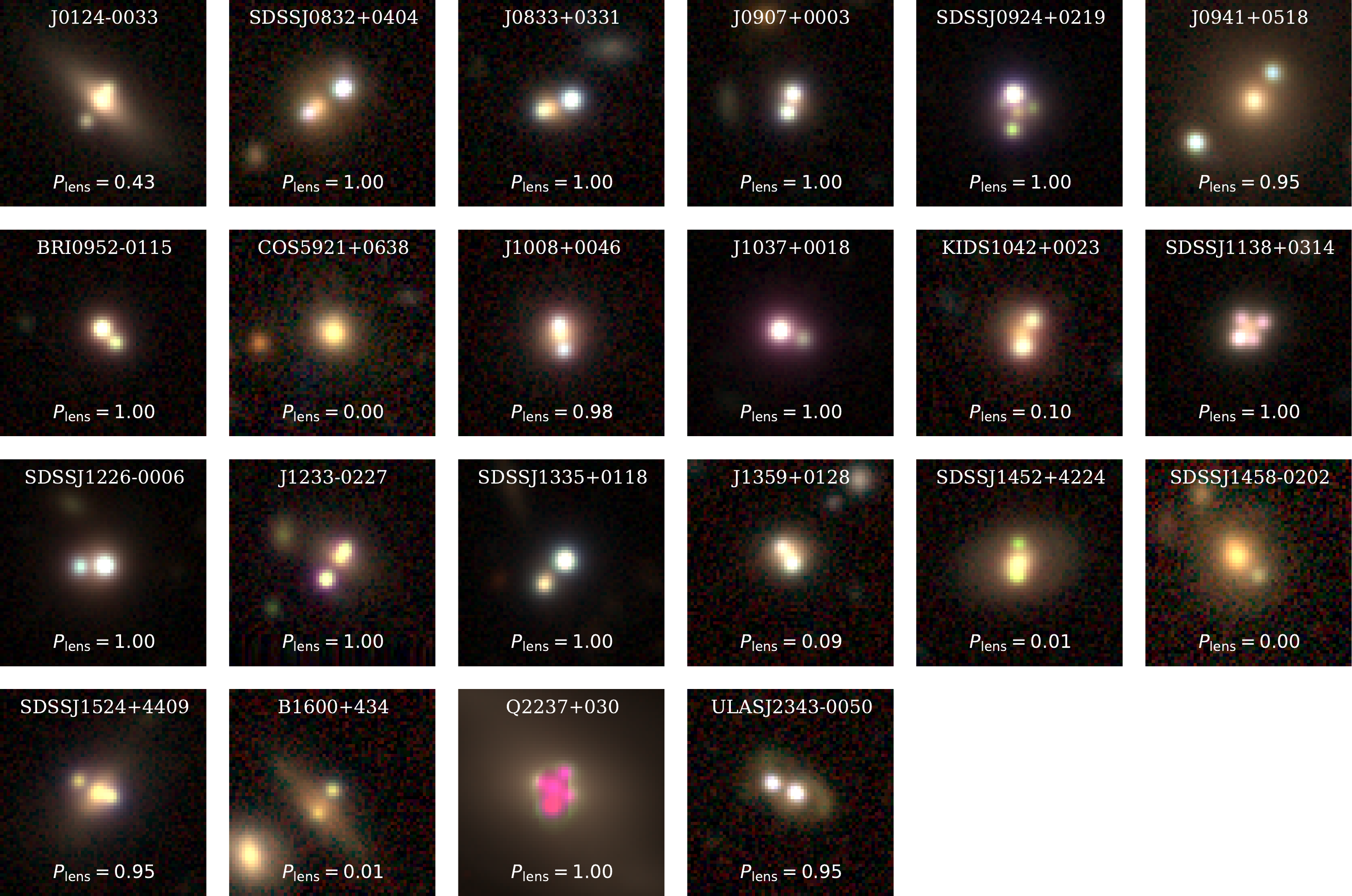}}
	\caption{
		Lensed quasar samples compiled from GLQD (refer to the main text). 
		The images are generated from HSC $grz$-band cutouts with dimensions of $64\times64$ pixels (approximately 10\farcs8 on each side), colorized, and stretched using a square-root scale. 
		The lens probabilities, determined by the VariLens classifier module for each system, are also displayed in the figure.
	}
	\label{fig:CLemon_test}
\end{figure*}

We conduct further examination using a supplementary dataset containing confirmed lensed quasars provided by the GLQD \citep{1992Gemin..36....1M,1995MNRAS.274L..25J,2003AJ....126..666I,2004PASJ...56..399O,2007AJ....133..214M,2008AJ....135..520O,2008AJ....135..496I,2008MNRAS.387..741J,2012AJ....143..119I,2015MNRAS.448.1446A,2016MNRAS.456.1595M,2018MNRAS.475.2086A,2018MNRAS.477L..70W,2018MNRAS.480.1163S,2018MNRAS.479.5060L,2019MNRAS.483.4242L,2019arXiv191208977K,2020A&A...636A..87C,2020MNRAS.494.3491L,2021MNRAS.502.1487J,2023MNRAS.520.3305L}.

HSC images are accessible for 22 of the 220 lenses cataloged in the database (see Figure~\ref{fig:CLemon_test}). 
To evaluate the completeness and purity of our classifier, we combine these confirmed lenses with a sample of contaminants outlined in Section~\ref{sec:hsc_data}, consisting of galaxies, stars, and quasars.
The non-lensed sources are anticipated to vastly outnumber the lens population by several thousand. 
Consequently, our model successfully identifies 16 of the known lensed quasars, resulting in a TPR (or completeness) of 73\% and an FPR of 5\%. 
Furthermore, we determine that the classifier achieves a purity\footnote{
	Assuming the proportion of lens systems to other sources in the Universe is approximately $\rm S=10^{-3}$, or 1 lens per 1000 objects, we express purity as $\rm AP = TPR \times S / (TPR \times S + FPR \times (1-S))$.
} of $\approx 1$\% in detecting lens candidates.
There are several possible reasons why VariLens did not recover these known lensed quasars. In the case of COS5921+0638, the multiple images of the quasars are too faint to be detected by eyes unless the lens galaxy light is removed. 
Similarly, for SDSSJ1458-0202 and B1600+434, the quasar counterimages are also not easily visible. 
In the case of SDSSJ1452+4224, the colors of the quasar
images appear slightly different, possibly due to reddening caused by dust in the foreground galaxy.
Factors like this, among others, may influence the performance of our model.
For KIDS1042+0023 and J1359+0128 the reasons behind the network's failure to detect them are less clear. 
Further investigation is needed to better understand the network's behavior, possibly using explainable artificial intelligence tools  \citep[e.g.,][]{2016arXiv161002391S,2013arXiv1312.6034S,2017arXiv170301365S}.

The VariLens classifier discussed in Section~\ref{sec:classifier} demonstrates outstanding performance on the test dataset, with almost perfect AUROC and low FPR. 
Despite this, the classifier encounters difficulties in generalizing to real-world data, as indicated by the numerous spurious sources detected by the network. 
This limitation significantly increases the time required for the necessary visual inspection of lens candidates during later stages. 
A promising approach to mitigate this issue is probably to combine multiple networks with diverse architectures through model averaging, which could greatly reduce the number of candidates while maintaining high completeness \citep[e.g.,][]{2023A&A...678A.103A,2023arXiv230603136C,2024MNRAS.533..525M}.

\subsubsection{Regression performance}

\begin{figure*}[htb!]
	\centering
	\resizebox{\hsize}{!}{\includegraphics{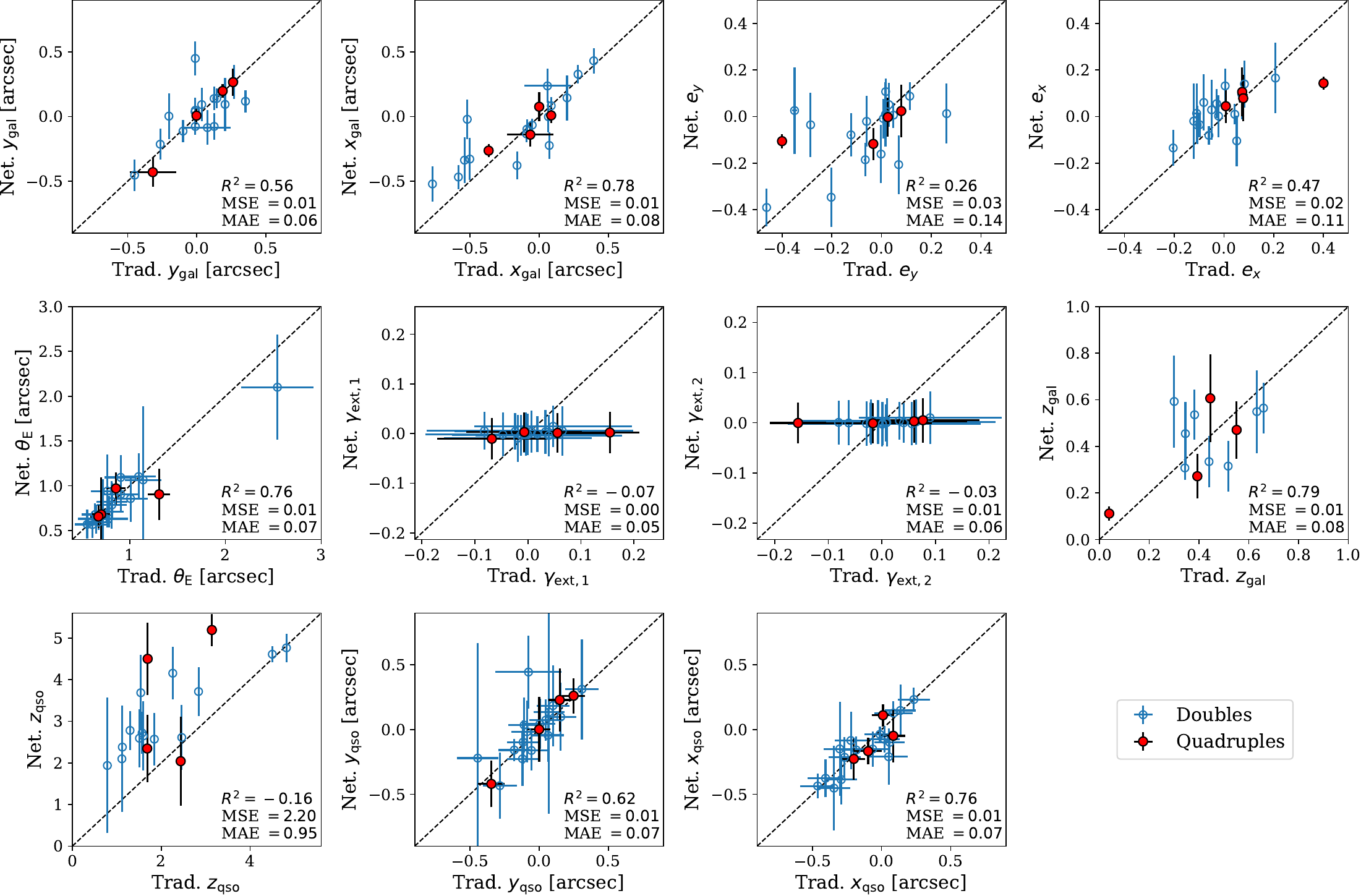}}
	\caption{
		Comparison of the derived SIE+$\gamma_\text{ext}$ parameters from traditional lens modeling using \texttt{PyAutoLens} (labeled ``Trad.'') and the deep learning-based approach with VariLens (``Net.'').
		Quadruply lensed quasars are indicated by red circles, while doubly lensed quasars are marked with blue circles.
		We also compare the redshifts inferred by VariLens with spectroscopic measurements.
	}
	\label{fig:model_comparison}
\end{figure*}

To evaluate the performance of our regression module compared to traditional, non-machine learning-based methods, we will utilize \texttt{PyAutoLens} for modeling the lensed quasars from GLQD. 
This process involves two stages: (1) performing light deblending to accurately determine the locations of the lensed source and foreground deflector in the image plane, along with the relevant morphological parameters of the deflector galaxy, and (2) modeling the lens system with an SIE+$\gamma_\text{ext}$ mass approximation.

While quasar emission is typically modeled as a point source, the PSF provided by the HSC database may exhibit slight off-centering or higher-order distortions, albeit minimal \citep{2024MNRAS.527.6253C}. 
To address this issue, we approximate the PSF using an elliptical Moffat profile \citep{1969A&A.....3..455M}.
The number of required point source models is first manually determined by visually inspecting the multiband images of each lensed quasar.
These point sources will have different positions and fluxes, but we restrict that their Moffat shapes should be the same within the observed image.

After that, we employ  a S\'{e}rsic profile to fit the lens galaxy's light \citep{1963BAAA....6...41S}, assuming each system only contains one deflector galaxy.
To simplify the optimization and prevent parameter degeneracy, we fix the S\'{e}rsic index to 4 (i.e., de Vaucouleurs profile), which approximates the profile of typical elliptical galaxies.
For further details on these light profiles, we refer to Appendix~\ref{sec:lights}. 
Light profile fitting for each system is performed simultaneously across the $grizy$ bands, assuming their positions and intrinsic shapes do not vary across different wavelengths. 
The main objective of this step is to infer the structural parameters of the system.
Following the previous step, we perform SIE+$\gamma_\text{ext}$ lens modeling for the lensed quasars. 
To streamline the process, we fix the ellipticity and center of mass to match those of the galaxy's light profile. 
Optimization is achieved by minimizing the $\chi^2$ statistics, comparing the predicted and observed positions of the sources \citep{2010PASJ...62.1017O, 2020A&A...636A..87C, 2015ApJ...807..138C}.
It is important to note that the number of data points for doubly imaged quasars in point-source lens modeling is insufficient to constrain the SIE+$\gamma_\text{ext}$ parameters, resulting in fitting degeneracy. 
Nonetheless, we adopt this approach for consistency in matching the output parameters predicted by VariLens.

The parameters derived from traditional lens modeling with \texttt{PyAutoLens} and those predicted by VariLens are reported in the tables provided in Appendix~\ref{sec:glq_model}.
This comparison is further illustrated in Figure~\ref{fig:model_comparison}.
Overall, key features like the Einstein radius, lens center, and source position are well-recovered, as these are relatively straightforward to infer directly from the multiband images. 
Consistent with tests on mock data, the Einstein radius estimated by VariLens closely matches that derived by the traditional approach for $\theta_\mathrm{E} \lesssim 2\arcsec$. 
While our network significantly underestimates the Einstein radius for only one object with a large value, the remaining estimates fall within $2\sigma$ uncertainties. 
Hence, despite using simplistic light models and priors to deblend the foreground galaxy and background quasar, our traditional lens modeling method still produced a good result for determining the positions of both the deflector and the source.
However, the simplicity of the \texttt{PyAutoLens} model may have led to imperfect deblending and inaccuracies in the estimation of morphological parameters, resulting in significant scatter in the predicted ellipticities of the lens galaxy.
Improving this strategy might require higher-resolution data, which would allow us to relax the priors while avoiding fitting degeneracies.
In addition, since VariLens is unable to accurately predict the external shear components, it becomes hard to compare its results with those obtained from traditional modeling.
Nevertheless, the majority of the inferred parameters are consistent with each other within $2\sigma$.

Our network provides adequate results in predicting the lens galaxy redshifts, although a noticeable scatter in $z_\mathrm{gal}$ is observed. 
On the other hand, the network lacks precision in predicting the redshifts of the lensed sources, with a systematic overestimation of $z_\mathrm{qso}$ being observed.
One explanation for this might be the same reason presented in Figure~\ref{fig:regression}. 
Our redshift predictions for the lensed quasars may only become accurate at $z \gtrsim 3$, where the Ly$\alpha$ forest absorption and break are more prominent, thereby improving the constraint on $z_\mathrm{qso}$.
However, most of the GLQD samples are located at lower redshifts, where estimating redshifts is challenging without distinct features such as the Lyman-break effect. 
Thus, acquiring more samples of high-$z$ lensed quasars will be critical for evaluating the accuracy of VariLens redshift estimates with real data.
Overall, VariLens and \texttt{PyAutoLens} provide relatively consistent SIE lens models and source positions. 
In the future, integrating both approaches to refine the lens models could enhance the fitting process and yield more accurate estimates.


\subsection{Lensed quasar candidates} \label{sec:lenscand}

\subsubsection{Catalog-level preselection and image-based analysis}

To identify lensed quasars, we begin with the approximately 80 million sources from our parent HSC catalog compiled in Section~\ref{sec:hsc_data}. 
Since most kpc-scale quasar pairs and lensed quasars typically have separations of $\lesssim3\arcsec$ \citep[e.g.,][]{2023AJ....165..191Y}, we narrow our selection to sources that indicate nearby companions within a 2\arcsec\ radius. 
We acknowledge that this criterion might be overly stringent. 
For example, among 22 optically bright lensed quasars with spectroscopic confirmation listed in the HSC catalog \citep{2024MNRAS.527.6253C}, our preselection process only identifies 15, resulting in a recovery rate of 68\%. 
The seven missed quasars likely lack detectable neighboring sources due to their faintness or issues with the HSC object deblending process. 
For this reason, we also take into account the spectroscopic classification of candidates, ensuring that known galaxies and quasars from the literature are not excluded, as detailed in Section~\ref{sec:hsc_data}.
By selecting sources with a neighboring object within 2\arcsec\ or with available spectroscopic data, we effectively narrow the candidate pool to a few million sources.
This strategy reduces the number of candidates while retaining many known lenses, thereby minimizing contamination and computational requirements for subsequent analyses.

Subsequently, we apply a brightness cut across all HSC bands to ensure that each source is robustly detected with S/N levels of at least 5. 
Given our multiwavelength data, we then prioritize candidates that exhibit X-ray, IR, or radio detections.
Adding multiwavelength information is helpful in eliminating unwanted sources, such as cosmic rays or moving objects that appear in one survey but not in others \citep[e.g.,][]{2022PhDT.........1A}.
Following previous studies \citep[e.g.,][]{2021AandA...653L...6C,2022AandA...662A...4S}, we focus on foreground deflectors with an extended shape, specifically those with an $i$-band Kron radius > 0\farcs5. 
At this stage, we have identified 710\,966 candidates that have successfully passed our catalog-level preselection.

Next, we apply image-based analysis to evaluate the lens probability for each candidate using VariLens. 
This process yields 18\,300 candidates with $P_\mathrm{lens} > 0.3$.
To further refine our selection, we incorporate Gaia Data Release 3 \citep{2023A&A...674A...1G} catalog to filter out multiple stars that might mimic multiply imaged quasars.
In this case, we evaluate the astrometric excess noise \citep[AEN;][]{2016A&A...595A...1G} and proper motion significance \citep[PMSIG;][]{2019MNRAS.483.4242L} when available. 
Elevated AEN numbers ($>10$~mas) may indicate a probable star-forming galaxy, while high PMSIG values ($>10\sigma$) strongly suggest the presence of a star within the system \citep{2019MNRAS.483.4242L}.
It is important to note that only about 30\% of our candidates have these astrometric measurements. 
Nevertheless, we successfully narrow the selection to 13\,831 sources.

\subsubsection{Visual inspection}

For the final selection step, we perform a visual inspection of the remaining targets, following a four-step procedure similar to the method detailed in \cite{2024arXiv240520383S}. 
In the first step, all candidates are inspected by two people through binary classification, where any systems flagged as a potential lens by at least one reviewer advance to the next stage. 
A calibration round is then conducted in the second step using 200 systems, including a subset of known lenses and contaminants, to align reviewer expectations and facilitate discussions on specific cases. 
Here, each system is graded on a scale of $G=0$, 1, 2, or 3, corresponding to the classifications: not a lens, possible lens, probable lens, and definite lens. 
In the third step, four reviewers inspect all systems flagged as potential lenses in the initial round. 
Finally, in the fourth step, four additional reviewers inspect systems with an average grade of \(G_\mathrm{N} > 1\) or \(G_\mathrm{N} \leq 1\) but with a standard deviation of $\sigma_{G_\mathrm{N}} > 0.75$, based on the assessments in round three. 

To be specific, definite lenses ($G=3$)  are candidates that clearly exhibit multiple images, signs of a counter-image, a cross, or a complete Einstein ring.
In other words, these systems have an obvious lensing configuration that a model would fit easily and do not require higher-resolution imaging for confirmation. 
Probable lenses ($G=2$) are candidates displaying significant tangential distortion and a lensing-like configuration but require additional evidence. 
This is due to factors such as the absence of multiple images, arcs/images being clustered on one side of the galaxy, lack of signs of a counter-image, need for central galaxy removal, unusual lensing configurations, or the necessity for higher-resolution imaging to resolve uncertainties. 
These systems could also be mistaken for ring galaxies or spirals.
Possible lenses ($G=1$) are systems likely to consist of a single image with potential tangential flexion or distortion or a single arc situated far from the central galaxy.
Finally, non-lenses ($G=0$) are systems that lack convincing lensing features and are most likely single images with tangential flexion or distortion that could be mistaken for lensing features but are unlikely to be true lenses.
This includes ring galaxies, spirals, merging galaxies, irregular systems, companions, or other unrelated sources. 

Our visual examination results in 338 possible, probable, and definite lens candidates voted by eight independent reviewer, while all other systems are rejected during earlier stages of the process.
We then define grade A lens candidates as systems with \( G_\mathrm{N} \geq 2.5 \) and grade B lens candidates as those with \( 1.5 \leq G_\mathrm{N} < 2.5 \). 
In total, we identify 8 grade A candidates and 34 grade B candidates, while the remaining 296 systems are classified as grade C.
The tables describing the properties of these candidates are provided in Appendix~\ref{sec:lens_candidates}.
As shown in Figure~\ref{fig:lens_candidates}, grade A candidates display a clear strong lensing configuration, such as detectable multiple images or a counter-image, without requiring higher-resolution imaging. 
Grade B candidates exhibit a lensing-like structure but lack visible multiple images or a counter-image. 
They often include single arcs or objects on one side of the central galaxy and are challenging to confirm without spectroscopic data, as chance alignments are possible.
The majority of doubly lensed quasar candidates also fall into this category.
Grade C candidates, as displayed in Figure~\ref{fig:lens_candidates_c}, resemble lenses but are likely false positives, such as ring galaxies, merging systems, double stars, or other ambiguous configurations.
A detailed summary of our selection process, along with the number of candidates identified at each step, is provided in Table~\ref{tab:preselection}.


\section{Summary and conclusion} \label{sec:conclusion}
In this research, we conduct a comprehensive search for lensed quasars within the redshift range $1.5 \leq z \leq 7.2$ using data from the HSC PDR3. 
Our approach is organized into two primary phases. 
Initially, we filter potential candidates based on their photometric colors derived from catalog information, narrowing the pool from around 80 million sources to 710\,966. 
Subsequently, we apply a physics-informed VAE network to assess the likelihood of each candidate being a lens or a contaminant, which yields 18\,300 top candidates. 
It is noteworthy that the training data is created by overlaying deflected point-source light on actual HSC galaxy images. 
This method enables the generation of realistic strong-lens simulations and concentrates on identifying systems with Einstein radii of $\theta_\mathrm{E}<5\arcsec $. 
After reviewing astrometric data when available and visually inspecting objects with a lens probability $P_\mathrm{lens} > 0.3$, we identify 42 grade A and B lens candidates. 
These findings illustrate that automated neural network-based classifiers, with minimal human supervision, are promising for detecting lensed quasars in extensive datasets.

The method described in this paper is highly adaptable for detecting galaxy-quasar lenses across a wide range of redshifts. 
It is particularly well-suited for upcoming surveys such as Euclid \citep{2011arXiv1110.3193L,2022A&A...662A.112E}, which will deliver high-resolution NIR images covering extensive regions of the extragalactic sky, and the Rubin Observatory Legacy Survey of Space and Time \citep[LSST;][]{2019ApJ...873..111I}, which will provide comprehensive optical multiband data. 
In order to achieve the best results on the network performance, adjustments to the bandpass profiles, seeing conditions, and image scales will be necessary.
Furthermore, incorporating more sophisticated galaxy mass profiles beyond the SIE+$\gamma_\text{ext}$ model could enhance the performance of the classifier. 
To fully capitalize on the scientific value of our lens catalog, it is essential to perform spectroscopic observations to verify the redshifts of both deflectors and sources, along with high-resolution imaging to enable precise lens modeling.

\section{Data availability}

Additional appendices, which include the tables referenced in Appendix~\ref{sec:glq_model} and \ref{sec:lens_candidates}, are available on the Zenodo repository  at:  \href{https://doi.org/10.5281/zenodo.14718063}{https://doi.org/10.5281/zenodo.14718063}.

\begin{acknowledgements}

We thank Stefan Taubenberger for his participation in the visual inspection of lens candidates.
SHS thanks the Max Planck Society for the support through the Max Planck Fellowship. 
SB acknowledges the funding provided by the Alexander von Humboldt Foundation.
CG acknowledges support through grant MUR2020 SKSTHZ.
Support for JHHC was provided by Schmidt Sciences.
This research is supported in part by the Excellence Cluster ORIGINS which is funded by the Deutsche Forschungsgemeinschaft (DFG, German Research Foundation) under Germany's Excellence Strategy -- EXC-2094 -- 390783311. 
SS has been funded by the European Union's Horizon 2022 research and innovation program under the Marie Skłodowska-Curie grant agreement No 101105167 -- FASTIDIoUS.

This work is based on data collected at the Subaru Telescope and retrieved from the HSC data archive system, operated by the Subaru Telescope and Astronomy Data Center at the National Astronomical Observatory of Japan.
The Hyper Suprime-Cam (HSC) collaboration includes the astronomical communities of Japan, Taiwan, and Princeton University. 
The HSC instrumentation and software were developed by the National Astronomical Observatory of Japan (NAOJ), the Kavli Institute for the Physics and Mathematics of the Universe (Kavli IPMU), the University of Tokyo, the High Energy Accelerator Research Organization (KEK), the Academia Sinica Institute for Astronomy and Astrophysics in Taiwan (ASIAA), and Princeton University. 
Funding was contributed by the FIRST program from the Japanese Cabinet Office, the Ministry of Education, Culture, Sports, Science and Technology (MEXT), the Japan Society for the Promotion of Science (JSPS), the Japan Science and Technology Agency (JST), the Toray Science Foundation, NAOJ, Kavli IPMU, KEK, ASIAA, and Princeton University. 
This paper makes use of software developed for the Large Synoptic Survey Telescope. 
We thank the LSST Project for making their code available as free software at \url{http://dm.lsst.org}.

This project has included data from the Sloan Digital Sky Survey (SDSS).
Funding for SDSS-IV has been provided by the Alfred P. Sloan Foundation, the U.S. Department of Energy Office of Science, and the Participating Institutions. 
SDSS-IV acknowledges support and resources from the Center for High-Performance Computing at the University of Utah. 
The SDSS website is \url{https://www.sdss.org/}.
SDSS-IV is managed by the Astrophysical Research Consortium for the Participating Institutions of the SDSS Collaboration.

The unWISE catalog utilized in this paper is based on data products from the Wide-field Infrared Survey Explorer, which is a joint project of the University of California, Los Angeles, and NEOWISE, which is a project of the Jet Propulsion Laboratory/California Institute of Technology. 
WISE and NEOWISE are funded by the National Aeronautics and Space Administration.

This work has made use of data from the European Space Agency (ESA) mission Gaia (\url{https://www.cosmos.esa.int/gaia}), processed by the Gaia Data Processing and Analysis Consortium (DPAC, \url{https://www.cosmos.esa.int/web/gaia/dpac/consortium}). 
Funding for the DPAC has been provided by national institutions, in particular, the institutions participating in the Gaia Multilateral Agreement.

This work is based on data from eROSITA, the soft X-ray instrument aboard SRG, a joint Russian-German science mission supported by the Russian Space Agency (Roskosmos), in the interests of the Russian Academy of Sciences represented by its Space Research Institute (IKI), and the Deutsches Zentrum für Luft- und Raumfahrt (DLR). The SRG spacecraft was built by Lavochkin Association (NPOL) and its subcontractors, and is operated by NPOL with support from the Max Planck Institute for Extraterrestrial Physics (MPE). The development and construction of the eROSITA X-ray instrument was led by MPE, with contributions from the Dr. Karl Remeis Observatory Bamberg \& ECAP (FAU Erlangen-Nuernberg), the University of Hamburg Observatory, the Leibniz Institute for Astrophysics Potsdam (AIP), and the Institute for Astronomy and Astrophysics of the University of Tübingen, with the support of DLR and the Max Planck Society. The Argelander Institute for Astronomy of the University of Bonn and the Ludwig Maximilians Universität Munich also participated in the science preparation for eROSITA.

We acknowledge the use of the VHS, VIKING, UKIDSS, and UHS data.

\end{acknowledgements}

\tiny{
	\noindent
	\textit{Facilities.} eROSITA, ESO:VISTA (VIRCAM), Gaia, Sloan (eBOSS/BOSS), Subaru (HSC), UKIRT (WFCAM), WISE.
}

\vspace{1mm}

\tiny{
	\noindent
	\textit{Software.}
	Astropy \citep{2013A&A...558A..33A,2018AJ....156..123A},
	Dask \citep{matthew_rocklin-proc-scipy-2015},
	EAZY \citep{2008ApJ...686.1503B},	
	Matplotlib \citep{2021zndo....592536C},
	NumPy \citep{2020Natur.585..357H},
	Pandas \citep{2022zndo...3509134R},
	PyAutoLens \citep{2021JOSS....6.2825N},	
	Seaborn \citep{2021JOSS....6.3021W},
	SIMQSO \citep{2013ApJ...768..105M},
	TensorFlow \citep{2016arXiv160508695A,2022zndo...4724125D}.
	TOPCAT \citep{2005ASPC..347...29T}.
}

%
   \bibliographystyle{aa} 
   \bibliography{biblio} 

\begin{thebibliography}{142}
\expandafter\ifx\csname natexlab\endcsname\relax\def\natexlab#1{#1}\fi

\bibitem[{{Abadi} {et~al.}(2016){Abadi}, {Barham}, {Chen}, {Chen}, {Davis}, {Dean}, {Devin}, {Ghemawat}, {Irving}, {Isard}, {Kudlur}, {Levenberg}, {Monga}, {Moore}, {Murray}, {Steiner}, {Tucker}, {Vasudevan}, {Warden}, {Wicke}, {Yu}, \& {Zheng}}]{2016arXiv160508695A}
{Abadi}, M., {Barham}, P., {Chen}, J., {et~al.} 2016, arXiv e-prints, arXiv:1605.08695

\bibitem[{{Agnello} {et~al.}(2015){Agnello}, {Kelly}, {Treu}, \& {Marshall}}]{2015MNRAS.448.1446A}
{Agnello}, A., {Kelly}, B.~C., {Treu}, T., \& {Marshall}, P.~J. 2015, \mnras, 448, 1446

\bibitem[{{Agnello} {et~al.}(2018){Agnello}, {Schechter}, {Morgan}, {Treu}, {Grillo}, {Malesani}, {Anguita}, {Apostolovski}, {Rusu}, {Motta}, {Rojas}, {Chehade}, \& {Shanks}}]{2018MNRAS.475.2086A}
{Agnello}, A., {Schechter}, P.~L., {Morgan}, N.~D., {et~al.} 2018, \mnras, 475, 2086

\bibitem[{{Aihara} {et~al.}(2019){Aihara}, {AlSayyad}, {Ando}, {Armstrong}, {Bosch}, {Egami}, {Furusawa}, {Furusawa}, {Goulding}, {Harikane}, {Hikage}, {Ho}, {Hsieh}, {Huang}, {Ikeda}, {Imanishi}, {Ito}, {Iwata}, {Jaelani}, {Kakuma}, {Kawana}, {Kikuta}, {Kobayashi}, {Koike}, {Komiyama}, {Li}, {Liang}, {Lin}, {Luo}, {Lupton}, {Lust}, {MacArthur}, {Matsuoka}, {Mineo}, {Miyatake}, {Miyazaki}, {More}, {Murata}, {Namiki}, {Nishizawa}, {Oguri}, {Okabe}, {Okamoto}, {Okura}, {Ono}, {Onodera}, {Onoue}, {Osato}, {Ouchi}, {Shibuya}, {Strauss}, {Sugiyama}, {Suto}, {Takada}, {Takagi}, {Takata}, {Takita}, {Tanaka}, {Terai}, {Toba}, {Uchiyama}, {Utsumi}, {Wang}, {Wang}, \& {Yamada}}]{2019PASJ...71..114A}
{Aihara}, H., {AlSayyad}, Y., {Ando}, M., {et~al.} 2019, \pasj, 71, 114

\bibitem[{{Aihara} {et~al.}(2022){Aihara}, {AlSayyad}, {Ando}, {Armstrong}, {Bosch}, {Egami}, {Furusawa}, {Furusawa}, {Harasawa}, {Harikane}, {Hsieh}, {Ikeda}, {Ito}, {Iwata}, {Kodama}, {Koike}, {Kokubo}, {Komiyama}, {Li}, {Liang}, {Lin}, {Lupton}, {Lust}, {MacArthur}, {Mawatari}, {Mineo}, {Miyatake}, {Miyazaki}, {More}, {Morishima}, {Murayama}, {Nakajima}, {Nakata}, {Nishizawa}, {Oguri}, {Okabe}, {Okura}, {Ono}, {Osato}, {Ouchi}, {Pan}, {Plazas Malag{\'o}n}, {Price}, {Reed}, {Rykoff}, {Shibuya}, {Simunovic}, {Strauss}, {Sugimori}, {Suto}, {Suzuki}, {Takada}, {Takagi}, {Takata}, {Takita}, {Tanaka}, {Tang}, {Taranu}, {Terai}, {Toba}, {Turner}, {Uchiyama}, {Vijarnwannaluk}, {Waters}, {Yamada}, {Yamamoto}, \& {Yamashita}}]{2022PASJ...74..247A}
{Aihara}, H., {AlSayyad}, Y., {Ando}, M., {et~al.} 2022, \pasj, 74, 247

\bibitem[{{Aihara} {et~al.}(2018){Aihara}, {Armstrong}, {Bickerton}, {Bosch}, {Coupon}, {Furusawa}, {Hayashi}, {Ikeda}, {Kamata}, {Karoji}, {Kawanomoto}, {Koike}, {Komiyama}, {Lang}, {Lupton}, {Mineo}, {Miyatake}, {Miyazaki}, {Morokuma}, {Obuchi}, {Oishi}, {Okura}, {Price}, {Takata}, {Tanaka}, {Tanaka}, {Tanaka}, {Uchida}, {Uraguchi}, {Utsumi}, {Wang}, {Yamada}, {Yamanoi}, {Yasuda}, {Arimoto}, {Chiba}, {Finet}, {Fujimori}, {Fujimoto}, {Furusawa}, {Goto}, {Goulding}, {Gunn}, {Harikane}, {Hattori}, {Hayashi}, {He{\l}miniak}, {Higuchi}, {Hikage}, {Ho}, {Hsieh}, {Huang}, {Huang}, {Imanishi}, {Iwata}, {Jaelani}, {Jian}, {Kashikawa}, {Katayama}, {Kojima}, {Konno}, {Koshida}, {Kusakabe}, {Leauthaud}, {Lee}, {Lin}, {Lin}, {Mandelbaum}, {Matsuoka}, {Medezinski}, {Miyama}, {Momose}, {More}, {More}, {Mukae}, {Murata}, {Murayama}, {Nagao}, {Nakata}, {Niida}, {Niikura}, {Nishizawa}, {Oguri}, {Okabe}, {Ono}, {Onodera}, {Onoue}, {Ouchi}, {Pyo}, {Shibuya}, {Shimasaku}, {Simet}, {Speagle}, {Spergel}, {Strauss}, {Sugahara},
  {Sugiyama}, {Suto}, {Suzuki}, {Tait}, {Takada}, {Terai}, {Toba}, {Turner}, {Uchiyama}, {Umetsu}, {Urata}, {Usuda}, {Yeh}, \& {Yuma}}]{2018PASJ...70S...8A}
{Aihara}, H., {Armstrong}, R., {Bickerton}, S., {et~al.} 2018, \pasj, 70, S8

\bibitem[{{Almeida} {et~al.}(2023){Almeida}, {Anderson}, {Argudo-Fern{\'a}ndez}, {Badenes}, {Barger}, {Barrera-Ballesteros}, {Bender}, {Benitez}, {Besser}, {Bird}, {Bizyaev}, {Blanton}, {Bochanski}, {Bovy}, {Brandt}, {Brownstein}, {Buchner}, {Bulbul}, {Burchett}, {Cano D{\'\i}az}, {Carlberg}, {Casey}, {Chandra}, {Cherinka}, {Chiappini}, {Coker}, {Comparat}, {Conroy}, {Contardo}, {Cortes}, {Covey}, {Crane}, {Cunha}, {Dabbieri}, {Davidson}, {Davis}, {de Andrade Queiroz}, {De Lee}, {M{\'e}ndez Delgado}, {Demasi}, {Di Mille}, {Donor}, {Dow}, {Dwelly}, {Eracleous}, {Eriksen}, {Fan}, {Farr}, {Frederick}, {Fries}, {Frinchaboy}, {G{\"a}nsicke}, {Ge}, {Gonz{\'a}lez {\'A}vila}, {Grabowski}, {Grier}, {Guiglion}, {Gupta}, {Hall}, {Hawkins}, {Hayes}, {Hermes}, {Hern{\'a}ndez-Garc{\'\i}a}, {Hogg}, {Holtzman}, {Ibarra-Medel}, {Ji}, {Jofre}, {Johnson}, {Jones}, {Kinemuchi}, {Kluge}, {Koekemoer}, {Kollmeier}, {Kounkel}, {Krishnarao}, {Krumpe}, {Lacerna}, {Lago}, {Laporte}, {Liu}, {Liu}, {Liu}, {Lopes}, {Macktoobian},
  {Majewski}, {Malanushenko}, {Maoz}, {Masseron}, {Masters}, {Matijevic}, {McBride}, {Medan}, {Merloni}, {Morrison}, {Myers}, {M{\'e}sz{\'a}ros}, {Negrete}, {Nidever}, {Nitschelm}, {Oravetz}, {Oravetz}, {Pan}, {Peng}, {Pinsonneault}, {Pogge}, {Qiu}, {Ramirez}, {Rix}, {Fern{\'a}ndez Rosso}, {Runnoe}, {Salvato}, {Sanchez}, {Santana}, {Saydjari}, {Sayres}, {Schlaufman}, {Schneider}, {Schwope}, {Serna}, {Shen}, {Sobeck}, {Song}, {Souto}, {Spoo}, {Stassun}, {Steinmetz}, {Straumit}, {Stringfellow}, {S{\'a}nchez-Gallego}, {Taghizadeh-Popp}, {Tayar}, {Thakar}, {Tissera}, {Tkachenko}, {Hernandez Toledo}, {Trakhtenbrot}, {Fern{\'a}ndez-Trincado}, {Troup}, {Trump}, {Tuttle}, {Ulloa}, {Vazquez-Mata}, {Vera Alfaro}, {Villanova}, {Wachter}, {Weijmans}, {Wheeler}, {Wilson}, {Wojno}, {Wolf}, {Xue}, {Ybarra}, {Zari}, \& {Zasowski}}]{2023arXiv230107688A}
{Almeida}, A., {Anderson}, S.~F., {Argudo-Fern{\'a}ndez}, M., {et~al.} 2023, \apjs, 267, 44

\bibitem[{{Andika}(2022)}]{2022PhDT.........1A}
{Andika}, I.~T. 2022, PhD thesis, Max-Planck-Institute for Astronomy, Heidelberg

\bibitem[{{Andika} {et~al.}(2022){Andika}, {Jahnke}, {Ba{\~n}ados}, {Bosman}, {Davies}, {Eilers}, {Farina}, {Onoue}, \& {van der Wel}}]{2022AJ....163..251A}
{Andika}, I.~T., {Jahnke}, K., {Ba{\~n}ados}, E., {et~al.} 2022, \aj, 163, 251

\bibitem[{{Andika} {et~al.}(2020){Andika}, {Jahnke}, {Onoue}, {Ba{\~n}ados}, {Mazzucchelli}, {Novak}, {Eilers}, {Venemans}, {Schindler}, {Walter}, {Neeleman}, {Simcoe}, {Decarli}, {Farina}, {Marian}, {Pensabene}, {Cooper}, \& {Rojas}}]{2020ApJ...903...34A}
{Andika}, I.~T., {Jahnke}, K., {Onoue}, M., {et~al.} 2020, \apj, 903, 34

\bibitem[{{Andika} {et~al.}(2023{\natexlab{a}}){Andika}, {Jahnke}, {van der Wel}, {Ba{\~n}ados}, {Bosman}, {Davies}, {Eilers}, {Jaelani}, {Mazzucchelli}, {Onoue}, \& {Schindler}}]{2023ApJ...943..150A}
{Andika}, I.~T., {Jahnke}, K., {van der Wel}, A., {et~al.} 2023{\natexlab{a}}, \apj, 943, 150

\bibitem[{{Andika} {et~al.}(2023{\natexlab{b}}){Andika}, {Suyu}, {Ca{\~n}ameras}, {Melo}, {Schuldt}, {Shu}, {Eilers}, {Jaelani}, \& {Yue}}]{2023A&A...678A.103A}
{Andika}, I.~T., {Suyu}, S.~H., {Ca{\~n}ameras}, R., {et~al.} 2023{\natexlab{b}}, \aap, 678, A103

\bibitem[{{Arcelin} {et~al.}(2021){Arcelin}, {Doux}, {Aubourg}, {Roucelle}, \& {LSST Dark Energy Science Collaboration}}]{2021MNRAS.500..531A}
{Arcelin}, B., {Doux}, C., {Aubourg}, E., {Roucelle}, C., \& {LSST Dark Energy Science Collaboration}. 2021, \mnras, 500, 531

\bibitem[{{Astropy Collaboration} {et~al.}(2018){Astropy Collaboration}, {Price-Whelan}, {Sip{\H{o}}cz}, {G{\"u}nther}, {Lim}, {Crawford}, {Conseil}, {Shupe}, {Craig}, {Dencheva}, {Ginsburg}, {VanderPlas}, {Bradley}, {P{\'e}rez-Su{\'a}rez}, {de Val-Borro}, {Aldcroft}, {Cruz}, {Robitaille}, {Tollerud}, {Ardelean}, {Babej}, {Bach}, {Bachetti}, {Bakanov}, {Bamford}, {Barentsen}, {Barmby}, {Baumbach}, {Berry}, {Biscani}, {Boquien}, {Bostroem}, {Bouma}, {Brammer}, {Bray}, {Breytenbach}, {Buddelmeijer}, {Burke}, {Calderone}, {Cano Rodr{\'\i}guez}, {Cara}, {Cardoso}, {Cheedella}, {Copin}, {Corrales}, {Crichton}, {D'Avella}, {Deil}, {Depagne}, {Dietrich}, {Donath}, {Droettboom}, {Earl}, {Erben}, {Fabbro}, {Ferreira}, {Finethy}, {Fox}, {Garrison}, {Gibbons}, {Goldstein}, {Gommers}, {Greco}, {Greenfield}, {Groener}, {Grollier}, {Hagen}, {Hirst}, {Homeier}, {Horton}, {Hosseinzadeh}, {Hu}, {Hunkeler}, {Ivezi{\'c}}, {Jain}, {Jenness}, {Kanarek}, {Kendrew}, {Kern}, {Kerzendorf}, {Khvalko}, {King}, {Kirkby}, {Kulkarni},
  {Kumar}, {Lee}, {Lenz}, {Littlefair}, {Ma}, {Macleod}, {Mastropietro}, {McCully}, {Montagnac}, {Morris}, {Mueller}, {Mumford}, {Muna}, {Murphy}, {Nelson}, {Nguyen}, {Ninan}, {N{\"o}the}, {Ogaz}, {Oh}, {Parejko}, {Parley}, {Pascual}, {Patil}, {Patil}, {Plunkett}, {Prochaska}, {Rastogi}, {Reddy Janga}, {Sabater}, {Sakurikar}, {Seifert}, {Sherbert}, {Sherwood-Taylor}, {Shih}, {Sick}, {Silbiger}, {Singanamalla}, {Singer}, {Sladen}, {Sooley}, {Sornarajah}, {Streicher}, {Teuben}, {Thomas}, {Tremblay}, {Turner}, {Terr{\'o}n}, {van Kerkwijk}, {de la Vega}, {Watkins}, {Weaver}, {Whitmore}, {Woillez}, {Zabalza}, \& {Astropy Contributors}}]{2018AJ....156..123A}
{Astropy Collaboration}, {Price-Whelan}, A.~M., {Sip{\H{o}}cz}, B.~M., {et~al.} 2018, \aj, 156, 123

\bibitem[{{Astropy Collaboration} {et~al.}(2013){Astropy Collaboration}, {Robitaille}, {Tollerud}, {Greenfield}, {Droettboom}, {Bray}, {Aldcroft}, {Davis}, {Ginsburg}, {Price-Whelan}, {Kerzendorf}, {Conley}, {Crighton}, {Barbary}, {Muna}, {Ferguson}, {Grollier}, {Parikh}, {Nair}, {Unther}, {Deil}, {Woillez}, {Conseil}, {Kramer}, {Turner}, {Singer}, {Fox}, {Weaver}, {Zabalza}, {Edwards}, {Azalee Bostroem}, {Burke}, {Casey}, {Crawford}, {Dencheva}, {Ely}, {Jenness}, {Labrie}, {Lim}, {Pierfederici}, {Pontzen}, {Ptak}, {Refsdal}, {Servillat}, \& {Streicher}}]{2013A&A...558A..33A}
{Astropy Collaboration}, {Robitaille}, T.~P., {Tollerud}, E.~J., {et~al.} 2013, \aap, 558, A33

\bibitem[{{Bank} {et~al.}(2020){Bank}, {Koenigstein}, \& {Giryes}}]{2020arXiv200305991B}
{Bank}, D., {Koenigstein}, N., \& {Giryes}, R. 2020, arXiv e-prints, arXiv:2003.05991

\bibitem[{{Barkana}(1998)}]{1998ApJ...502..531B}
{Barkana}, R. 1998, \apj, 502, 531

\bibitem[{{Belokurov} {et~al.}(2007){Belokurov}, {Evans}, {Moiseev}, {King}, {Hewett}, {Pettini}, {Wyrzykowski}, {McMahon}, {Smith}, {Gilmore}, {Sanchez}, {Udalski}, {Koposov}, {Zucker}, \& {Walcher}}]{2007ApJ...671L...9B}
{Belokurov}, V., {Evans}, N.~W., {Moiseev}, A., {et~al.} 2007, \apjl, 671, L9

\bibitem[{{Best} {et~al.}(2018){Best}, {Magnier}, {Liu}, {Aller}, {Zhang}, {Burgett}, {Chambers}, {Draper}, {Flewelling}, {Kaiser}, {Kudritzki}, {Metcalfe}, {Tonry}, {Wainscoat}, \& {Waters}}]{2018ApJS..234....1B}
{Best}, W. M.~J., {Magnier}, E.~A., {Liu}, M.~C., {et~al.} 2018, \apjs, 234, 1

\bibitem[{{Blanton} {et~al.}(2017){Blanton}, {Bershady}, {Abolfathi}, {Albareti}, {Allende Prieto}, {Almeida}, {Alonso-Garc{\'\i}a}, {Anders}, {Anderson}, {Andrews}, {Aquino-Ort{\'\i}z}, {Arag{\'o}n-Salamanca}, {Argudo-Fern{\'a}ndez}, {Armengaud}, {Aubourg}, {Avila-Reese}, {Badenes}, {Bailey}, {Barger}, {Barrera-Ballesteros}, {Bartosz}, {Bates}, {Baumgarten}, {Bautista}, {Beaton}, {Beers}, {Belfiore}, {Bender}, {Berlind}, {Bernardi}, {Beutler}, {Bird}, {Bizyaev}, {Blanc}, {Blomqvist}, {Bolton}, {Boquien}, {Borissova}, {van den Bosch}, {Bovy}, {Brandt}, {Brinkmann}, {Brownstein}, {Bundy}, {Burgasser}, {Burtin}, {Busca}, {Cappellari}, {Delgado Carigi}, {Carlberg}, {Carnero Rosell}, {Carrera}, {Chanover}, {Cherinka}, {Cheung}, {G{\'o}mez Maqueo Chew}, {Chiappini}, {Choi}, {Chojnowski}, {Chuang}, {Chung}, {Cirolini}, {Clerc}, {Cohen}, {Comparat}, {da Costa}, {Cousinou}, {Covey}, {Crane}, {Croft}, {Cruz-Gonzalez}, {Garrido Cuadra}, {Cunha}, {Damke}, {Darling}, {Davies}, {Dawson}, {de la Macorra}, {Dell'Agli}, {De
  Lee}, {Delubac}, {Di Mille}, {Diamond-Stanic}, {Cano-D{\'\i}az}, {Donor}, {Downes}, {Drory}, {du Mas des Bourboux}, {Duckworth}, {Dwelly}, {Dyer}, {Ebelke}, {Eigenbrot}, {Eisenstein}, {Emsellem}, {Eracleous}, {Escoffier}, {Evans}, {Fan}, {Fern{\'a}ndez-Alvar}, {Fernandez-Trincado}, {Feuillet}, {Finoguenov}, {Fleming}, {Font-Ribera}, {Fredrickson}, {Freischlad}, {Frinchaboy}, {Fuentes}, {Galbany}, {Garcia-Dias}, {Garc{\'\i}a-Hern{\'a}ndez}, {Gaulme}, {Geisler}, {Gelfand}, {Gil-Mar{\'\i}n}, {Gillespie}, {Goddard}, {Gonzalez-Perez}, {Grabowski}, {Green}, {Grier}, {Gunn}, {Guo}, {Guy}, {Hagen}, {Hahn}, {Hall}, {Harding}, {Hasselquist}, {Hawley}, {Hearty}, {Gonzalez Hern{\'a}ndez}, {Ho}, {Hogg}, {Holley-Bockelmann}, {Holtzman}, {Holzer}, {Huehnerhoff}, {Hutchinson}, {Hwang}, {Ibarra-Medel}, {da Silva Ilha}, {Ivans}, {Ivory}, {Jackson}, {Jensen}, {Johnson}, {Jones}, {J{\"o}nsson}, {Jullo}, {Kamble}, {Kinemuchi}, {Kirkby}, {Kitaura}, {Klaene}, {Knapp}, {Kneib}, {Kollmeier}, {Lacerna}, {Lane}, {Lang}, {Law},
  {Lazarz}, {Lee}, {Le Goff}, {Liang}, {Li}, {Li}, {Lian}, {Lima}, {Lin}, {Lin}, {Bertran de Lis}, {Liu}, {de Icaza Lizaola}, {Long}, {Lucatello}, {Lundgren}, {MacDonald}, {Deconto Machado}, {MacLeod}, {Mahadevan}, {Geimba Maia}, {Maiolino}, {Majewski}, {Malanushenko}, {Malanushenko}, {Manchado}, {Mao}, {Maraston}, {Marques-Chaves}, {Masseron}, {Masters}, {McBride}, {McDermid}, {McGrath}, {McGreer}, {Medina Pe{\~n}a}, {Melendez}, {Merloni}, {Merrifield}, {Meszaros}, {Meza}, {Minchev}, {Minniti}, {Miyaji}, {More}, {Mulchaey}, {M{\"u}ller-S{\'a}nchez}, {Muna}, {Munoz}, {Myers}, {Nair}, {Nandra}, {Correa do Nascimento}, {Negrete}, {Ness}, {Newman}, {Nichol}, {Nidever}, {Nitschelm}, {Ntelis}, {O'Connell}, {Oelkers}, {Oravetz}, {Oravetz}, {Pace}, {Padilla}, {Palanque-Delabrouille}, {Alonso Palicio}, {Pan}, {Parejko}, {Parikh}, {P{\^a}ris}, {Park}, {Patten}, {Peirani}, {Pellejero-Ibanez}, {Penny}, {Percival}, {Perez-Fournon}, {Petitjean}, {Pieri}, {Pinsonneault}, {Pisani}, {Poleski}, {Prada}, {Prakash}, {Queiroz},
  {Raddick}, {Raichoor}, {Barboza Rembold}, {Richstein}, {Riffel}, {Riffel}, {Rix}, {Robin}, {Rockosi}, {Rodr{\'\i}guez-Torres}, {Roman-Lopes}, {Rom{\'a}n-Z{\'u}{\~n}iga}, {Rosado}, {Ross}, {Rossi}, {Ruan}, {Ruggeri}, {Rykoff}, {Salazar-Albornoz}, {Salvato}, {S{\'a}nchez}, {Aguado}, {S{\'a}nchez-Gallego}, {Santana}, {Santiago}, {Sayres}, {Schiavon}, {da Silva Schimoia}, {Schlafly}, {Schlegel}, {Schneider}, {Schultheis}, {Schuster}, {Schwope}, {Seo}, {Shao}, {Shen}, {Shetrone}, {Shull}, {Simon}, {Skinner}, {Skrutskie}, {Slosar}, {Smith}, {Sobeck}, {Sobreira}, {Somers}, {Souto}, {Stark}, {Stassun}, {Stauffer}, {Steinmetz}, {Storchi-Bergmann}, {Streblyanska}, {Stringfellow}, {Su{\'a}rez}, {Sun}, {Suzuki}, {Szigeti}, {Taghizadeh-Popp}, {Tang}, {Tao}, {Tayar}, {Tembe}, {Teske}, {Thakar}, {Thomas}, {Thompson}, {Tinker}, {Tissera}, {Tojeiro}, {Hernandez Toledo}, {de la Torre}, {Tremonti}, {Troup}, {Valenzuela}, {Martinez Valpuesta}, {Vargas-Gonz{\'a}lez}, {Vargas-Maga{\~n}a}, {Vazquez}, {Villanova}, {Vivek}, {Vogt},
  {Wake}, {Walterbos}, {Wang}, {Weaver}, {Weijmans}, {Weinberg}, {Westfall}, {Whelan}, {Wild}, {Wilson}, {Wood-Vasey}, {Wylezalek}, {Xiao}, {Yan}, {Yang}, {Ybarra}, {Y{\`e}che}, {Zakamska}, {Zamora}, {Zarrouk}, {Zasowski}, {Zhang}, {Zhao}, {Zheng}, {Zheng}, {Zhou}, {Zhou}, {Zhu}, {Zoccali}, \& {Zou}}]{2017AJ....154...28B}
{Blanton}, M.~R., {Bershady}, M.~A., {Abolfathi}, B., {et~al.} 2017, \aj, 154, 28

\bibitem[{{Bonvin} {et~al.}(2017){Bonvin}, {Courbin}, {Suyu}, {Marshall}, {Rusu}, {Sluse}, {Tewes}, {Wong}, {Collett}, {Fassnacht}, {Treu}, {Auger}, {Hilbert}, {Koopmans}, {Meylan}, {Rumbaugh}, {Sonnenfeld}, \& {Spiniello}}]{2017MNRAS.465.4914B}
{Bonvin}, V., {Courbin}, F., {Suyu}, S.~H., {et~al.} 2017, \mnras, 465, 4914

\bibitem[{{Boroson} \& {Green}(1992)}]{1992ApJS...80..109B}
{Boroson}, T.~A. \& {Green}, R.~F. 1992, \apjs, 80, 109

\bibitem[{{Brammer} {et~al.}(2008){Brammer}, {van Dokkum}, \& {Coppi}}]{2008ApJ...686.1503B}
{Brammer}, G.~B., {van Dokkum}, P.~G., \& {Coppi}, P. 2008, \apj, 686, 1503

\bibitem[{{Ca{\~n}ameras} {et~al.}(2024){Ca{\~n}ameras}, {Schuldt}, {Shu}, {Suyu}, {Taubenberger}, {Andika}, {Bag}, {Inoue}, {Jaelani}, {Leal-Taix{\'e}}, {Meinhardt}, {Melo}, \& {More}}]{2023arXiv230603136C}
{Ca{\~n}ameras}, R., {Schuldt}, S., {Shu}, Y., {et~al.} 2024, \aap, 692, A72

\bibitem[{{Ca{\~n}ameras} {et~al.}(2021){Ca{\~n}ameras}, {Schuldt}, {Shu}, {Suyu}, {Taubenberger}, {Meinhardt}, {Leal-Taix{\'e}}, {Chao}, {Inoue}, {Jaelani}, \& {More}}]{2021AandA...653L...6C}
{Ca{\~n}ameras}, R., {Schuldt}, S., {Shu}, Y., {et~al.} 2021, \aap, 653, L6

\bibitem[{{Calzetti} {et~al.}(2000){Calzetti}, {Armus}, {Bohlin}, {Kinney}, {Koornneef}, \& {Storchi-Bergmann}}]{2000ApJ...533..682C}
{Calzetti}, D., {Armus}, L., {Bohlin}, R.~C., {et~al.} 2000, \apj, 533, 682

\bibitem[{{Carnero Rosell} {et~al.}(2019){Carnero Rosell}, {Santiago}, {dal Ponte}, {Burningham}, {da Costa}, {James}, {Marshall}, {McMahon}, {Bechtol}, {De Paris}, {Li}, {Pieres}, {Garc{\'\i}a-Bellido}, {Abbott}, {Annis}, {Avila}, {Bernstein}, {Brooks}, {Burke}, {Carrasco Kind}, {Carretero}, {De Vicente}, {Drlica-Wagner}, {Fosalba}, {Frieman}, {Gaztanaga}, {Gruendl}, {Gschwend}, {Gutierrez}, {Hollowood}, {Maia}, {Menanteau}, {Miquel}, {Plazas}, {Roodman}, {Sanchez}, {Scarpine}, {Schindler}, {Serrano}, {Sevilla-Noarbe}, {Smith}, {Sobreira}, {Soares-Santos}, {Suchyta}, {Swanson}, {Tarle}, {Vikram}, {Walker}, \& {DES Collaboration}}]{2019MNRAS.489.5301C}
{Carnero Rosell}, A., {Santiago}, B., {dal Ponte}, M., {et~al.} 2019, \mnras, 489, 5301

\bibitem[{{Caswell} {et~al.}(2021){Caswell}, {Droettboom}, {Lee}, {Sales De Andrade}, {Hoffmann}, {Hunter}, {Klymak}, {Firing}, {Stansby}, {Varoquaux}, {Hedegaard Nielsen}, {Root}, {May}, {Elson}, {Sepp{\"a}nen}, {Dale}, {Lee}, {McDougall}, {Straw}, {Hobson}, {Hannah}, {Gohlke}, {Vincent}, {Yu}, {Ma}, {Silvester}, {Moad}, {Kniazev}, {Ernest}, \& {Ivanov}}]{2021zndo....592536C}
{Caswell}, T.~A., {Droettboom}, M., {Lee}, A., {et~al.} 2021, {matplotlib/matplotlib: REL: v3.5.1}, Zenodo

\bibitem[{{Chan} {et~al.}(2022){Chan}, {Lemon}, {Courbin}, {Gavazzi}, {Cl{\'e}ment}, {Millon}, {Paic}, {Rojas}, {Savary}, {Vernardos}, {Cuillandre}, {Fabbro}, {Gwyn}, {Hudson}, {Kilbinger}, \& {McConnachie}}]{2022A&A...659A.140C}
{Chan}, J.~H.~H., {Lemon}, C., {Courbin}, F., {et~al.} 2022, \aap, 659, A140

\bibitem[{{Chan} {et~al.}(2015){Chan}, {Suyu}, {Chiueh}, {More}, {Marshall}, {Coupon}, {Oguri}, \& {Price}}]{2015ApJ...807..138C}
{Chan}, J. H.~H., {Suyu}, S.~H., {Chiueh}, T., {et~al.} 2015, \apj, 807, 138

\bibitem[{{Chan} {et~al.}(2020){Chan}, {Suyu}, {Sonnenfeld}, {Jaelani}, {More}, {Yonehara}, {Kubota}, {Coupon}, {Lee}, {Oguri}, {Rusu}, \& {Wong}}]{2020A&A...636A..87C}
{Chan}, J. H.~H., {Suyu}, S.~H., {Sonnenfeld}, A., {et~al.} 2020, \aap, 636, A87

\bibitem[{{Chan} {et~al.}(2024){Chan}, {Wong}, {Ding}, {Chao}, {Chiu}, {Jaelani}, {Kayo}, {More}, {Oguri}, \& {Suyu}}]{2024MNRAS.527.6253C}
{Chan}, J. H.~H., {Wong}, K.~C., {Ding}, X., {et~al.} 2024, \mnras, 527, 6253

\bibitem[{{Dawson} {et~al.}(2016){Dawson}, {Kneib}, {Percival}, {Alam}, {Albareti}, {Anderson}, {Armengaud}, {Aubourg}, {Bailey}, {Bautista}, {Berlind}, {Bershady}, {Beutler}, {Bizyaev}, {Blanton}, {Blomqvist}, {Bolton}, {Bovy}, {Brandt}, {Brinkmann}, {Brownstein}, {Burtin}, {Busca}, {Cai}, {Chuang}, {Clerc}, {Comparat}, {Cope}, {Croft}, {Cruz-Gonzalez}, {da Costa}, {Cousinou}, {Darling}, {de la Macorra}, {de la Torre}, {Delubac}, {du Mas des Bourboux}, {Dwelly}, {Ealet}, {Eisenstein}, {Eracleous}, {Escoffier}, {Fan}, {Finoguenov}, {Font-Ribera}, {Frinchaboy}, {Gaulme}, {Georgakakis}, {Green}, {Guo}, {Guy}, {Ho}, {Holder}, {Huehnerhoff}, {Hutchinson}, {Jing}, {Jullo}, {Kamble}, {Kinemuchi}, {Kirkby}, {Kitaura}, {Klaene}, {Laher}, {Lang}, {Laurent}, {Le Goff}, {Li}, {Liang}, {Lima}, {Lin}, {Lin}, {Lin}, {Long}, {Lundgren}, {MacDonald}, {Geimba Maia}, {Malanushenko}, {Malanushenko}, {Mariappan}, {McBride}, {McGreer}, {M{\'e}nard}, {Merloni}, {Meza}, {Montero-Dorta}, {Muna}, {Myers}, {Nandra}, {Naugle},
  {Newman}, {Noterdaeme}, {Nugent}, {Ogando}, {Olmstead}, {Oravetz}, {Oravetz}, {Padmanabhan}, {Palanque-Delabrouille}, {Pan}, {Parejko}, {P{\^a}ris}, {Peacock}, {Petitjean}, {Pieri}, {Pisani}, {Prada}, {Prakash}, {Raichoor}, {Reid}, {Rich}, {Ridl}, {Rodriguez-Torres}, {Carnero Rosell}, {Ross}, {Rossi}, {Ruan}, {Salvato}, {Sayres}, {Schneider}, {Schlegel}, {Seljak}, {Seo}, {Sesar}, {Shandera}, {Shu}, {Slosar}, {Sobreira}, {Streblyanska}, {Suzuki}, {Taylor}, {Tao}, {Tinker}, {Tojeiro}, {Vargas-Maga{\~n}a}, {Wang}, {Weaver}, {Weinberg}, {White}, {Wood-Vasey}, {Yeche}, {Zhai}, {Zhao}, {Zhao}, {Zheng}, {Ben Zhu}, \& {Zou}}]{2016AJ....151...44D}
{Dawson}, K.~S., {Kneib}, J.-P., {Percival}, W.~J., {et~al.} 2016, \aj, 151, 44

\bibitem[{{Dawson} {et~al.}(2013){Dawson}, {Schlegel}, {Ahn}, {Anderson}, {Aubourg}, {Bailey}, {Barkhouser}, {Bautista}, {Beifiori}, {Berlind}, {Bhardwaj}, {Bizyaev}, {Blake}, {Blanton}, {Blomqvist}, {Bolton}, {Borde}, {Bovy}, {Brandt}, {Brewington}, {Brinkmann}, {Brown}, {Brownstein}, {Bundy}, {Busca}, {Carithers}, {Carnero}, {Carr}, {Chen}, {Comparat}, {Connolly}, {Cope}, {Croft}, {Cuesta}, {da Costa}, {Davenport}, {Delubac}, {de Putter}, {Dhital}, {Ealet}, {Ebelke}, {Eisenstein}, {Escoffier}, {Fan}, {Filiz Ak}, {Finley}, {Font-Ribera}, {G{\'e}nova-Santos}, {Gunn}, {Guo}, {Haggard}, {Hall}, {Hamilton}, {Harris}, {Harris}, {Ho}, {Hogg}, {Holder}, {Honscheid}, {Huehnerhoff}, {Jordan}, {Jordan}, {Kauffmann}, {Kazin}, {Kirkby}, {Klaene}, {Kneib}, {Le Goff}, {Lee}, {Long}, {Loomis}, {Lundgren}, {Lupton}, {Maia}, {Makler}, {Malanushenko}, {Malanushenko}, {Mandelbaum}, {Manera}, {Maraston}, {Margala}, {Masters}, {McBride}, {McDonald}, {McGreer}, {McMahon}, {Mena}, {Miralda-Escud{\'e}}, {Montero-Dorta},
  {Montesano}, {Muna}, {Myers}, {Naugle}, {Nichol}, {Noterdaeme}, {Nuza}, {Olmstead}, {Oravetz}, {Oravetz}, {Owen}, {Padmanabhan}, {Palanque-Delabrouille}, {Pan}, {Parejko}, {P{\^a}ris}, {Percival}, {P{\'e}rez-Fournon}, {P{\'e}rez-R{\`a}fols}, {Petitjean}, {Pfaffenberger}, {Pforr}, {Pieri}, {Prada}, {Price-Whelan}, {Raddick}, {Rebolo}, {Rich}, {Richards}, {Rockosi}, {Roe}, {Ross}, {Ross}, {Rossi}, {Rubi{\~n}o-Martin}, {Samushia}, {S{\'a}nchez}, {Sayres}, {Schmidt}, {Schneider}, {Sc{\'o}ccola}, {Seo}, {Shelden}, {Sheldon}, {Shen}, {Shu}, {Slosar}, {Smee}, {Snedden}, {Stauffer}, {Steele}, {Strauss}, {Streblyanska}, {Suzuki}, {Swanson}, {Tal}, {Tanaka}, {Thomas}, {Tinker}, {Tojeiro}, {Tremonti}, {Vargas Maga{\~n}a}, {Verde}, {Viel}, {Wake}, {Watson}, {Weaver}, {Weinberg}, {Weiner}, {West}, {White}, {Wood-Vasey}, {Yeche}, {Zehavi}, {Zhao}, \& {Zheng}}]{2013AJ....145...10D}
{Dawson}, K.~S., {Schlegel}, D.~J., {Ahn}, C.~P., {et~al.} 2013, \aj, 145, 10

\bibitem[{{Developers}(2022)}]{2022zndo...4724125D}
{Developers}, T. 2022, {TensorFlow}, Zenodo

\bibitem[{{Ding} {et~al.}(2021){Ding}, {Treu}, {Birrer}, {Agnello}, {Sluse}, {Fassnacht}, {Auger}, {Wong}, {Suyu}, {Morishita}, {Rusu}, \& {Galan}}]{2021MNRAS.501..269D}
{Ding}, X., {Treu}, T., {Birrer}, S., {et~al.} 2021, \mnras, 501, 269

\bibitem[{{Dux} {et~al.}(2024){Dux}, {Lemon}, {Courbin}, {Neira}, {Anguita}, {Galan}, {Kim}, {Hempel}, {Hempel}, \& {Lachaume}}]{2024A&A...682A..47D}
{Dux}, F., {Lemon}, C., {Courbin}, F., {et~al.} 2024, \aap, 682, A47

\bibitem[{{Dye} {et~al.}(2018){Dye}, {Lawrence}, {Read}, {Fan}, {Kerr}, {Varricatt}, {Furnell}, {Edge}, {Irwin}, {Hambly}, {Lucas}, {Almaini}, {Chambers}, {Green}, {Hewett}, {Liu}, {McGreer}, {Best}, {Zhang}, {Sutorius}, {Froebrich}, {Magnier}, {Hasinger}, {Lederer}, {Bold}, \& {Tedds}}]{2018MNRAS.473.5113D}
{Dye}, S., {Lawrence}, A., {Read}, M.~A., {et~al.} 2018, \mnras, 473, 5113

\bibitem[{{Edge} {et~al.}(2013){Edge}, {Sutherland}, {Kuijken}, {Driver}, {McMahon}, {Eales}, \& {Emerson}}]{2013Msngr.154...32E}
{Edge}, A., {Sutherland}, W., {Kuijken}, K., {et~al.} 2013, The Messenger, 154, 32

\bibitem[{{Ertl} {et~al.}(2023){Ertl}, {Schuldt}, {Suyu}, {Schmidt}, {Treu}, {Birrer}, {Shajib}, \& {Sluse}}]{2023A&A...672A...2E}
{Ertl}, S., {Schuldt}, S., {Suyu}, S.~H., {et~al.} 2023, \aap, 672, A2

\bibitem[{{Etherington} {et~al.}(2022){Etherington}, {Nightingale}, {Massey}, {Cao}, {Robertson}, {Amorisco}, {Amvrosiadis}, {Cole}, {Frenk}, {He}, {Li}, \& {Tam}}]{2022MNRAS.517.3275E}
{Etherington}, A., {Nightingale}, J.~W., {Massey}, R., {et~al.} 2022, \mnras, 517, 3275

\bibitem[{{Euclid Collaboration} {et~al.}(2019){Euclid Collaboration}, {Barnett}, {Warren}, {Mortlock}, {Cuby}, {Conselice}, {Hewett}, {Willott}, {Auricchio}, {Balaguera-Antol{\'\i}nez}, {Baldi}, {Bardelli}, {Bellagamba}, {Bender}, {Biviano}, {Bonino}, {Bozzo}, {Branchini}, {Brescia}, {Brinchmann}, {Burigana}, {Camera}, {Capobianco}, {Carbone}, {Carretero}, {Carvalho}, {Castander}, {Castellano}, {Cavuoti}, {Cimatti}, {Cl{\'e}dassou}, {Congedo}, {Conversi}, {Copin}, {Corcione}, {Coupon}, {Courtois}, {Cropper}, {Da Silva}, {Duncan}, {Dusini}, {Ealet}, {Farrens}, {Fosalba}, {Fotopoulou}, {Fourmanoit}, {Frailis}, {Fumana}, {Galeotta}, {Garilli}, {Gillard}, {Gillis}, {Graci{\'a}-Carpio}, {Grupp}, {Hoekstra}, {Hormuth}, {Israel}, {Jahnke}, {Kermiche}, {Kilbinger}, {Kirkpatrick}, {Kitching}, {Kohley}, {Kubik}, {Kunz}, {Kurki-Suonio}, {Laureijs}, {Ligori}, {Lilje}, {Lloro}, {Maiorano}, {Mansutti}, {Marggraf}, {Martinet}, {Marulli}, {Massey}, {Mauri}, {Medinaceli}, {Mei}, {Mellier}, {Metcalf}, {Metge}, {Meylan},
  {Moresco}, {Moscardini}, {Munari}, {Neissner}, {Niemi}, {Nutma}, {Padilla}, {Paltani}, {Pasian}, {Paykari}, {Percival}, {Pettorino}, {Polenta}, {Poncet}, {Pozzetti}, {Raison}, {Renzi}, {Rhodes}, {Rix}, {Romelli}, {Roncarelli}, {Rossetti}, {Saglia}, {Sapone}, {Scaramella}, {Schneider}, {Scottez}, {Secroun}, {Serrano}, {Sirri}, {Stanco}, {Sureau}, {Tallada-Cresp{\'\i}}, {Tavagnacco}, {Taylor}, {Tenti}, {Tereno}, {Toledo-Moreo}, {Torradeflot}, {Valenziano}, {Vassallo}, {Wang}, {Zacchei}, {Zamorani}, {Zoubian}, \& {Zucca}}]{2019A&A...631A..85E}
{Euclid Collaboration}, {Barnett}, R., {Warren}, S.~J., {et~al.} 2019, \aap, 631, A85

\bibitem[{{Euclid Collaboration} {et~al.}(2022){Euclid Collaboration}, {Scaramella}, {Amiaux}, {Mellier}, {Burigana}, {Carvalho}, {Cuillandre}, {Da Silva}, {Derosa}, {Dinis}, {Maiorano}, {Maris}, {Tereno}, {Laureijs}, {Boenke}, {Buenadicha}, {Dupac}, {Gaspar Venancio}, {G{\'o}mez-{\'A}lvarez}, {Hoar}, {Lorenzo Alvarez}, {Racca}, {Saavedra-Criado}, {Schwartz}, {Vavrek}, {Schirmer}, {Aussel}, {Azzollini}, {Cardone}, {Cropper}, {Ealet}, {Garilli}, {Gillard}, {Granett}, {Guzzo}, {Hoekstra}, {Jahnke}, {Kitching}, {Maciaszek}, {Meneghetti}, {Miller}, {Nakajima}, {Niemi}, {Pasian}, {Percival}, {Pottinger}, {Sauvage}, {Scodeggio}, {Wachter}, {Zacchei}, {Aghanim}, {Amara}, {Auphan}, {Auricchio}, {Awan}, {Balestra}, {Bender}, {Bodendorf}, {Bonino}, {Branchini}, {Brau-Nogue}, {Brescia}, {Candini}, {Capobianco}, {Carbone}, {Carlberg}, {Carretero}, {Casas}, {Castander}, {Castellano}, {Cavuoti}, {Cimatti}, {Cledassou}, {Congedo}, {Conselice}, {Conversi}, {Copin}, {Corcione}, {Costille}, {Courbin}, {Degaudenzi}, {Douspis},
  {Dubath}, {Duncan}, {Dusini}, {Farrens}, {Ferriol}, {Fosalba}, {Fourmanoit}, {Frailis}, {Franceschi}, {Franzetti}, {Fumana}, {Gillis}, {Giocoli}, {Grazian}, {Grupp}, {Haugan}, {Holmes}, {Hormuth}, {Hudelot}, {Kermiche}, {Kiessling}, {Kilbinger}, {Kohley}, {Kubik}, {K{\"u}mmel}, {Kunz}, {Kurki-Suonio}, {Lahav}, {Ligori}, {Lilje}, {Lloro}, {Mansutti}, {Marggraf}, {Markovic}, {Marulli}, {Massey}, {Maurogordato}, {Melchior}, {Merlin}, {Meylan}, {Mohr}, {Moresco}, {Morin}, {Moscardini}, {Munari}, {Nichol}, {Padilla}, {Paltani}, {Peacock}, {Pedersen}, {Pettorino}, {Pires}, {Poncet}, {Popa}, {Pozzetti}, {Raison}, {Rebolo}, {Rhodes}, {Rix}, {Roncarelli}, {Rossetti}, {Saglia}, {Schneider}, {Schrabback}, {Secroun}, {Seidel}, {Serrano}, {Sirignano}, {Sirri}, {Skottfelt}, {Stanco}, {Starck}, {Tallada-Cresp{\'\i}}, {Tavagnacco}, {Taylor}, {Teplitz}, {Toledo-Moreo}, {Torradeflot}, {Trifoglio}, {Valentijn}, {Valenziano}, {Verdoes Kleijn}, {Wang}, {Welikala}, {Weller}, {Wetzstein}, {Zamorani}, {Zoubian}, {Andreon},
  {Baldi}, {Bardelli}, {Boucaud}, {Camera}, {Di Ferdinando}, {Fabbian}, {Farinelli}, {Galeotta}, {Graci{\'a}-Carpio}, {Maino}, {Medinaceli}, {Mei}, {Neissner}, {Polenta}, {Renzi}, {Romelli}, {Rosset}, {Sureau}, {Tenti}, {Vassallo}, {Zucca}, {Baccigalupi}, {Balaguera-Antol{\'\i}nez}, {Battaglia}, {Biviano}, {Borgani}, {Bozzo}, {Cabanac}, {Cappi}, {Casas}, {Castignani}, {Colodro-Conde}, {Coupon}, {Courtois}, {Cuby}, {de la Torre}, {Desai}, {Dole}, {Fabricius}, {Farina}, {Ferreira}, {Finelli}, {Flose-Reimberg}, {Fotopoulou}, {Ganga}, {Gozaliasl}, {Hook}, {Keihanen}, {Kirkpatrick}, {Liebing}, {Lindholm}, {Mainetti}, {Martinelli}, {Martinet}, {Maturi}, {McCracken}, {Metcalf}, {Morgante}, {Nightingale}, {Nucita}, {Patrizii}, {Potter}, {Riccio}, {S{\'a}nchez}, {Sapone}, {Schewtschenko}, {Schultheis}, {Scottez}, {Teyssier}, {Tutusaus}, {Valiviita}, {Viel}, {Vriend}, \& {Whittaker}}]{2022A&A...662A.112E}
{Euclid Collaboration}, {Scaramella}, R., {Amiaux}, J., {et~al.} 2022, \aap, 662, A112

\bibitem[{{Fan} {et~al.}(2023){Fan}, {Ba{\~n}ados}, \& {Simcoe}}]{2023ARA&A..61..373F}
{Fan}, X., {Ba{\~n}ados}, E., \& {Simcoe}, R.~A. 2023, \araa, 61, 373

\bibitem[{{Fitzpatrick}(1999)}]{1999PASP..111...63F}
{Fitzpatrick}, E.~L. 1999, \pasp, 111, 63

\bibitem[{{Flesch}(2021)}]{2021arXiv210512985F}
{Flesch}, E.~W. 2021, arXiv e-prints, arXiv:2105.12985

\bibitem[{{Gaia Collaboration} {et~al.}(2016){Gaia Collaboration}, {Prusti}, {de Bruijne}, {Brown}, {Vallenari}, {Babusiaux}, {Bailer-Jones}, {Bastian}, {Biermann}, {Evans}, {Eyer}, {Jansen}, {Jordi}, {Klioner}, {Lammers}, {Lindegren}, {Luri}, {Mignard}, {Milligan}, {Panem}, {Poinsignon}, {Pourbaix}, {Randich}, {Sarri}, {Sartoretti}, {Siddiqui}, {Soubiran}, {Valette}, {van Leeuwen}, {Walton}, {Aerts}, {Arenou}, {Cropper}, {Drimmel}, {H{\o}g}, {Katz}, {Lattanzi}, {O'Mullane}, {Grebel}, {Holland}, {Huc}, {Passot}, {Bramante}, {Cacciari}, {Casta{\~n}eda}, {Chaoul}, {Cheek}, {De Angeli}, {Fabricius}, {Guerra}, {Hern{\'a}ndez}, {Jean-Antoine-Piccolo}, {Masana}, {Messineo}, {Mowlavi}, {Nienartowicz}, {Ord{\'o}{\~n}ez-Blanco}, {Panuzzo}, {Portell}, {Richards}, {Riello}, {Seabroke}, {Tanga}, {Th{\'e}venin}, {Torra}, {Els}, {Gracia-Abril}, {Comoretto}, {Garcia-Reinaldos}, {Lock}, {Mercier}, {Altmann}, {Andrae}, {Astraatmadja}, {Bellas-Velidis}, {Benson}, {Berthier}, {Blomme}, {Busso}, {Carry}, {Cellino}, {Clementini},
  {Cowell}, {Creevey}, {Cuypers}, {Davidson}, {De Ridder}, {de Torres}, {Delchambre}, {Dell'Oro}, {Ducourant}, {Fr{\'e}mat}, {Garc{\'\i}a-Torres}, {Gosset}, {Halbwachs}, {Hambly}, {Harrison}, {Hauser}, {Hestroffer}, {Hodgkin}, {Huckle}, {Hutton}, {Jasniewicz}, {Jordan}, {Kontizas}, {Korn}, {Lanzafame}, {Manteiga}, {Moitinho}, {Muinonen}, {Osinde}, {Pancino}, {Pauwels}, {Petit}, {Recio-Blanco}, {Robin}, {Sarro}, {Siopis}, {Smith}, {Smith}, {Sozzetti}, {Thuillot}, {van Reeven}, {Viala}, {Abbas}, {Abreu Aramburu}, {Accart}, {Aguado}, {Allan}, {Allasia}, {Altavilla}, {{\'A}lvarez}, {Alves}, {Anderson}, {Andrei}, {Anglada Varela}, {Antiche}, {Antoja}, {Ant{\'o}n}, {Arcay}, {Atzei}, {Ayache}, {Bach}, {Baker}, {Balaguer-N{\'u}{\~n}ez}, {Barache}, {Barata}, {Barbier}, {Barblan}, {Baroni}, {Barrado y Navascu{\'e}s}, {Barros}, {Barstow}, {Becciani}, {Bellazzini}, {Bellei}, {Bello Garc{\'\i}a}, {Belokurov}, {Bendjoya}, {Berihuete}, {Bianchi}, {Bienaym{\'e}}, {Billebaud}, {Blagorodnova}, {Blanco-Cuaresma}, {Boch},
  {Bombrun}, {Borrachero}, {Bouquillon}, {Bourda}, {Bouy}, {Bragaglia}, {Breddels}, {Brouillet}, {Br{\"u}semeister}, {Bucciarelli}, {Budnik}, {Burgess}, {Burgon}, {Burlacu}, {Busonero}, {Buzzi}, {Caffau}, {Cambras}, {Campbell}, {Cancelliere}, {Cantat-Gaudin}, {Carlucci}, {Carrasco}, {Castellani}, {Charlot}, {Charnas}, {Charvet}, {Chassat}, {Chiavassa}, {Clotet}, {Cocozza}, {Collins}, {Collins}, {Costigan}, {Crifo}, {Cross}, {Crosta}, {Crowley}, {Dafonte}, {Damerdji}, {Dapergolas}, {David}, {David}, {De Cat}, {de Felice}, {de Laverny}, {De Luise}, {De March}, {de Martino}, {de Souza}, {Debosscher}, {del Pozo}, {Delbo}, {Delgado}, {Delgado}, {di Marco}, {Di Matteo}, {Diakite}, {Distefano}, {Dolding}, {Dos Anjos}, {Drazinos}, {Dur{\'a}n}, {Dzigan}, {Ecale}, {Edvardsson}, {Enke}, {Erdmann}, {Escolar}, {Espina}, {Evans}, {Eynard Bontemps}, {Fabre}, {Fabrizio}, {Faigler}, {Falc{\~a}o}, {Farr{\`a}s Casas}, {Faye}, {Federici}, {Fedorets}, {Fern{\'a}ndez-Hern{\'a}ndez}, {Fernique}, {Fienga}, {Figueras}, {Filippi},
  {Findeisen}, {Fonti}, {Fouesneau}, {Fraile}, {Fraser}, {Fuchs}, {Furnell}, {Gai}, {Galleti}, {Galluccio}, {Garabato}, {Garc{\'\i}a-Sedano}, {Gar{\'e}}, {Garofalo}, {Garralda}, {Gavras}, {Gerssen}, {Geyer}, {Gilmore}, {Girona}, {Giuffrida}, {Gomes}, {Gonz{\'a}lez-Marcos}, {Gonz{\'a}lez-N{\'u}{\~n}ez}, {Gonz{\'a}lez-Vidal}, {Granvik}, {Guerrier}, {Guillout}, {Guiraud}, {G{\'u}rpide}, {Guti{\'e}rrez-S{\'a}nchez}, {Guy}, {Haigron}, {Hatzidimitriou}, {Haywood}, {Heiter}, {Helmi}, {Hobbs}, {Hofmann}, {Holl}, {Holland}, {Hunt}, {Hypki}, {Icardi}, {Irwin}, {Jevardat de Fombelle}, {Jofr{\'e}}, {Jonker}, {Jorissen}, {Julbe}, {Karampelas}, {Kochoska}, {Kohley}, {Kolenberg}, {Kontizas}, {Koposov}, {Kordopatis}, {Koubsky}, {Kowalczyk}, {Krone-Martins}, {Kudryashova}, {Kull}, {Bachchan}, {Lacoste-Seris}, {Lanza}, {Lavigne}, {Le Poncin-Lafitte}, {Lebreton}, {Lebzelter}, {Leccia}, {Leclerc}, {Lecoeur-Taibi}, {Lemaitre}, {Lenhardt}, {Leroux}, {Liao}, {Licata}, {Lindstr{\o}m}, {Lister}, {Livanou}, {Lobel}, {L{\"o}ffler},
  {L{\'o}pez}, {Lopez-Lozano}, {Lorenz}, {Loureiro}, {MacDonald}, {Magalh{\~a}es Fernandes}, {Managau}, {Mann}, {Mantelet}, {Marchal}, {Marchant}, {Marconi}, {Marie}, {Marinoni}, {Marrese}, {Marschalk{\'o}}, {Marshall}, {Mart{\'\i}n-Fleitas}, {Martino}, {Mary}, {Matijevi{\v{c}}}, {Mazeh}, {McMillan}, {Messina}, {Mestre}, {Michalik}, {Millar}, {Miranda}, {Molina}, {Molinaro}, {Molinaro}, {Moln{\'a}r}, {Moniez}, {Montegriffo}, {Monteiro}, {Mor}, {Mora}, {Morbidelli}, {Morel}, {Morgenthaler}, {Morley}, {Morris}, {Mulone}, {Muraveva}, {Musella}, {Narbonne}, {Nelemans}, {Nicastro}, {Noval}, {Ord{\'e}novic}, {Ordieres-Mer{\'e}}, {Osborne}, {Pagani}, {Pagano}, {Pailler}, {Palacin}, {Palaversa}, {Parsons}, {Paulsen}, {Pecoraro}, {Pedrosa}, {Pentik{\"a}inen}, {Pereira}, {Pichon}, {Piersimoni}, {Pineau}, {Plachy}, {Plum}, {Poujoulet}, {Pr{\v{s}}a}, {Pulone}, {Ragaini}, {Rago}, {Rambaux}, {Ramos-Lerate}, {Ranalli}, {Rauw}, {Read}, {Regibo}, {Renk}, {Reyl{\'e}}, {Ribeiro}, {Rimoldini}, {Ripepi}, {Riva}, {Rixon},
  {Roelens}, {Romero-G{\'o}mez}, {Rowell}, {Royer}, {Rudolph}, {Ruiz-Dern}, {Sadowski}, {Sagrist{\`a} Sell{\'e}s}, {Sahlmann}, {Salgado}, {Salguero}, {Sarasso}, {Savietto}, {Schnorhk}, {Schultheis}, {Sciacca}, {Segol}, {Segovia}, {Segransan}, {Serpell}, {Shih}, {Smareglia}, {Smart}, {Smith}, {Solano}, {Solitro}, {Sordo}, {Soria Nieto}, {Souchay}, {Spagna}, {Spoto}, {Stampa}, {Steele}, {Steidelm{\"u}ller}, {Stephenson}, {Stoev}, {Suess}, {S{\"u}veges}, {Surdej}, {Szabados}, {Szegedi-Elek}, {Tapiador}, {Taris}, {Tauran}, {Taylor}, {Teixeira}, {Terrett}, {Tingley}, {Trager}, {Turon}, {Ulla}, {Utrilla}, {Valentini}, {van Elteren}, {Van Hemelryck}, {van Leeuwen}, {Varadi}, {Vecchiato}, {Veljanoski}, {Via}, {Vicente}, {Vogt}, {Voss}, {Votruba}, {Voutsinas}, {Walmsley}, {Weiler}, {Weingrill}, {Werner}, {Wevers}, {Whitehead}, {Wyrzykowski}, {Yoldas}, {{\v{Z}}erjal}, {Zucker}, {Zurbach}, {Zwitter}, {Alecu}, {Allen}, {Allende Prieto}, {Amorim}, {Anglada-Escud{\'e}}, {Arsenijevic}, {Azaz}, {Balm}, {Beck}, {Bernstein},
  {Bigot}, {Bijaoui}, {Blasco}, {Bonfigli}, {Bono}, {Boudreault}, {Bressan}, {Brown}, {Brunet}, {Bunclark}, {Buonanno}, {Butkevich}, {Carret}, {Carrion}, {Chemin}, {Ch{\'e}reau}, {Corcione}, {Darmigny}, {de Boer}, {de Teodoro}, {de Zeeuw}, {Delle Luche}, {Domingues}, {Dubath}, {Fodor}, {Fr{\'e}zouls}, {Fries}, {Fustes}, {Fyfe}, {Gallardo}, {Gallegos}, {Gardiol}, {Gebran}, {Gomboc}, {G{\'o}mez}, {Grux}, {Gueguen}, {Heyrovsky}, {Hoar}, {Iannicola}, {Isasi Parache}, {Janotto}, {Joliet}, {Jonckheere}, {Keil}, {Kim}, {Klagyivik}, {Klar}, {Knude}, {Kochukhov}, {Kolka}, {Kos}, {Kutka}, {Lainey}, {LeBouquin}, {Liu}, {Loreggia}, {Makarov}, {Marseille}, {Martayan}, {Martinez-Rubi}, {Massart}, {Meynadier}, {Mignot}, {Munari}, {Nguyen}, {Nordlander}, {Ocvirk}, {O'Flaherty}, {Olias Sanz}, {Ortiz}, {Osorio}, {Oszkiewicz}, {Ouzounis}, {Palmer}, {Park}, {Pasquato}, {Peltzer}, {Peralta}, {P{\'e}turaud}, {Pieniluoma}, {Pigozzi}, {Poels}, {Prat}, {Prod'homme}, {Raison}, {Rebordao}, {Risquez}, {Rocca-Volmerange}, {Rosen},
  {Ruiz-Fuertes}, {Russo}, {Sembay}, {Serraller Vizcaino}, {Short}, {Siebert}, {Silva}, {Sinachopoulos}, {Slezak}, {Soffel}, {Sosnowska}, {Strai{\v{z}}ys}, {ter Linden}, {Terrell}, {Theil}, {Tiede}, {Troisi}, {Tsalmantza}, {Tur}, {Vaccari}, {Vachier}, {Valles}, {Van Hamme}, {Veltz}, {Virtanen}, {Wallut}, {Wichmann}, {Wilkinson}, {Ziaeepour}, \& {Zschocke}}]{2016A&A...595A...1G}
{Gaia Collaboration}, {Prusti}, T., {de Bruijne}, J.~H.~J., {et~al.} 2016, \aap, 595, A1

\bibitem[{{Gaia Collaboration} {et~al.}(2023){Gaia Collaboration}, {Vallenari}, {Brown}, {Prusti}, {de Bruijne}, {Arenou}, {Babusiaux}, {Biermann}, {Creevey}, {Ducourant}, {Evans}, {Eyer}, {Guerra}, {Hutton}, {Jordi}, {Klioner}, {Lammers}, {Lindegren}, {Luri}, {Mignard}, {Panem}, {Pourbaix}, {Randich}, {Sartoretti}, {Soubiran}, {Tanga}, {Walton}, {Bailer-Jones}, {Bastian}, {Drimmel}, {Jansen}, {Katz}, {Lattanzi}, {van Leeuwen}, {Bakker}, {Cacciari}, {Casta{\~n}eda}, {De Angeli}, {Fabricius}, {Fouesneau}, {Fr{\'e}mat}, {Galluccio}, {Guerrier}, {Heiter}, {Masana}, {Messineo}, {Mowlavi}, {Nicolas}, {Nienartowicz}, {Pailler}, {Panuzzo}, {Riclet}, {Roux}, {Seabroke}, {Sordo}, {Th{\'e}venin}, {Gracia-Abril}, {Portell}, {Teyssier}, {Altmann}, {Andrae}, {Audard}, {Bellas-Velidis}, {Benson}, {Berthier}, {Blomme}, {Burgess}, {Busonero}, {Busso}, {C{\'a}novas}, {Carry}, {Cellino}, {Cheek}, {Clementini}, {Damerdji}, {Davidson}, {de Teodoro}, {Nu{\~n}ez Campos}, {Delchambre}, {Dell'Oro}, {Esquej},
  {Fern{\'a}ndez-Hern{\'a}ndez}, {Fraile}, {Garabato}, {Garc{\'\i}a-Lario}, {Gosset}, {Haigron}, {Halbwachs}, {Hambly}, {Harrison}, {Hern{\'a}ndez}, {Hestroffer}, {Hodgkin}, {Holl}, {Jan{\ss}en}, {Jevardat de Fombelle}, {Jordan}, {Krone-Martins}, {Lanzafame}, {L{\"o}ffler}, {Marchal}, {Marrese}, {Moitinho}, {Muinonen}, {Osborne}, {Pancino}, {Pauwels}, {Recio-Blanco}, {Reyl{\'e}}, {Riello}, {Rimoldini}, {Roegiers}, {Rybizki}, {Sarro}, {Siopis}, {Smith}, {Sozzetti}, {Utrilla}, {van Leeuwen}, {Abbas}, {{\'A}brah{\'a}m}, {Abreu Aramburu}, {Aerts}, {Aguado}, {Ajaj}, {Aldea-Montero}, {Altavilla}, {{\'A}lvarez}, {Alves}, {Anders}, {Anderson}, {Anglada Varela}, {Antoja}, {Baines}, {Baker}, {Balaguer-N{\'u}{\~n}ez}, {Balbinot}, {Balog}, {Barache}, {Barbato}, {Barros}, {Barstow}, {Bartolom{\'e}}, {Bassilana}, {Bauchet}, {Becciani}, {Bellazzini}, {Berihuete}, {Bernet}, {Bertone}, {Bianchi}, {Binnenfeld}, {Blanco-Cuaresma}, {Blazere}, {Boch}, {Bombrun}, {Bossini}, {Bouquillon}, {Bragaglia}, {Bramante}, {Breedt},
  {Bressan}, {Brouillet}, {Brugaletta}, {Bucciarelli}, {Burlacu}, {Butkevich}, {Buzzi}, {Caffau}, {Cancelliere}, {Cantat-Gaudin}, {Carballo}, {Carlucci}, {Carnerero}, {Carrasco}, {Casamiquela}, {Castellani}, {Castro-Ginard}, {Chaoul}, {Charlot}, {Chemin}, {Chiaramida}, {Chiavassa}, {Chornay}, {Comoretto}, {Contursi}, {Cooper}, {Cornez}, {Cowell}, {Crifo}, {Cropper}, {Crosta}, {Crowley}, {Dafonte}, {Dapergolas}, {David}, {David}, {de Laverny}, {De Luise}, {De March}, {De Ridder}, {de Souza}, {de Torres}, {del Peloso}, {del Pozo}, {Delbo}, {Delgado}, {Delisle}, {Demouchy}, {Dharmawardena}, {Di Matteo}, {Diakite}, {Diener}, {Distefano}, {Dolding}, {Edvardsson}, {Enke}, {Fabre}, {Fabrizio}, {Faigler}, {Fedorets}, {Fernique}, {Fienga}, {Figueras}, {Fournier}, {Fouron}, {Fragkoudi}, {Gai}, {Garcia-Gutierrez}, {Garcia-Reinaldos}, {Garc{\'\i}a-Torres}, {Garofalo}, {Gavel}, {Gavras}, {Gerlach}, {Geyer}, {Giacobbe}, {Gilmore}, {Girona}, {Giuffrida}, {Gomel}, {Gomez}, {Gonz{\'a}lez-N{\'u}{\~n}ez},
  {Gonz{\'a}lez-Santamar{\'\i}a}, {Gonz{\'a}lez-Vidal}, {Granvik}, {Guillout}, {Guiraud}, {Guti{\'e}rrez-S{\'a}nchez}, {Guy}, {Hatzidimitriou}, {Hauser}, {Haywood}, {Helmer}, {Helmi}, {Sarmiento}, {Hidalgo}, {Hilger}, {H{\l}adczuk}, {Hobbs}, {Holland}, {Huckle}, {Jardine}, {Jasniewicz}, {Jean-Antoine Piccolo}, {Jim{\'e}nez-Arranz}, {Jorissen}, {Juaristi Campillo}, {Julbe}, {Karbevska}, {Kervella}, {Khanna}, {Kontizas}, {Kordopatis}, {Korn}, {K{\'o}sp{\'a}l}, {Kostrzewa-Rutkowska}, {Kruszy{\'n}ska}, {Kun}, {Laizeau}, {Lambert}, {Lanza}, {Lasne}, {Le Campion}, {Lebreton}, {Lebzelter}, {Leccia}, {Leclerc}, {Lecoeur-Taibi}, {Liao}, {Licata}, {Lindstr{\o}m}, {Lister}, {Livanou}, {Lobel}, {Lorca}, {Loup}, {Madrero Pardo}, {Magdaleno Romeo}, {Managau}, {Mann}, {Manteiga}, {Marchant}, {Marconi}, {Marcos}, {Marcos Santos}, {Mar{\'\i}n Pina}, {Marinoni}, {Marocco}, {Marshall}, {Martin Polo}, {Mart{\'\i}n-Fleitas}, {Marton}, {Mary}, {Masip}, {Massari}, {Mastrobuono-Battisti}, {Mazeh}, {McMillan}, {Messina}, {Michalik},
  {Millar}, {Mints}, {Molina}, {Molinaro}, {Moln{\'a}r}, {Monari}, {Mongui{\'o}}, {Montegriffo}, {Montero}, {Mor}, {Mora}, {Morbidelli}, {Morel}, {Morris}, {Muraveva}, {Murphy}, {Musella}, {Nagy}, {Noval}, {Oca{\~n}a}, {Ogden}, {Ordenovic}, {Osinde}, {Pagani}, {Pagano}, {Palaversa}, {Palicio}, {Pallas-Quintela}, {Panahi}, {Payne-Wardenaar}, {Pe{\~n}alosa Esteller}, {Penttil{\"a}}, {Pichon}, {Piersimoni}, {Pineau}, {Plachy}, {Plum}, {Poggio}, {Pr{\v{s}}a}, {Pulone}, {Racero}, {Ragaini}, {Rainer}, {Raiteri}, {Rambaux}, {Ramos}, {Ramos-Lerate}, {Re Fiorentin}, {Regibo}, {Richards}, {Rios Diaz}, {Ripepi}, {Riva}, {Rix}, {Rixon}, {Robichon}, {Robin}, {Robin}, {Roelens}, {Rogues}, {Rohrbasser}, {Romero-G{\'o}mez}, {Rowell}, {Royer}, {Ruz Mieres}, {Rybicki}, {Sadowski}, {S{\'a}ez N{\'u}{\~n}ez}, {Sagrist{\`a} Sell{\'e}s}, {Sahlmann}, {Salguero}, {Samaras}, {Sanchez Gimenez}, {Sanna}, {Santove{\~n}a}, {Sarasso}, {Schultheis}, {Sciacca}, {Segol}, {Segovia}, {S{\'e}gransan}, {Semeux}, {Shahaf}, {Siddiqui}, {Siebert},
  {Siltala}, {Silvelo}, {Slezak}, {Slezak}, {Smart}, {Snaith}, {Solano}, {Solitro}, {Souami}, {Souchay}, {Spagna}, {Spina}, {Spoto}, {Steele}, {Steidelm{\"u}ller}, {Stephenson}, {S{\"u}veges}, {Surdej}, {Szabados}, {Szegedi-Elek}, {Taris}, {Taylor}, {Teixeira}, {Tolomei}, {Tonello}, {Torra}, {Torra}, {Torralba Elipe}, {Trabucchi}, {Tsounis}, {Turon}, {Ulla}, {Unger}, {Vaillant}, {van Dillen}, {van Reeven}, {Vanel}, {Vecchiato}, {Viala}, {Vicente}, {Voutsinas}, {Weiler}, {Wevers}, {Wyrzykowski}, {Yoldas}, {Yvard}, {Zhao}, {Zorec}, {Zucker}, \& {Zwitter}}]{2023A&A...674A...1G}
{Gaia Collaboration}, {Vallenari}, A., {Brown}, A.~G.~A., {et~al.} 2023, \aap, 674, A1

\bibitem[{{Glikman} {et~al.}(2023){Glikman}, {Rusu}, {Chen}, {Chan}, {Spingola}, {Stacey}, {McKean}, {Berghea}, {Djorgovski}, {Graham}, {Stern}, {Urrutia}, {Lacy}, {Secrest}, \& {O'Meara}}]{2023ApJ...943...25G}
{Glikman}, E., {Rusu}, C.~E., {Chen}, G. C.~F., {et~al.} 2023, \apj, 943, 25

\bibitem[{{Gordon} {et~al.}(2020){Gordon}, {Boyce}, {O'Dea}, {Rudnick}, {Andernach}, {Vantyghem}, {Baum}, {Bui}, \& {Dionyssiou}}]{2020RNAAS...4..175G}
{Gordon}, Y.~A., {Boyce}, M.~M., {O'Dea}, C.~P., {et~al.} 2020, Research Notes of the American Astronomical Society, 4, 175

\bibitem[{{Gordon} {et~al.}(2021){Gordon}, {Boyce}, {O'Dea}, {Rudnick}, {Andernach}, {Vantyghem}, {Baum}, {Bui}, {Dionyssiou}, {Safi-Harb}, \& {Sander}}]{2021ApJS..255...30G}
{Gordon}, Y.~A., {Boyce}, M.~M., {O'Dea}, C.~P., {et~al.} 2021, \apjs, 255, 30

\bibitem[{{Gordon} {et~al.}(2023){Gordon}, {Rudnick}, {Andernach}, {Morabito}, {O'Dea}, {Achong}, {Baum}, {Bayona-Figueroa}, {Hooper}, {Mingo}, {Morris}, \& {Vantyghem}}]{2023ApJS..267...37G}
{Gordon}, Y.~A., {Rudnick}, L., {Andernach}, H., {et~al.} 2023, \apjs, 267, 37

\bibitem[{{Green}(2018)}]{2018JOSS....3..695G}
{Green}, G.~M. 2018, The Journal of Open Source Software, 3, 695

\bibitem[{{Gu} {et~al.}(2022){Gu}, {Huang}, {Sheu}, {Aldering}, {Bolton}, {Boone}, {Dey}, {Filipp}, {Jullo}, {Perlmutter}, {Rubin}, {Schlafly}, {Schlegel}, {Shu}, \& {Suyu}}]{2022ApJ...935...49G}
{Gu}, A., {Huang}, X., {Sheu}, W., {et~al.} 2022, \apj, 935, 49

\bibitem[{{Harris} {et~al.}(2020){Harris}, {Millman}, {van der Walt}, {Gommers}, {Virtanen}, {Cournapeau}, {Wieser}, {Taylor}, {Berg}, {Smith}, {Kern}, {Picus}, {Hoyer}, {van Kerkwijk}, {Brett}, {Haldane}, {del R{\'\i}o}, {Wiebe}, {Peterson}, {G{\'e}rard-Marchant}, {Sheppard}, {Reddy}, {Weckesser}, {Abbasi}, {Gohlke}, \& {Oliphant}}]{2020Natur.585..357H}
{Harris}, C.~R., {Millman}, K.~J., {van der Walt}, S.~J., {et~al.} 2020, \nat, 585, 357

\bibitem[{{Hezaveh} {et~al.}(2017){Hezaveh}, {Perreault Levasseur}, \& {Marshall}}]{2017Natur.548..555H}
{Hezaveh}, Y.~D., {Perreault Levasseur}, L., \& {Marshall}, P.~J. 2017, \nat, 548, 555

\bibitem[{{Inada} {et~al.}(2003){Inada}, {Becker}, {Burles}, {Castander}, {Eisenstein}, {Hall}, {Johnston}, {Pindor}, {Richards}, {Schechter}, {Sekiguchi}, {White}, {Brinkmann}, {Frieman}, {Kleinman}, {Krzesi{\'n}ski}, {Long}, {Neilsen}, {Newman}, {Nitta}, {Schneider}, {Snedden}, \& {York}}]{2003AJ....126..666I}
{Inada}, N., {Becker}, R.~H., {Burles}, S., {et~al.} 2003, \aj, 126, 666

\bibitem[{{Inada} {et~al.}(2008){Inada}, {Oguri}, {Becker}, {Shin}, {Richards}, {Hennawi}, {White}, {Pindor}, {Strauss}, {Kochanek}, {Johnston}, {Gregg}, {Kayo}, {Eisenstein}, {Hall}, {Castander}, {Clocchiatti}, {Anderson}, {Schneider}, {York}, {Lupton}, {Chiu}, {Kawano}, {Scranton}, {Frieman}, {Keeton}, {Morokuma}, {Rix}, {Turner}, {Burles}, {Brunner}, {Sheldon}, {Bahcall}, \& {Masataka}}]{2008AJ....135..496I}
{Inada}, N., {Oguri}, M., {Becker}, R.~H., {et~al.} 2008, \aj, 135, 496

\bibitem[{{Inada} {et~al.}(2012){Inada}, {Oguri}, {Shin}, {Kayo}, {Strauss}, {Morokuma}, {Rusu}, {Fukugita}, {Kochanek}, {Richards}, {Schneider}, {York}, {Bahcall}, {Frieman}, {Hall}, \& {White}}]{2012AJ....143..119I}
{Inada}, N., {Oguri}, M., {Shin}, M.-S., {et~al.} 2012, \aj, 143, 119

\bibitem[{{Inayoshi} {et~al.}(2020){Inayoshi}, {Visbal}, \& {Haiman}}]{2020ARA&A..58...27I}
{Inayoshi}, K., {Visbal}, E., \& {Haiman}, Z. 2020, \araa, 58, 27

\bibitem[{{Ivezi{\'c}} {et~al.}(2019){Ivezi{\'c}}, {Kahn}, {Tyson}, {Abel}, {Acosta}, {Allsman}, {Alonso}, {AlSayyad}, {Anderson}, {Andrew}, {Angel}, {Angeli}, {Ansari}, {Antilogus}, {Araujo}, {Armstrong}, {Arndt}, {Astier}, {Aubourg}, {Auza}, {Axelrod}, {Bard}, {Barr}, {Barrau}, {Bartlett}, {Bauer}, {Bauman}, {Baumont}, {Bechtol}, {Bechtol}, {Becker}, {Becla}, {Beldica}, {Bellavia}, {Bianco}, {Biswas}, {Blanc}, {Blazek}, {Blandford}, {Bloom}, {Bogart}, {Bond}, {Booth}, {Borgland}, {Borne}, {Bosch}, {Boutigny}, {Brackett}, {Bradshaw}, {Brandt}, {Brown}, {Bullock}, {Burchat}, {Burke}, {Cagnoli}, {Calabrese}, {Callahan}, {Callen}, {Carlin}, {Carlson}, {Chandrasekharan}, {Charles-Emerson}, {Chesley}, {Cheu}, {Chiang}, {Chiang}, {Chirino}, {Chow}, {Ciardi}, {Claver}, {Cohen-Tanugi}, {Cockrum}, {Coles}, {Connolly}, {Cook}, {Cooray}, {Covey}, {Cribbs}, {Cui}, {Cutri}, {Daly}, {Daniel}, {Daruich}, {Daubard}, {Daues}, {Dawson}, {Delgado}, {Dellapenna}, {de Peyster}, {de Val-Borro}, {Digel}, {Doherty}, {Dubois},
  {Dubois-Felsmann}, {Durech}, {Economou}, {Eifler}, {Eracleous}, {Emmons}, {Fausti Neto}, {Ferguson}, {Figueroa}, {Fisher-Levine}, {Focke}, {Foss}, {Frank}, {Freemon}, {Gangler}, {Gawiser}, {Geary}, {Gee}, {Geha}, {Gessner}, {Gibson}, {Gilmore}, {Glanzman}, {Glick}, {Goldina}, {Goldstein}, {Goodenow}, {Graham}, {Gressler}, {Gris}, {Guy}, {Guyonnet}, {Haller}, {Harris}, {Hascall}, {Haupt}, {Hernandez}, {Herrmann}, {Hileman}, {Hoblitt}, {Hodgson}, {Hogan}, {Howard}, {Huang}, {Huffer}, {Ingraham}, {Innes}, {Jacoby}, {Jain}, {Jammes}, {Jee}, {Jenness}, {Jernigan}, {Jevremovi{\'c}}, {Johns}, {Johnson}, {Johnson}, {Jones}, {Juramy-Gilles}, {Juri{\'c}}, {Kalirai}, {Kallivayalil}, {Kalmbach}, {Kantor}, {Karst}, {Kasliwal}, {Kelly}, {Kessler}, {Kinnison}, {Kirkby}, {Knox}, {Kotov}, {Krabbendam}, {Krughoff}, {Kub{\'a}nek}, {Kuczewski}, {Kulkarni}, {Ku}, {Kurita}, {Lage}, {Lambert}, {Lange}, {Langton}, {Le Guillou}, {Levine}, {Liang}, {Lim}, {Lintott}, {Long}, {Lopez}, {Lotz}, {Lupton}, {Lust}, {MacArthur}, {Mahabal},
  {Mandelbaum}, {Markiewicz}, {Marsh}, {Marshall}, {Marshall}, {May}, {McKercher}, {McQueen}, {Meyers}, {Migliore}, {Miller}, {Mills}, {Miraval}, {Moeyens}, {Moolekamp}, {Monet}, {Moniez}, {Monkewitz}, {Montgomery}, {Morrison}, {Mueller}, {Muller}, {Mu{\~n}oz Arancibia}, {Neill}, {Newbry}, {Nief}, {Nomerotski}, {Nordby}, {O'Connor}, {Oliver}, {Olivier}, {Olsen}, {O'Mullane}, {Ortiz}, {Osier}, {Owen}, {Pain}, {Palecek}, {Parejko}, {Parsons}, {Pease}, {Peterson}, {Peterson}, {Petravick}, {Libby Petrick}, {Petry}, {Pierfederici}, {Pietrowicz}, {Pike}, {Pinto}, {Plante}, {Plate}, {Plutchak}, {Price}, {Prouza}, {Radeka}, {Rajagopal}, {Rasmussen}, {Regnault}, {Reil}, {Reiss}, {Reuter}, {Ridgway}, {Riot}, {Ritz}, {Robinson}, {Roby}, {Roodman}, {Rosing}, {Roucelle}, {Rumore}, {Russo}, {Saha}, {Sassolas}, {Schalk}, {Schellart}, {Schindler}, {Schmidt}, {Schneider}, {Schneider}, {Schoening}, {Schumacher}, {Schwamb}, {Sebag}, {Selvy}, {Sembroski}, {Seppala}, {Serio}, {Serrano}, {Shaw}, {Shipsey}, {Sick}, {Silvestri},
  {Slater}, {Smith}, {Smith}, {Sobhani}, {Soldahl}, {Storrie-Lombardi}, {Stover}, {Strauss}, {Street}, {Stubbs}, {Sullivan}, {Sweeney}, {Swinbank}, {Szalay}, {Takacs}, {Tether}, {Thaler}, {Thayer}, {Thomas}, {Thornton}, {Thukral}, {Tice}, {Trilling}, {Turri}, {Van Berg}, {Vanden Berk}, {Vetter}, {Virieux}, {Vucina}, {Wahl}, {Walkowicz}, {Walsh}, {Walter}, {Wang}, {Wang}, {Warner}, {Wiecha}, {Willman}, {Winters}, {Wittman}, {Wolff}, {Wood-Vasey}, {Wu}, {Xin}, {Yoachim}, \& {Zhan}}]{2019ApJ...873..111I}
{Ivezi{\'c}}, {\v{Z}}., {Kahn}, S.~M., {Tyson}, J.~A., {et~al.} 2019, \apj, 873, 111

\bibitem[{{Jackson} {et~al.}(1995){Jackson}, {de Bruyn}, {Myers}, {Bremer}, {Miley}, {Schilizzi}, {Browne}, {Nair}, {Wilkinson}, {Blandford}, {Pearson}, \& {Readhead}}]{1995MNRAS.274L..25J}
{Jackson}, N., {de Bruyn}, A.~G., {Myers}, S., {et~al.} 1995, \mnras, 274, L25

\bibitem[{{Jackson} {et~al.}(2008){Jackson}, {Ofek}, \& {Oguri}}]{2008MNRAS.387..741J}
{Jackson}, N., {Ofek}, E.~O., \& {Oguri}, M. 2008, \mnras, 387, 741

\bibitem[{{Jaelani} {et~al.}(2021){Jaelani}, {Rusu}, {Kayo}, {More}, {Sonnenfeld}, {Silverman}, {Schramm}, {Anguita}, {Inada}, {Kondo}, {Schechter}, {Lee}, {Oguri}, {Chan}, {Wong}, \& {Inoue}}]{2021MNRAS.502.1487J}
{Jaelani}, A.~T., {Rusu}, C.~E., {Kayo}, I., {et~al.} 2021, \mnras, 502, 1487

\bibitem[{{Kingma} \& {Ba}(2014)}]{2014arXiv1412.6980K}
{Kingma}, D.~P. \& {Ba}, J. 2014, arXiv e-prints, arXiv:1412.6980

\bibitem[{{Kingma} \& {Welling}(2019)}]{2019arXiv190602691K}
{Kingma}, D.~P. \& {Welling}, M. 2019, arXiv e-prints, arXiv:1906.02691

\bibitem[{{Krone-Martins} {et~al.}(2019){Krone-Martins}, {Graham}, {Stern}, {Djorgovski}, {Delchambre}, {Ducourant}, {Teixeira}, {Drake}, {Scarano}, {Surdej}, {Galluccio}, {Jalan}, {Wertz}, {Kl{\"u}ter}, {Mignard}, {Spindola-Duarte}, {Dobie}, {Slezak}, {Sluse}, {Murphy}, {Boehm}, {Nierenberg}, {Bastian}, {Wambsganss}, \& {LeCampion}}]{2019arXiv191208977K}
{Krone-Martins}, A., {Graham}, M.~J., {Stern}, D., {et~al.} 2019, arXiv e-prints, arXiv:1912.08977

\bibitem[{{Lacy} {et~al.}(2020){Lacy}, {Baum}, {Chandler}, {Chatterjee}, {Clarke}, {Deustua}, {English}, {Farnes}, {Gaensler}, {Gugliucci}, {Hallinan}, {Kent}, {Kimball}, {Law}, {Lazio}, {Marvil}, {Mao}, {Medlin}, {Mooley}, {Murphy}, {Myers}, {Osten}, {Richards}, {Rosolowsky}, {Rudnick}, {Schinzel}, {Sivakoff}, {Sjouwerman}, {Taylor}, {White}, {Wrobel}, {Andernach}, {Beasley}, {Berger}, {Bhatnager}, {Birkinshaw}, {Bower}, {Brandt}, {Brown}, {Burke-Spolaor}, {Butler}, {Comerford}, {Demorest}, {Fu}, {Giacintucci}, {Golap}, {G{\"u}th}, {Hales}, {Hiriart}, {Hodge}, {Horesh}, {Ivezi{\'c}}, {Jarvis}, {Kamble}, {Kassim}, {Liu}, {Loinard}, {Lyons}, {Masters}, {Mezcua}, {Moellenbrock}, {Mroczkowski}, {Nyland}, {O'Dea}, {O'Sullivan}, {Peters}, {Radford}, {Rao}, {Robnett}, {Salcido}, {Shen}, {Sobotka}, {Witz}, {Vaccari}, {van Weeren}, {Vargas}, {Williams}, \& {Yoon}}]{2020PASP..132c5001L}
{Lacy}, M., {Baum}, S.~A., {Chandler}, C.~J., {et~al.} 2020, \pasp, 132, 035001

\bibitem[{{Laureijs} {et~al.}(2011){Laureijs}, {Amiaux}, {Arduini}, {Augu{\`e}res}, {Brinchmann}, {Cole}, {Cropper}, {Dabin}, {Duvet}, {Ealet}, {Garilli}, {Gondoin}, {Guzzo}, {Hoar}, {Hoekstra}, {Holmes}, {Kitching}, {Maciaszek}, {Mellier}, {Pasian}, {Percival}, {Rhodes}, {Saavedra Criado}, {Sauvage}, {Scaramella}, {Valenziano}, {Warren}, {Bender}, {Castander}, {Cimatti}, {Le F{\`e}vre}, {Kurki-Suonio}, {Levi}, {Lilje}, {Meylan}, {Nichol}, {Pedersen}, {Popa}, {Rebolo Lopez}, {Rix}, {Rottgering}, {Zeilinger}, {Grupp}, {Hudelot}, {Massey}, {Meneghetti}, {Miller}, {Paltani}, {Paulin-Henriksson}, {Pires}, {Saxton}, {Schrabback}, {Seidel}, {Walsh}, {Aghanim}, {Amendola}, {Bartlett}, {Baccigalupi}, {Beaulieu}, {Benabed}, {Cuby}, {Elbaz}, {Fosalba}, {Gavazzi}, {Helmi}, {Hook}, {Irwin}, {Kneib}, {Kunz}, {Mannucci}, {Moscardini}, {Tao}, {Teyssier}, {Weller}, {Zamorani}, {Zapatero Osorio}, {Boulade}, {Foumond}, {Di Giorgio}, {Guttridge}, {James}, {Kemp}, {Martignac}, {Spencer}, {Walton}, {Bl{\"u}mchen}, {Bonoli},
  {Bortoletto}, {Cerna}, {Corcione}, {Fabron}, {Jahnke}, {Ligori}, {Madrid}, {Martin}, {Morgante}, {Pamplona}, {Prieto}, {Riva}, {Toledo}, {Trifoglio}, {Zerbi}, {Abdalla}, {Douspis}, {Grenet}, {Borgani}, {Bouwens}, {Courbin}, {Delouis}, {Dubath}, {Fontana}, {Frailis}, {Grazian}, {Koppenh{\"o}fer}, {Mansutti}, {Melchior}, {Mignoli}, {Mohr}, {Neissner}, {Noddle}, {Poncet}, {Scodeggio}, {Serrano}, {Shane}, {Starck}, {Surace}, {Taylor}, {Verdoes-Kleijn}, {Vuerli}, {Williams}, {Zacchei}, {Altieri}, {Escudero Sanz}, {Kohley}, {Oosterbroek}, {Astier}, {Bacon}, {Bardelli}, {Baugh}, {Bellagamba}, {Benoist}, {Bianchi}, {Biviano}, {Branchini}, {Carbone}, {Cardone}, {Clements}, {Colombi}, {Conselice}, {Cresci}, {Deacon}, {Dunlop}, {Fedeli}, {Fontanot}, {Franzetti}, {Giocoli}, {Garcia-Bellido}, {Gow}, {Heavens}, {Hewett}, {Heymans}, {Holland}, {Huang}, {Ilbert}, {Joachimi}, {Jennins}, {Kerins}, {Kiessling}, {Kirk}, {Kotak}, {Krause}, {Lahav}, {van Leeuwen}, {Lesgourgues}, {Lombardi}, {Magliocchetti}, {Maguire},
  {Majerotto}, {Maoli}, {Marulli}, {Maurogordato}, {McCracken}, {McLure}, {Melchiorri}, {Merson}, {Moresco}, {Nonino}, {Norberg}, {Peacock}, {Pello}, {Penny}, {Pettorino}, {Di Porto}, {Pozzetti}, {Quercellini}, {Radovich}, {Rassat}, {Roche}, {Ronayette}, {Rossetti}, {Sartoris}, {Schneider}, {Semboloni}, {Serjeant}, {Simpson}, {Skordis}, {Smadja}, {Smartt}, {Spano}, {Spiro}, {Sullivan}, {Tilquin}, {Trotta}, {Verde}, {Wang}, {Williger}, {Zhao}, {Zoubian}, \& {Zucca}}]{2011arXiv1110.3193L}
{Laureijs}, R., {Amiaux}, J., {Arduini}, S., {et~al.} 2011, arXiv e-prints, arXiv:1110.3193

\bibitem[{{Lawrence} {et~al.}(2007){Lawrence}, {Warren}, {Almaini}, {Edge}, {Hambly}, {Jameson}, {Lucas}, {Casali}, {Adamson}, {Dye}, {Emerson}, {Foucaud}, {Hewett}, {Hirst}, {Hodgkin}, {Irwin}, {Lodieu}, {McMahon}, {Simpson}, {Smail}, {Mortlock}, \& {Folger}}]{2007MNRAS.379.1599L}
{Lawrence}, A., {Warren}, S.~J., {Almaini}, O., {et~al.} 2007, \mnras, 379, 1599

\bibitem[{{Lemon} {et~al.}(2023){Lemon}, {Anguita}, {Auger-Williams}, {Courbin}, {Galan}, {McMahon}, {Neira}, {Oguri}, {Schechter}, {Shajib}, {Treu}, {Agnello}, \& {Spiniello}}]{2023MNRAS.520.3305L}
{Lemon}, C., {Anguita}, T., {Auger-Williams}, M.~W., {et~al.} 2023, \mnras, 520, 3305

\bibitem[{{Lemon} {et~al.}(2020){Lemon}, {Auger}, {McMahon}, {Anguita}, {Apostolovski}, {Chen}, {Fassnacht}, {Melo}, {Motta}, {Shajib}, {Treu}, {Agnello}, {Buckley-Geer}, {Schechter}, {Birrer}, {Collett}, {Courbin}, {Rusu}, {Abbott}, {Allam}, {Annis}, {Avila}, {Bertin}, {Brooks}, {Burke}, {Carnero Rosell}, {Carrasco Kind}, {Carretero}, {Costanzi}, {da Costa}, {De Vicente}, {Desai}, {Eifler}, {Flaugher}, {Frieman}, {Garc{\'\i}a-Bellido}, {Gaztanaga}, {Gerdes}, {Gruen}, {Gruendl}, {Gschwend}, {Gutierrez}, {Honscheid}, {James}, {Kim}, {Krause}, {Kuehn}, {Kuropatkin}, {Lahav}, {Lima}, {Lin}, {Maia}, {March}, {Marshall}, {Menanteau}, {Miquel}, {Palmese}, {Paz-Chinch{\'o}n}, {Plazas}, {Roodman}, {Sanchez}, {Schubnell}, {Serrano}, {Smith}, {Soares-Santos}, {Suchyta}, {Tarle}, \& {Walker}}]{2020MNRAS.494.3491L}
{Lemon}, C., {Auger}, M.~W., {McMahon}, R., {et~al.} 2020, \mnras, 494, 3491

\bibitem[{{Lemon} {et~al.}(2019){Lemon}, {Auger}, \& {McMahon}}]{2019MNRAS.483.4242L}
{Lemon}, C.~A., {Auger}, M.~W., \& {McMahon}, R.~G. 2019, \mnras, 483, 4242

\bibitem[{{Lemon} {et~al.}(2018){Lemon}, {Auger}, {McMahon}, \& {Ostrovski}}]{2018MNRAS.479.5060L}
{Lemon}, C.~A., {Auger}, M.~W., {McMahon}, R.~G., \& {Ostrovski}, F. 2018, \mnras, 479, 5060

\bibitem[{{Madireddy} {et~al.}(2019){Madireddy}, {Ramachandra}, {Li}, {Butler}, {Balaprakash}, {Habib}, {Heitmann}, \& {The LSST Dark Energy Science Collaboration}}]{2019arXiv191103867M}
{Madireddy}, S., {Ramachandra}, N., {Li}, N., {et~al.} 2019, arXiv e-prints, arXiv:1911.03867

\bibitem[{{Mason} {et~al.}(2015){Mason}, {Treu}, {Schmidt}, {Collett}, {Trenti}, {Marshall}, {Barone-Nugent}, {Bradley}, {Stiavelli}, \& {Wyithe}}]{2015ApJ...805...79M}
{Mason}, C.~A., {Treu}, T., {Schmidt}, K.~B., {et~al.} 2015, \apj, 805, 79

\bibitem[{{McGreer} {et~al.}(2013){McGreer}, {Jiang}, {Fan}, {Richards}, {Strauss}, {Ross}, {White}, {Shen}, {Schneider}, {Myers}, {Brandt}, {DeGraf}, {Glikman}, {Ge}, \& {Streblyanska}}]{2013ApJ...768..105M}
{McGreer}, I.~D., {Jiang}, L., {Fan}, X., {et~al.} 2013, \apj, 768, 105

\bibitem[{{McMahon} {et~al.}(1992){McMahon}, {Irwin}, \& {Hazard}}]{1992Gemin..36....1M}
{McMahon}, R., {Irwin}, M., \& {Hazard}, C. 1992, GEMINI Newsletter Royal Greenwich Observatory, 36, 1

\bibitem[{{McMahon} {et~al.}(2013){McMahon}, {Banerji}, {Gonzalez}, {Koposov}, {Bejar}, {Lodieu}, {Rebolo}, \& {VHS Collaboration}}]{2013Msngr.154...35M}
{McMahon}, R.~G., {Banerji}, M., {Gonzalez}, E., {et~al.} 2013, The Messenger, 154, 35

\bibitem[{{Merloni} {et~al.}(2024){Merloni}, {Lamer}, {Liu}, {Ramos-Ceja}, {Brunner}, {Bulbul}, {Dennerl}, {Doroshenko}, {Freyberg}, {Friedrich}, {Gatuzz}, {Georgakakis}, {Haberl}, {Igo}, {Kreykenbohm}, {Liu}, {Maitra}, {Malyali}, {Mayer}, {Nandra}, {Predehl}, {Robrade}, {Salvato}, {Sanders}, {Stewart}, {Tub{\'\i}n-Arenas}, {Weber}, {Wilms}, {Arcodia}, {Artis}, {Aschersleben}, {Avakyan}, {Aydar}, {Bahar}, {Balzer}, {Becker}, {Berger}, {Boller}, {Bornemann}, {Br{\"u}ggen}, {Brusa}, {Buchner}, {Burwitz}, {Camilloni}, {Clerc}, {Comparat}, {Coutinho}, {Czesla}, {Dannhauer}, {Dauner}, {Dauser}, {Dietl}, {Dolag}, {Dwelly}, {Egg}, {Ehl}, {Freund}, {Friedrich}, {Gaida}, {Garrel}, {Ghirardini}, {Gokus}, {Gr{\"u}nwald}, {Grandis}, {Grotova}, {Gruen}, {Gueguen}, {H{\"a}mmerich}, {Hamaus}, {Hasinger}, {Haubner}, {Homan}, {Ider Chitham}, {Joseph}, {Joyce}, {K{\"o}nig}, {Kaltenbrunner}, {Khokhriakova}, {Kink}, {Kirsch}, {Kluge}, {Knies}, {Krippendorf}, {Krumpe}, {Kurpas}, {Li}, {Liu}, {Locatelli}, {Lorenz}, {M{\"u}ller},
  {Magaudda}, {Mannes}, {McCall}, {Meidinger}, {Michailidis}, {Migkas}, {Mu{\~n}oz-Giraldo}, {Musiimenta}, {Nguyen-Dang}, {Ni}, {Olechowska}, {Ota}, {Pacaud}, {Pasini}, {Perinati}, {Pires}, {Pommranz}, {Ponti}, {Poppenhaeger}, {P{\"u}hlhofer}, {Rau}, {Reh}, {Reiprich}, {Roster}, {Saeedi}, {Santangelo}, {Sasaki}, {Schmitt}, {Schneider}, {Schrabback}, {Schuster}, {Schwope}, {Seppi}, {Serim}, {Shreeram}, {Sokolova-Lapa}, {Starck}, {Stelzer}, {Stierhof}, {Suleimanov}, {Tenzer}, {Traulsen}, {Tr{\"u}mper}, {Tsuge}, {Urrutia}, {Veronica}, {Waddell}, {Willer}, {Wolf}, {Yeung}, {Zainab}, {Zangrandi}, {Zhang}, {Zhang}, \& {Zheng}}]{2024A&A...682A..34M}
{Merloni}, A., {Lamer}, G., {Liu}, T., {et~al.} 2024, \aap, 682, A34

\bibitem[{{Millon} {et~al.}(2020){Millon}, {Courbin}, {Bonvin}, {Paic}, {Meylan}, {Tewes}, {Sluse}, {Magain}, {Chan}, {Galan}, {Joseph}, {Lemon}, {Tihhonova}, {Anderson}, {Marmier}, {Chazelas}, {Lendl}, {Triaud}, \& {Wyttenbach}}]{2020A&A...640A.105M}
{Millon}, M., {Courbin}, F., {Bonvin}, V., {et~al.} 2020, \aap, 640, A105

\bibitem[{{Miralda-Escud{\'e}}(1998)}]{1998ApJ...501...15M}
{Miralda-Escud{\'e}}, J. 1998, \apj, 501, 15

\bibitem[{{Moffat}(1969)}]{1969A&A.....3..455M}
{Moffat}, A.~F.~J. 1969, \aap, 3, 455

\bibitem[{{More} {et~al.}(2024){More}, {Ca{\~n}ameras}, {Jaelani}, {Shu}, {Ishida}, {Wong}, {Inoue}, {Schuldt}, \& {Sonnenfeld}}]{2024MNRAS.533..525M}
{More}, A., {Ca{\~n}ameras}, R., {Jaelani}, A.~T., {et~al.} 2024, \mnras, 533, 525

\bibitem[{{More} {et~al.}(2016){More}, {Oguri}, {Kayo}, {Zinn}, {Strauss}, {Santiago}, {Mosquera}, {Inada}, {Kochanek}, {Rusu}, {Brownstein}, {da Costa}, {Kneib}, {Maia}, {Quimby}, {Schneider}, {Streblyanska}, \& {York}}]{2016MNRAS.456.1595M}
{More}, A., {Oguri}, M., {Kayo}, I., {et~al.} 2016, \mnras, 456, 1595

\bibitem[{{Morningstar} {et~al.}(2018){Morningstar}, {Hezaveh}, {Perreault Levasseur}, {Blandford}, {Marshall}, {Putzky}, \& {Wechsler}}]{2018arXiv180800011M}
{Morningstar}, W.~R., {Hezaveh}, Y.~D., {Perreault Levasseur}, L., {et~al.} 2018, arXiv e-prints, arXiv:1808.00011

\bibitem[{{Morokuma} {et~al.}(2007){Morokuma}, {Inada}, {Oguri}, {Ichikawa}, {Kawano}, {Tokita}, {Kayo}, {Hall}, {Kochanek}, {Richards}, {York}, \& {Schneider}}]{2007AJ....133..214M}
{Morokuma}, T., {Inada}, N., {Oguri}, M., {et~al.} 2007, \aj, 133, 214

\bibitem[{{Nightingale} {et~al.}(2023){Nightingale}, {Amvrosiadis}, {Hayes}, {He}, {Etherington}, {Cao}, {Cole}, {Frawley}, {Frenk}, {Lange}, {Li}, {Massey}, {Negrello}, \& {Robertson}}]{2023JOSS....8.4475N}
{Nightingale}, J., {Amvrosiadis}, A., {Hayes}, R., {et~al.} 2023, The Journal of Open Source Software, 8, 4475

\bibitem[{{Nightingale} {et~al.}(2021){Nightingale}, {Hayes}, {Kelly}, {Amvrosiadis}, {Etherington}, {He}, {Li}, {Cao}, {Frawley}, {Cole}, {Enia}, {Frenk}, {Harvey}, {Li}, {Massey}, {Negrello}, \& {Robertson}}]{2021JOSS....6.2825N}
{Nightingale}, J., {Hayes}, R., {Kelly}, A., {et~al.} 2021, The Journal of Open Source Software, 6, 2825

\bibitem[{{Nightingale} {et~al.}(2018){Nightingale}, {Dye}, \& {Massey}}]{2018MNRAS.478.4738N}
{Nightingale}, J.~W., {Dye}, S., \& {Massey}, R.~J. 2018, \mnras, 478, 4738

\bibitem[{{Oguri}(2010)}]{2010PASJ...62.1017O}
{Oguri}, M. 2010, \pasj, 62, 1017

\bibitem[{{Oguri} {et~al.}(2004){Oguri}, {Inada}, {Castander}, {Gregg}, {Becker}, {Ichikawa}, {Pindor}, {Brinkmann}, {Eisenstein}, {Frieman}, {Hall}, {Johnston}, {Richards}, {Schechter}, {Schneider}, \& {Szalay}}]{2004PASJ...56..399O}
{Oguri}, M., {Inada}, N., {Castander}, F.~J., {et~al.} 2004, \pasj, 56, 399

\bibitem[{{Oguri} {et~al.}(2008){Oguri}, {Inada}, {Clocchiatti}, {Kayo}, {Shin}, {Hennawi}, {Strauss}, {Morokuma}, {Schneider}, \& {York}}]{2008AJ....135..520O}
{Oguri}, M., {Inada}, N., {Clocchiatti}, A., {et~al.} 2008, \aj, 135, 520

\bibitem[{{Oguri} \& {Marshall}(2010)}]{2010MNRAS.405.2579O}
{Oguri}, M. \& {Marshall}, P.~J. 2010, \mnras, 405, 2579

\bibitem[{{Pacucci} \& {Loeb}(2019)}]{2019ApJ...870L..12P}
{Pacucci}, F. \& {Loeb}, A. 2019, \apjl, 870, L12

\bibitem[{{Pearson} {et~al.}(2021){Pearson}, {Maresca}, {Li}, \& {Dye}}]{2021MNRAS.505.4362P}
{Pearson}, J., {Maresca}, J., {Li}, N., \& {Dye}, S. 2021, \mnras, 505, 4362

\bibitem[{{Perreault Levasseur} {et~al.}(2017){Perreault Levasseur}, {Hezaveh}, \& {Wechsler}}]{2017ApJ...850L...7P}
{Perreault Levasseur}, L., {Hezaveh}, Y.~D., \& {Wechsler}, R.~H. 2017, \apjl, 850, L7

\bibitem[{{Planck Collaboration} {et~al.}(2020){Planck Collaboration}, {Aghanim}, {Akrami}, {Ashdown}, {Aumont}, {Baccigalupi}, {Ballardini}, {Banday}, {Barreiro}, {Bartolo}, {Basak}, {Battye}, {Benabed}, {Bernard}, {Bersanelli}, {Bielewicz}, {Bock}, {Bond}, {Borrill}, {Bouchet}, {Boulanger}, {Bucher}, {Burigana}, {Butler}, {Calabrese}, {Cardoso}, {Carron}, {Challinor}, {Chiang}, {Chluba}, {Colombo}, {Combet}, {Contreras}, {Crill}, {Cuttaia}, {de Bernardis}, {de Zotti}, {Delabrouille}, {Delouis}, {Di Valentino}, {Diego}, {Dor{\'e}}, {Douspis}, {Ducout}, {Dupac}, {Dusini}, {Efstathiou}, {Elsner}, {En{\ss}lin}, {Eriksen}, {Fantaye}, {Farhang}, {Fergusson}, {Fernandez-Cobos}, {Finelli}, {Forastieri}, {Frailis}, {Fraisse}, {Franceschi}, {Frolov}, {Galeotta}, {Galli}, {Ganga}, {G{\'e}nova-Santos}, {Gerbino}, {Ghosh}, {Gonz{\'a}lez-Nuevo}, {G{\'o}rski}, {Gratton}, {Gruppuso}, {Gudmundsson}, {Hamann}, {Handley}, {Hansen}, {Herranz}, {Hildebrandt}, {Hivon}, {Huang}, {Jaffe}, {Jones}, {Karakci}, {Keih{\"a}nen},
  {Keskitalo}, {Kiiveri}, {Kim}, {Kisner}, {Knox}, {Krachmalnicoff}, {Kunz}, {Kurki-Suonio}, {Lagache}, {Lamarre}, {Lasenby}, {Lattanzi}, {Lawrence}, {Le Jeune}, {Lemos}, {Lesgourgues}, {Levrier}, {Lewis}, {Liguori}, {Lilje}, {Lilley}, {Lindholm}, {L{\'o}pez-Caniego}, {Lubin}, {Ma}, {Mac{\'\i}as-P{\'e}rez}, {Maggio}, {Maino}, {Mandolesi}, {Mangilli}, {Marcos-Caballero}, {Maris}, {Martin}, {Martinelli}, {Mart{\'\i}nez-Gonz{\'a}lez}, {Matarrese}, {Mauri}, {McEwen}, {Meinhold}, {Melchiorri}, {Mennella}, {Migliaccio}, {Millea}, {Mitra}, {Miville-Desch{\^e}nes}, {Molinari}, {Montier}, {Morgante}, {Moss}, {Natoli}, {N{\o}rgaard-Nielsen}, {Pagano}, {Paoletti}, {Partridge}, {Patanchon}, {Peiris}, {Perrotta}, {Pettorino}, {Piacentini}, {Polastri}, {Polenta}, {Puget}, {Rachen}, {Reinecke}, {Remazeilles}, {Renzi}, {Rocha}, {Rosset}, {Roudier}, {Rubi{\~n}o-Mart{\'\i}n}, {Ruiz-Granados}, {Salvati}, {Sandri}, {Savelainen}, {Scott}, {Shellard}, {Sirignano}, {Sirri}, {Spencer}, {Sunyaev}, {Suur-Uski}, {Tauber}, {Tavagnacco},
  {Tenti}, {Toffolatti}, {Tomasi}, {Trombetti}, {Valenziano}, {Valiviita}, {Van Tent}, {Vibert}, {Vielva}, {Villa}, {Vittorio}, {Wandelt}, {Wehus}, {White}, {White}, {Zacchei}, \& {Zonca}}]{2020A&A...641A...6P}
{Planck Collaboration}, {Aghanim}, N., {Akrami}, Y., {et~al.} 2020, \aap, 641, A6

\bibitem[{{Prakash} {et~al.}(2016){Prakash}, {Licquia}, {Newman}, {Ross}, {Myers}, {Dawson}, {Kneib}, {Percival}, {Bautista}, {Comparat}, {Tinker}, {Schlegel}, {Tojeiro}, {Ho}, {Lang}, {Rao}, {McBride}, {Ben Zhu}, {Brownstein}, {Bailey}, {Bolton}, {Delubac}, {Mariappan}, {Blanton}, {Reid}, {Schneider}, {Seo}, {Carnero Rosell}, \& {Prada}}]{2016ApJS..224...34P}
{Prakash}, A., {Licquia}, T.~C., {Newman}, J.~A., {et~al.} 2016, \apjs, 224, 34

\bibitem[{{Queirolo} {et~al.}(2023){Queirolo}, {Seitz}, {Riffeser}, {Kluge}, {Bender}, {G{\"o}ssl}, {Hopp}, {Ries}, {Schmidt}, \& {Z{\"o}ller}}]{2023arXiv231209311Q}
{Queirolo}, G., {Seitz}, S., {Riffeser}, A., {et~al.} 2023, arXiv e-prints, arXiv:2312.09311

\bibitem[{{Reback} {et~al.}(2022){Reback}, {jbrockmendel}, {McKinney}, {Van den Bossche}, {Augspurger}, {Roeschke}, {Hawkins}, {Cloud}, {gfyoung}, {Sinhrks}, {Hoefler}, {Klein}, {Petersen}, {Tratner}, {She}, {Ayd}, {Naveh}, {Darbyshire}, {Garcia}, {Shadrach}, {Schendel}, {Hayden}, {Saxton}, {Gorelli}, {Li}, {Zeitlin}, {Jancauskas}, {McMaster}, {W{\"o}rtwein}, \& {Battiston}}]{2022zndo...3509134R}
{Reback}, J., {jbrockmendel}, {McKinney}, W., {et~al.} 2022, {pandas-dev/pandas: Pandas 1.4.2}, Zenodo

\bibitem[{{Refsdal}(1964)}]{1964MNRAS.128..307R}
{Refsdal}, S. 1964, \mnras, 128, 307

\bibitem[{{Riess} {et~al.}(2022){Riess}, {Yuan}, {Macri}, {Scolnic}, {Brout}, {Casertano}, {Jones}, {Murakami}, {Anand}, {Breuval}, {Brink}, {Filippenko}, {Hoffmann}, {Jha}, {D'arcy Kenworthy}, {Mackenty}, {Stahl}, \& {Zheng}}]{2022ApJ...934L...7R}
{Riess}, A.~G., {Yuan}, W., {Macri}, L.~M., {et~al.} 2022, \apjl, 934, L7

\bibitem[{Rocklin(2015)}]{matthew_rocklin-proc-scipy-2015}
Rocklin, M. 2015, in Proceedings of the 14th Python in Science Conference, ed. K.~Huff \& J.~Bergstra, 130--136

\bibitem[{{Rojas} {et~al.}(2022{\natexlab{a}}){Rojas}, {Savary}, {Cl{\'e}ment}, {Maus}, {Courbin}, {Lemon}, {Chan}, {Vernardos}, {Joseph}, {Ca{\~n}ameras}, \& {Galan}}]{2022AandA...668A..73R}
{Rojas}, K., {Savary}, E., {Cl{\'e}ment}, B., {et~al.} 2022{\natexlab{a}}, \aap, 668, A73

\bibitem[{{Rojas} {et~al.}(2022{\natexlab{b}}){Rojas}, {Savary}, {Cl{\'e}ment}, {Maus}, {Courbin}, {Lemon}, {Chan}, {Vernardos}, {Joseph}, {Ca{\~n}ameras}, \& {Galan}}]{2022A&A...668A..73R}
{Rojas}, K., {Savary}, E., {Cl{\'e}ment}, B., {et~al.} 2022{\natexlab{b}}, \aap, 668, A73

\bibitem[{{Rumbaugh} {et~al.}(2015){Rumbaugh}, {Fassnacht}, {McKean}, {Koopmans}, {Auger}, \& {Suyu}}]{2015MNRAS.450.1042R}
{Rumbaugh}, N., {Fassnacht}, C.~D., {McKean}, J.~P., {et~al.} 2015, \mnras, 450, 1042

\bibitem[{{Savary} {et~al.}(2022){Savary}, {Rojas}, {Maus}, {Cl{\'e}ment}, {Courbin}, {Gavazzi}, {Chan}, {Lemon}, {Vernardos}, {Ca{\~n}ameras}, {Schuldt}, {Suyu}, {Cuillandre}, {Fabbro}, {Gwyn}, {Hudson}, {Kilbinger}, {Scott}, \& {Stone}}]{2022A&A...666A...1S}
{Savary}, E., {Rojas}, K., {Maus}, M., {et~al.} 2022, \aap, 666, A1

\bibitem[{{Schlafly} \& {Finkbeiner}(2011)}]{2011ApJ...737..103S}
{Schlafly}, E.~F. \& {Finkbeiner}, D.~P. 2011, \apj, 737, 103

\bibitem[{{Schlafly} {et~al.}(2019){Schlafly}, {Meisner}, \& {Green}}]{2019ApJS..240...30S}
{Schlafly}, E.~F., {Meisner}, A.~M., \& {Green}, G.~M. 2019, \apjs, 240, 30

\bibitem[{{Schlegel} {et~al.}(1998){Schlegel}, {Finkbeiner}, \& {Davis}}]{1998ApJ...500..525S}
{Schlegel}, D.~J., {Finkbeiner}, D.~P., \& {Davis}, M. 1998, \apj, 500, 525

\bibitem[{{Schmidt} {et~al.}(2023){Schmidt}, {Treu}, {Birrer}, {Shajib}, {Lemon}, {Millon}, {Sluse}, {Agnello}, {Anguita}, {Auger-Williams}, {McMahon}, {Motta}, {Schechter}, {Spiniello}, {Kayo}, {Courbin}, {Ertl}, {Fassnacht}, {Frieman}, {More}, {Schuldt}, {Suyu}, {Aguena}, {Andrade-Oliveira}, {Annis}, {Bacon}, {Bertin}, {Brooks}, {Burke}, {Carnero Rosell}, {Carrasco Kind}, {Carretero}, {Conselice}, {Costanzi}, {da Costa}, {Pereira}, {De Vicente}, {Desai}, {Doel}, {Everett}, {Ferrero}, {Friedel}, {Garc{\'\i}a-Bellido}, {Gaztanaga}, {Gruen}, {Gruendl}, {Gschwend}, {Gutierrez}, {Hinton}, {Hollowood}, {Honscheid}, {James}, {Kuehn}, {Lahav}, {Menanteau}, {Miquel}, {Palmese}, {Paz-Chinch{\'o}n}, {Pieres}, {Plazas Malag{\'o}n}, {Prat}, {Rodriguez-Monroy}, {Romer}, {Sanchez}, {Scarpine}, {Sevilla-Noarbe}, {Smith}, {Suchyta}, {Tarle}, {To}, {Varga}, \& {DES Collaboration}}]{2023MNRAS.518.1260S}
{Schmidt}, T., {Treu}, T., {Birrer}, S., {et~al.} 2023, \mnras, 518, 1260

\bibitem[{{Schuldt} {et~al.}(2023{\natexlab{a}}){Schuldt}, {Ca{\~n}ameras}, {Shu}, {Suyu}, {Taubenberger}, {Meinhardt}, \& {Leal-Taix{\'e}}}]{2023A&A...671A.147S}
{Schuldt}, S., {Ca{\~n}ameras}, R., {Shu}, Y., {et~al.} 2023{\natexlab{a}}, \aap, 671, A147

\bibitem[{{Schuldt} {et~al.}(2024){Schuldt}, {Canameras}, {Andika}, {Bag}, {Melo}, {Shu}, {Suyu}, {Taubenberger}, \& {Grillo}}]{2024arXiv240520383S}
{Schuldt}, S., {Canameras}, R., {Andika}, I.~T., {et~al.} 2024, arXiv e-prints, arXiv:2405.20383

\bibitem[{{Schuldt} {et~al.}(2019){Schuldt}, {Chiriv{\`\i}}, {Suyu}, {Y{\i}ld{\i}r{\i}m}, {Sonnenfeld}, {Halkola}, \& {Lewis}}]{2019A&A...631A..40S}
{Schuldt}, S., {Chiriv{\`\i}}, G., {Suyu}, S.~H., {et~al.} 2019, \aap, 631, A40

\bibitem[{{Schuldt} {et~al.}(2023{\natexlab{b}}){Schuldt}, {Suyu}, {Ca{\~n}ameras}, {Shu}, {Taubenberger}, {Ertl}, \& {Halkola}}]{2023A&A...673A..33S}
{Schuldt}, S., {Suyu}, S.~H., {Ca{\~n}ameras}, R., {et~al.} 2023{\natexlab{b}}, \aap, 673, A33

\bibitem[{{Schuldt} {et~al.}(2021){Schuldt}, {Suyu}, {Meinhardt}, {Leal-Taix{\'e}}, {Ca{\~n}ameras}, {Taubenberger}, \& {Halkola}}]{2021A&A...646A.126S}
{Schuldt}, S., {Suyu}, S.~H., {Meinhardt}, T., {et~al.} 2021, \aap, 646, A126

\bibitem[{{Selvaraju} {et~al.}(2016){Selvaraju}, {Cogswell}, {Das}, {Vedantam}, {Parikh}, \& {Batra}}]{2016arXiv161002391S}
{Selvaraju}, R.~R., {Cogswell}, M., {Das}, A., {et~al.} 2016, arXiv e-prints, arXiv:1610.02391

\bibitem[{{S{\'e}rsic}(1963)}]{1963BAAA....6...41S}
{S{\'e}rsic}, J.~L. 1963, Boletin de la Asociacion Argentina de Astronomia La Plata Argentina, 6, 41

\bibitem[{{Shajib} {et~al.}(2020){Shajib}, {Birrer}, {Treu}, {Agnello}, {Buckley-Geer}, {Chan}, {Christensen}, {Lemon}, {Lin}, {Millon}, {Poh}, {Rusu}, {Sluse}, {Spiniello}, {Chen}, {Collett}, {Courbin}, {Fassnacht}, {Frieman}, {Galan}, {Gilman}, {More}, {Anguita}, {Auger}, {Bonvin}, {McMahon}, {Meylan}, {Wong}, {Abbott}, {Annis}, {Avila}, {Bechtol}, {Brooks}, {Brout}, {Burke}, {Carnero Rosell}, {Carrasco Kind}, {Carretero}, {Castander}, {Costanzi}, {da Costa}, {De Vicente}, {Desai}, {Dietrich}, {Doel}, {Drlica-Wagner}, {Evrard}, {Finley}, {Flaugher}, {Fosalba}, {Garc{\'\i}a-Bellido}, {Gerdes}, {Gruen}, {Gruendl}, {Gschwend}, {Gutierrez}, {Hollowood}, {Honscheid}, {Huterer}, {James}, {Jeltema}, {Krause}, {Kuropatkin}, {Li}, {Lima}, {MacCrann}, {Maia}, {Marshall}, {Melchior}, {Miquel}, {Ogando}, {Palmese}, {Paz-Chinch{\'o}n}, {Plazas}, {Romer}, {Roodman}, {Sako}, {Sanchez}, {Santiago}, {Scarpine}, {Schubnell}, {Scolnic}, {Serrano}, {Sevilla-Noarbe}, {Smith}, {Soares-Santos}, {Suchyta}, {Tarle}, {Thomas},
  {Walker}, \& {Zhang}}]{2020MNRAS.494.6072S}
{Shajib}, A.~J., {Birrer}, S., {Treu}, T., {et~al.} 2020, \mnras, 494, 6072

\bibitem[{{Shajib} {et~al.}(2022){Shajib}, {Vernardos}, {Collett}, {Motta}, {Sluse}, {Williams}, {Saha}, {Birrer}, {Spiniello}, \& {Treu}}]{2022arXiv221010790S}
{Shajib}, A.~J., {Vernardos}, G., {Collett}, T.~E., {et~al.} 2022, arXiv e-prints, arXiv:2210.10790

\bibitem[{{Shu} {et~al.}(2022){Shu}, {Ca{\~n}ameras}, {Schuldt}, {Suyu}, {Taubenberger}, {Inoue}, \& {Jaelani}}]{2022AandA...662A...4S}
{Shu}, Y., {Ca{\~n}ameras}, R., {Schuldt}, S., {et~al.} 2022, \aap, 662, A4

\bibitem[{{Simonyan} {et~al.}(2013){Simonyan}, {Vedaldi}, \& {Zisserman}}]{2013arXiv1312.6034S}
{Simonyan}, K., {Vedaldi}, A., \& {Zisserman}, A. 2013, arXiv e-prints, arXiv:1312.6034

\bibitem[{{Songaila} \& {Cowie}(2010)}]{2010ApJ...721.1448S}
{Songaila}, A. \& {Cowie}, L.~L. 2010, \apj, 721, 1448

\bibitem[{{Spiniello} {et~al.}(2018){Spiniello}, {Agnello}, {Napolitano}, {Sergeyev}, {Getman}, {Tortora}, {Spavone}, {Bilicki}, {Buddelmeijer}, {Koopmans}, {Kuijken}, {Vernardos}, {Bannikova}, \& {Capaccioli}}]{2018MNRAS.480.1163S}
{Spiniello}, C., {Agnello}, A., {Napolitano}, N.~R., {et~al.} 2018, \mnras, 480, 1163

\bibitem[{{Stacey} {et~al.}(2022){Stacey}, {Costa}, {McKean}, {Sharon}, {Calistro Rivera}, {Glikman}, \& {van der Werf}}]{2022MNRAS.517.3377S}
{Stacey}, H.~R., {Costa}, T., {McKean}, J.~P., {et~al.} 2022, \mnras, 517, 3377

\bibitem[{{Sundararajan} {et~al.}(2017){Sundararajan}, {Taly}, \& {Yan}}]{2017arXiv170301365S}
{Sundararajan}, M., {Taly}, A., \& {Yan}, Q. 2017, arXiv e-prints, arXiv:1703.01365

\bibitem[{{Taylor}(2005)}]{2005ASPC..347...29T}
{Taylor}, M.~B. 2005, in Astronomical Society of the Pacific Conference Series, Vol. 347, Astronomical Data Analysis Software and Systems XIV, ed. P.~{Shopbell}, M.~{Britton}, \& R.~{Ebert}, 29

\bibitem[{{Treu} {et~al.}(2022){Treu}, {Suyu}, \& {Marshall}}]{2022A&ARv..30....8T}
{Treu}, T., {Suyu}, S.~H., \& {Marshall}, P.~J. 2022, \aapr, 30, 8

\bibitem[{{Verde} {et~al.}(2019){Verde}, {Treu}, \& {Riess}}]{2019NatAs...3..891V}
{Verde}, L., {Treu}, T., \& {Riess}, A.~G. 2019, Nature Astronomy, 3, 891

\bibitem[{{Vestergaard} \& {Wilkes}(2001)}]{2001ApJS..134....1V}
{Vestergaard}, M. \& {Wilkes}, B.~J. 2001, \apjs, 134, 1

\bibitem[{{Waskom}(2021)}]{2021JOSS....6.3021W}
{Waskom}, M. 2021, The Journal of Open Source Software, 6, 3021

\bibitem[{{Williams} {et~al.}(2018){Williams}, {Agnello}, {Treu}, {Abramson}, {Anguita}, {Apostolovski}, {Chen}, {Fassnacht}, {Hsueh}, {Lemaux}, {Motta}, {Oldham}, {Rojas}, {Rusu}, {Shajib}, \& {Wang}}]{2018MNRAS.477L..70W}
{Williams}, P.~R., {Agnello}, A., {Treu}, T., {et~al.} 2018, \mnras, 477, L70

\bibitem[{{Wong} {et~al.}(2020){Wong}, {Suyu}, {Chen}, {Rusu}, {Millon}, {Sluse}, {Bonvin}, {Fassnacht}, {Taubenberger}, {Auger}, {Birrer}, {Chan}, {Courbin}, {Hilbert}, {Tihhonova}, {Treu}, {Agnello}, {Ding}, {Jee}, {Komatsu}, {Shajib}, {Sonnenfeld}, {Blandford}, {Koopmans}, {Marshall}, \& {Meylan}}]{2020MNRAS.498.1420W}
{Wong}, K.~C., {Suyu}, S.~H., {Chen}, G. C.~F., {et~al.} 2020, \mnras, 498, 1420

\bibitem[{{Woodfinden} {et~al.}(2022){Woodfinden}, {Nadathur}, {Percival}, {Radinovic}, {Massara}, \& {Winther}}]{2022MNRAS.516.4307W}
{Woodfinden}, A., {Nadathur}, S., {Percival}, W.~J., {et~al.} 2022, \mnras, 516, 4307

\bibitem[{{Worseck} \& {Prochaska}(2011)}]{2011ApJ...728...23W}
{Worseck}, G. \& {Prochaska}, J.~X. 2011, \apj, 728, 23

\bibitem[{{Wright} {et~al.}(2010){Wright}, {Eisenhardt}, {Mainzer}, {Ressler}, {Cutri}, {Jarrett}, {Kirkpatrick}, {Padgett}, {McMillan}, {Skrutskie}, {Stanford}, {Cohen}, {Walker}, {Mather}, {Leisawitz}, {Gautier}, {McLean}, {Benford}, {Lonsdale}, {Blain}, {Mendez}, {Irace}, {Duval}, {Liu}, {Royer}, {Heinrichsen}, {Howard}, {Shannon}, {Kendall}, {Walsh}, {Larsen}, {Cardon}, {Schick}, {Schwalm}, {Abid}, {Fabinsky}, {Naes}, \& {Tsai}}]{2010AJ....140.1868W}
{Wright}, E.~L., {Eisenhardt}, P. R.~M., {Mainzer}, A.~K., {et~al.} 2010, \aj, 140, 1868

\bibitem[{{Yang} {et~al.}(2016){Yang}, {Wang}, {Wu}, {Fan}, {McGreer}, {Bian}, {Yi}, {Yang}, {Ai}, {Dong}, {Zuo}, {Green}, {Jiang}, {Wang}, {Wang}, \& {Yue}}]{2016ApJ...829...33Y}
{Yang}, J., {Wang}, F., {Wu}, X.-B., {et~al.} 2016, \apj, 829, 33

\bibitem[{{Yue} {et~al.}(2022){Yue}, {Fan}, {Yang}, \& {Wang}}]{2022AJ....163..139Y}
{Yue}, M., {Fan}, X., {Yang}, J., \& {Wang}, F. 2022, \aj, 163, 139

\bibitem[{{Yue} {et~al.}(2023){Yue}, {Fan}, {Yang}, \& {Wang}}]{2023AJ....165..191Y}
{Yue}, M., {Fan}, X., {Yang}, J., \& {Wang}, F. 2023, \aj, 165, 191

\bibitem[{{Yue} {et~al.}(2021){Yue}, {Yang}, {Fan}, {Wang}, {Spilker}, {Georgiev}, {Keeton}, {Litke}, {Marrone}, {Walter}, {Wang}, {Wu}, {Venemans}, \& {Zabludoff}}]{2021ApJ...917...99Y}
{Yue}, M., {Yang}, J., {Fan}, X., {et~al.} 2021, \apj, 917, 99

\bibitem[{{Zhao} {et~al.}(2022){Zhao}, {Variu}, {He}, {Forero-S{\'a}nchez}, {Tamone}, {Chuang}, {Kitaura}, {Tao}, {Yu}, {Kneib}, {Percival}, {Shan}, {Zhao}, {Burtin}, {Dawson}, {Rossi}, {Schneider}, \& {de la Macorra}}]{2022MNRAS.511.5492Z}
{Zhao}, C., {Variu}, A., {He}, M., {et~al.} 2022, \mnras, 511, 5492

\end{thebibliography}
%

\begin{appendix}
\normalsize

\section{Network architecture overview} \label{sec:full_architecture}

We present the detailed architecture of VariLens here, which is designed to process HSC images for detecting and modeling lensed quasars. 
The network is composed of distinct modules that handle regression and classification tasks through a combination of encoding and decoding techniques. 

\begin{figure}[htb!]
	\centering
	\resizebox{\hsize}{!}{\includegraphics{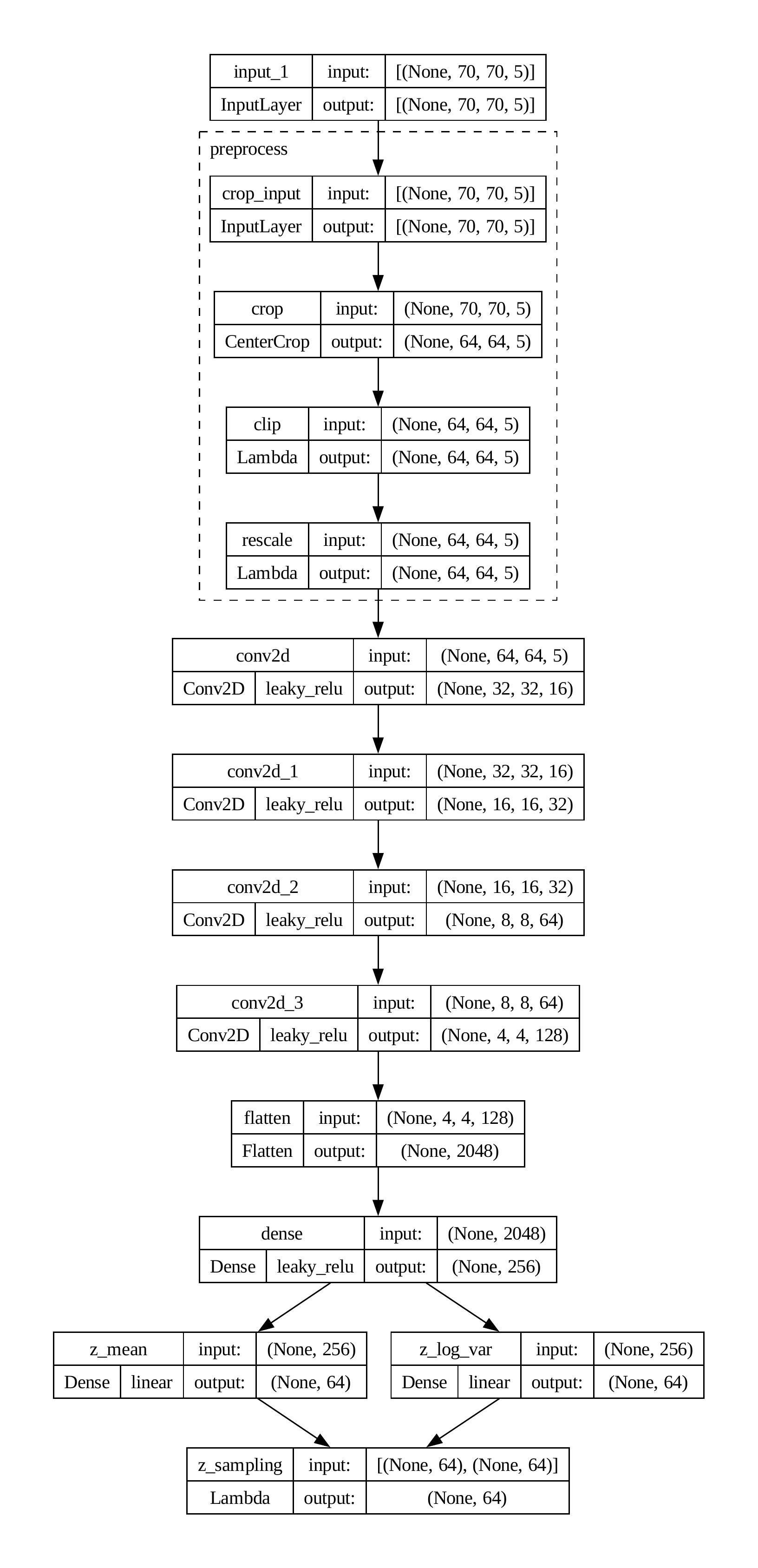}}
	\caption{
		The encoder of VariLens.
		This module processes a batch of 5-band HSC images, each measuring $70 \times 70$ pixels, and compresses them into a latent representation of 64 dimensions. 
		The batch size is flexible, with ``None'' meaning it can adapt based on user specifications. 
		The leftmost column in the architecture diagram denotes the custom label and layer type, as defined by the \texttt{TensorFlow} deep learning framework.
	}
	\label{fig:arch_enc}
\end{figure}

\begin{figure}[htb!]
	\centering
	\resizebox{\hsize}{!}{\includegraphics{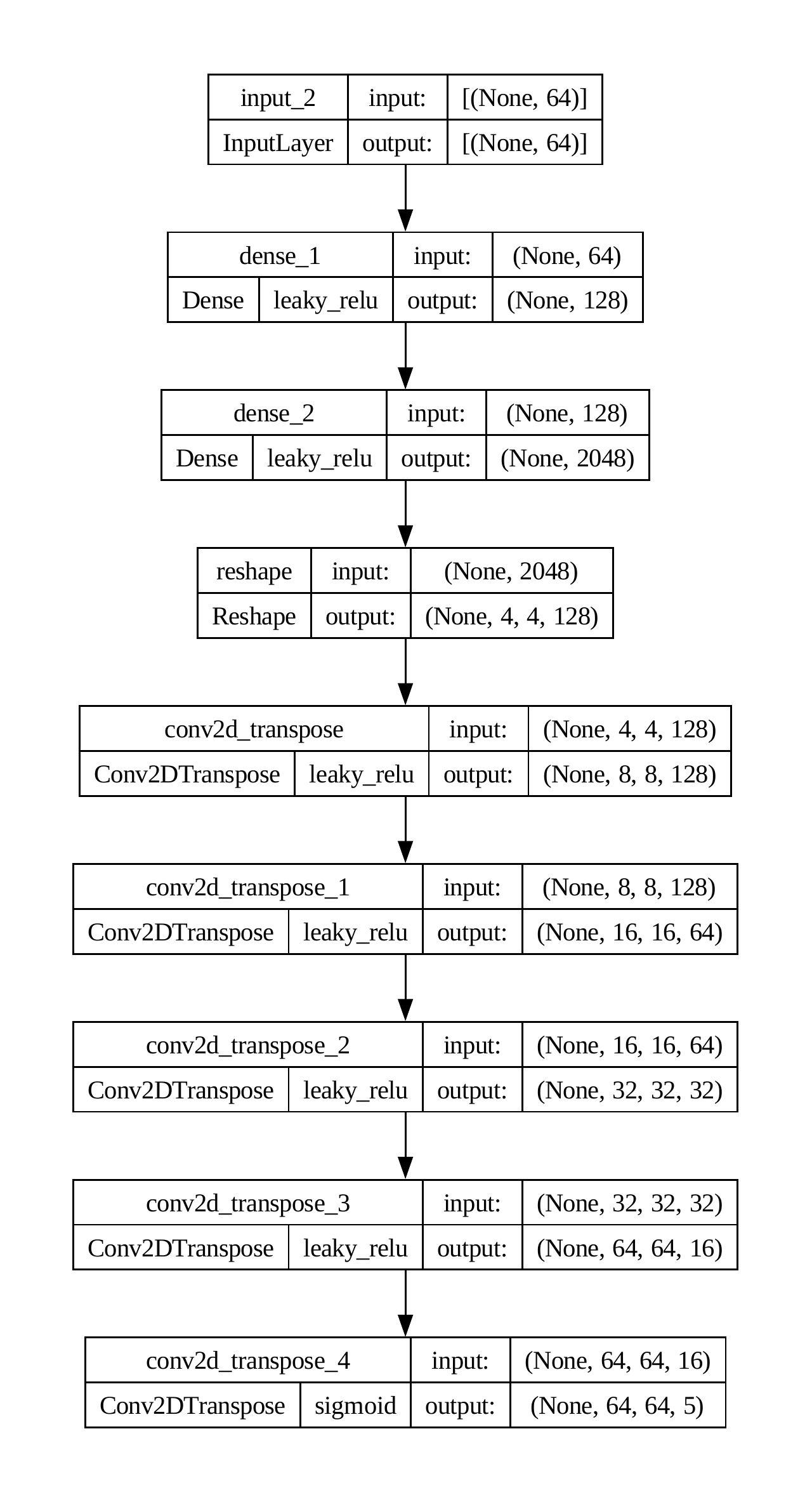}}
	\caption{
		The decoder of VariLens.
		This module takes the latent representation generated by the encoder and reconstructs them back into the original images. 
		This process retains only the most important information while smoothing out noise, effectively acting as a form of lossy compression-decompression that prioritizes essential features.	}
	\label{fig:arch_dec}
\end{figure}

\begin{figure}[htb!]
	\centering
	\resizebox{\hsize}{!}{\includegraphics{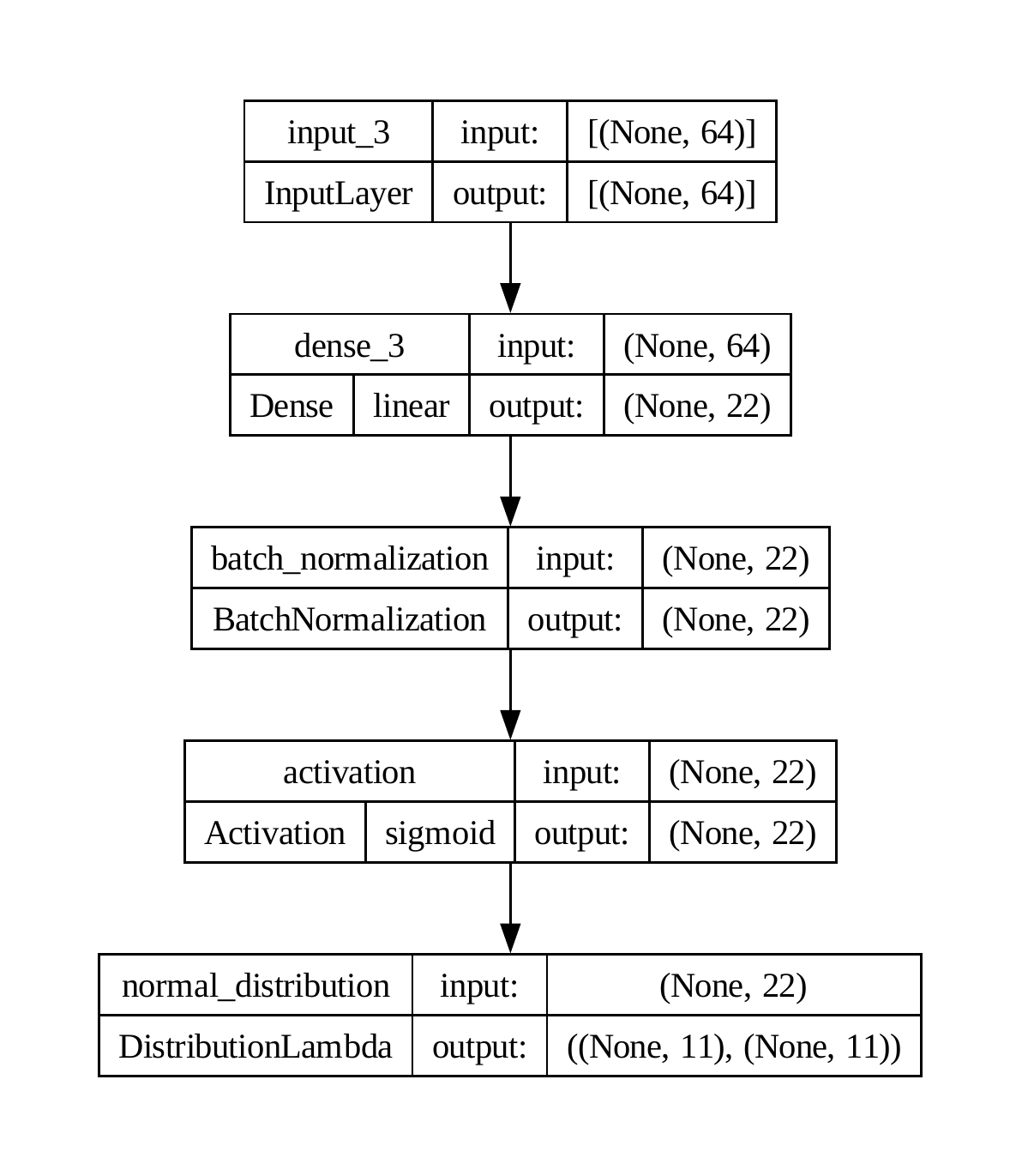}}
	\caption{
		The regression module of VariLens.
		This regressor processes the latent representations produced by the encoder to estimate key lens and source parameters. 
		Simultaneously, it helps the latent distribution align with physically meaningful constraints.
	}
	\label{fig:arch_reg}
\end{figure}

The overview of the VariLens architecture is already presented in Figure~\ref{fig:arch_compact}, which illustrates that the network consists of three key modules: the encoder, decoder, and regressor.
As shown in Figure~\ref{fig:arch_enc}, the encoder transforms the input images into a latent representation. 
The decoder, depicted in Figure~\ref{fig:arch_dec}, reconstructs the latent distribution back into the original data. 
Meanwhile, the regressor, illustrated in Figure~\ref{fig:arch_reg}, estimates the lens and source parameters while ensuring that the latent distribution remains physically meaningful.
Due to space limitations, the batch normalization layers between the convolution and activation operations are not displayed in the figures. 
However, they are indeed present in the network architecture, similar to what is seen in the regression module, for example.

\begin{figure}[htb!]
	\centering
	\resizebox{\hsize}{!}{\includegraphics{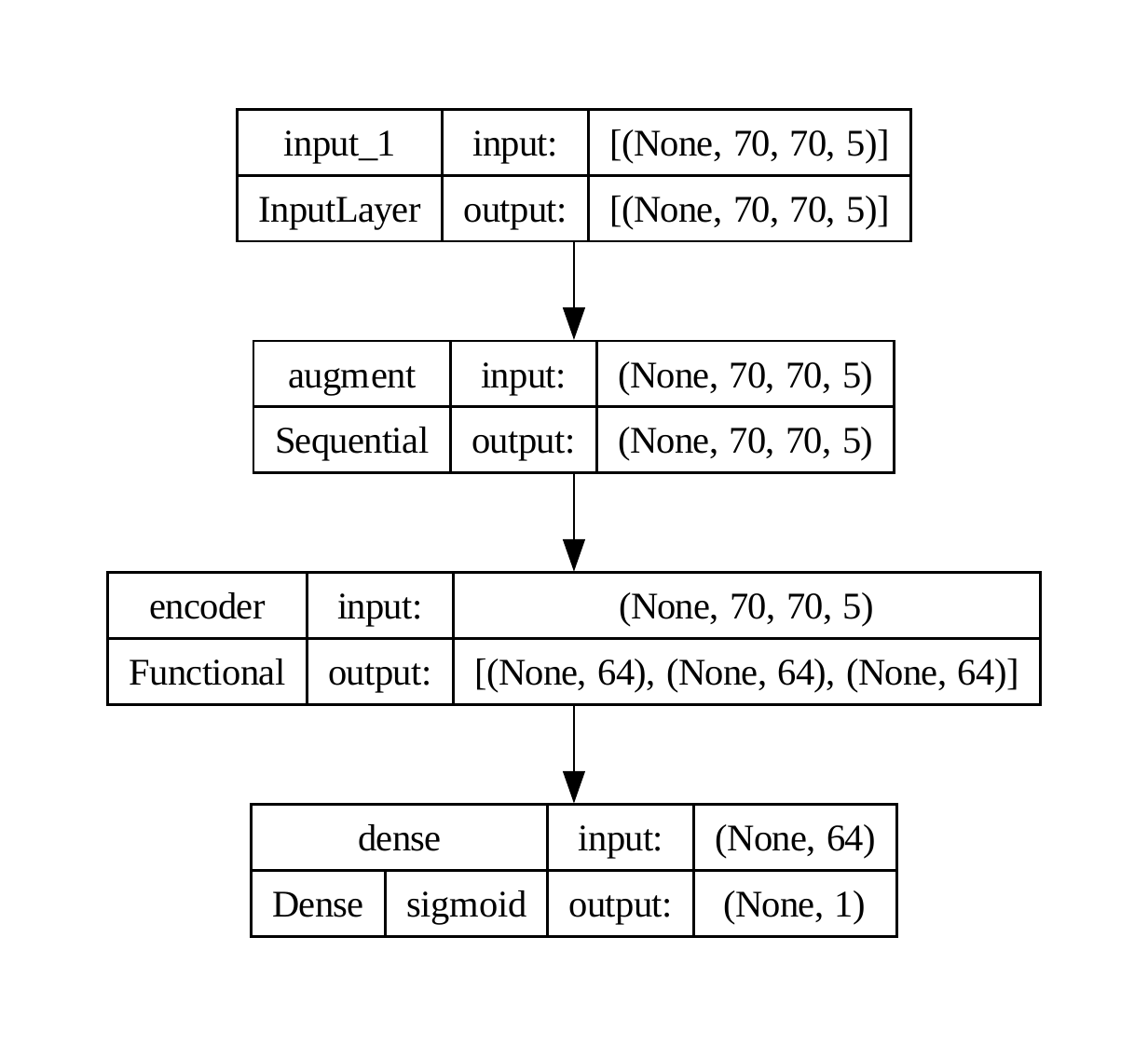}}
	\caption{
		The classification module of VariLens.
		This module leverages the pretrained encoder to process batches of 5-band HSC images, each with dimensions of $70 \times 70$ pixels. 
		It outputs the probability of a lens through a final dense layer equipped with a sigmoid activation function, enabling binary classification of lens candidates.
	}
	\label{fig:arch_cls}
\end{figure}

Figure~\ref{fig:arch_cls} displays the classification module, where the network processes 5-band HSC images using an encoder and outputs the lens probability via a final dense layer with a sigmoid activation function. 
We note that the augmentation module within the classifier includes random flips, shifts, and rotations, which are applied only during the training phase but are not used during inference.


\clearpage

\section{Regression metrics definition} \label{sec:regression_metrics}

The coefficient of determination, \( R^2 \), can be computed as follows:
\begin{equation}
	R^2 = 1 - \frac{\sum_{i=1}^{N} (y_i - \hat{y}_i)^2}{\sum_{i=1}^{N} (y_i - \bar{y})^2},
\end{equation}
where \(\hat{y}_i\) denotes the predicted value for the \(i\)-th sample, \( y_i \) represents the true value of the \(i\)-th sample, and \(\bar{y}\) is the mean of the true values across all samples, computed as \(\bar{y} = \frac{1}{N} \sum_{i=1}^{N} y_i\). 
The numerator, \(\sum_{i=1}^{N} (y_i - \hat{y}_i)^2\), is the residual sum of squares, which measures the variance of the prediction errors. 
The denominator, \(\sum_{i=1}^{N} (y_i - \bar{y})^2\), is the total sum of squares, indicating the variance in the true values.

Additionally, the mean squared error (MSE) can be computed utilizing:
\begin{equation}
	\text{MSE} = \frac{1}{N} \sum_{i=1}^{N} (y_i - \hat{y}_i)^2,
\end{equation}
while the mean absolute error (MAE) is given by the expression as follows:
\begin{equation}
	\text{MAE} = \frac{1}{N} \sum_{i=1}^{N} |y_i - \hat{y}_i|.
\end{equation}
Here, \( N \) denotes the total data points. 
MSE provides the average of the squared differences between the true and predicted values, thereby emphasizing larger errors. 
In contrast, MAE calculates the average of the absolute differences, providing a simple metric of prediction accuracy that is less affected by outliers compared to MSE.

\section{Quasar and galaxy light profiles} \label{sec:lights}

The elliptical Moffat profile, which we use to model the quasar's light by approximating the point spread function (PSF), is given by:
\begin{equation}
	I(x, y) = I_0 \left[ 1 + \left( \frac{r_{\text{ell}}}{\alpha / \sqrt{q}} \right)^2 \right]^{-\beta},
	\label{eq:moffat}
\end{equation}
where \(I(x, y)\) is the intensity at the location \((x, y)\), \(I_0\) is the intensity normalization, \(\alpha\) controls the overall size of the profile, and \(\beta\) adjusts the profile's wing shape. 
The parameter \(q\) is the axis ratio of the ellipse -- that is, expressed as \(q = b/a\) --  where \(a\) and \(b\) are the semi-major and semi-minor axes, respectively. 
The elliptical radius \(r_{\text{ell}}\) is given by:
\begin{equation}
	r_{\text{ell}}^2 = \frac{(x' - x_0)^2}{a'^2} + \frac{(y' - y_0)^2}{b'^2},
\end{equation}
with \((x', y')\) being the coordinates after rotation to align with the ellipse's principal axes, and \((x_0, y_0)\) representing the center of the profile. 
The quantities \(a'\) and \(b'\) denote the transformed semi-major and semi-minor axes, respectively, accounting for the ellipticity components \(e_1\) and \(e_2\). 
The components \(e_1\), \(e_2\), and \(q\) are connected to the orientation angle (\(\phi\)) through the relations:
\begin{equation}
	e_1 = \frac{a^2 - b^2}{a^2 + b^2} \cos(2\phi), \quad e_2 = \frac{a^2 - b^2}{a^2 + b^2} \sin(2\phi).
\end{equation}
It is important to note that Equation~\ref{eq:moffat} is a reparameterization of the original expression proposed by \cite{1969A&A.....3..455M}, as introduced by \cite{2021JOSS....6.2825N} to improve numerical stability in lens modeling.

On the other hand, the mathematical equation for the S\'{e}rsic light distribution, used to model the lens galaxy's light, is:
\begin{equation}
	I(r) = I_0 \exp \left( -b_n \left[ \left(\frac{r}{R_e}\right)^{\frac{1}{n}} - 1 \right] \right)
\end{equation}
where \( I(r) \) denotes the intensity at the elliptical radius \( r \), \( I_0 \) represents the intensity at the effective radius \( R_e \), and \( n \) is the S\'ersic index, which governs the concentration of the profile. 
The elliptical radius is defined as:
\begin{equation}
	r = \sqrt{\frac{x'^2}{a^2} + \frac{y'^2}{b^2}},
\end{equation} 
where \( a \) and \( b \) are the semi-major and semi-minor axes, and \( (x', y') \) are the coordinates in the rotated reference frame aligned with the galaxy's major axis. 
The constant \( b_n \) is a normalization factor that depends on \( n \) and is is typically approximated as:
\begin{equation}
	b_n = 2n - \frac{1}{3} + \frac{4}{405n} + \frac{46}{25515n^2} + \frac{131}{1148175n^3} - \frac{2194697}{30690717750n^4}	
\end{equation}
This constant \(b_n\) ensures that \(R_e\) contains half of the total intensity of the light profile.

\section{Lens modeling results for GLQD sources} \label{sec:glq_model}

The result of the lens modeling for GLQD sources are stored on the \href{https://doi.org/10.5281/zenodo.14718063}{Zenodo} repository.
This consists of tables containing the parameters obtained using the traditional approach with \texttt{PyAutoLens} and the ones inferred through the deep-learning based method with VariLens.
A comparison of these estimates is illustrated in Figure~\ref{fig:model_comparison}.

\section{Full List of Lens Candidates} \label{sec:lens_candidates}

\begin{figure*}[htb!]
	\centering
	\centering
	\resizebox{\hsize}{!}{\includegraphics{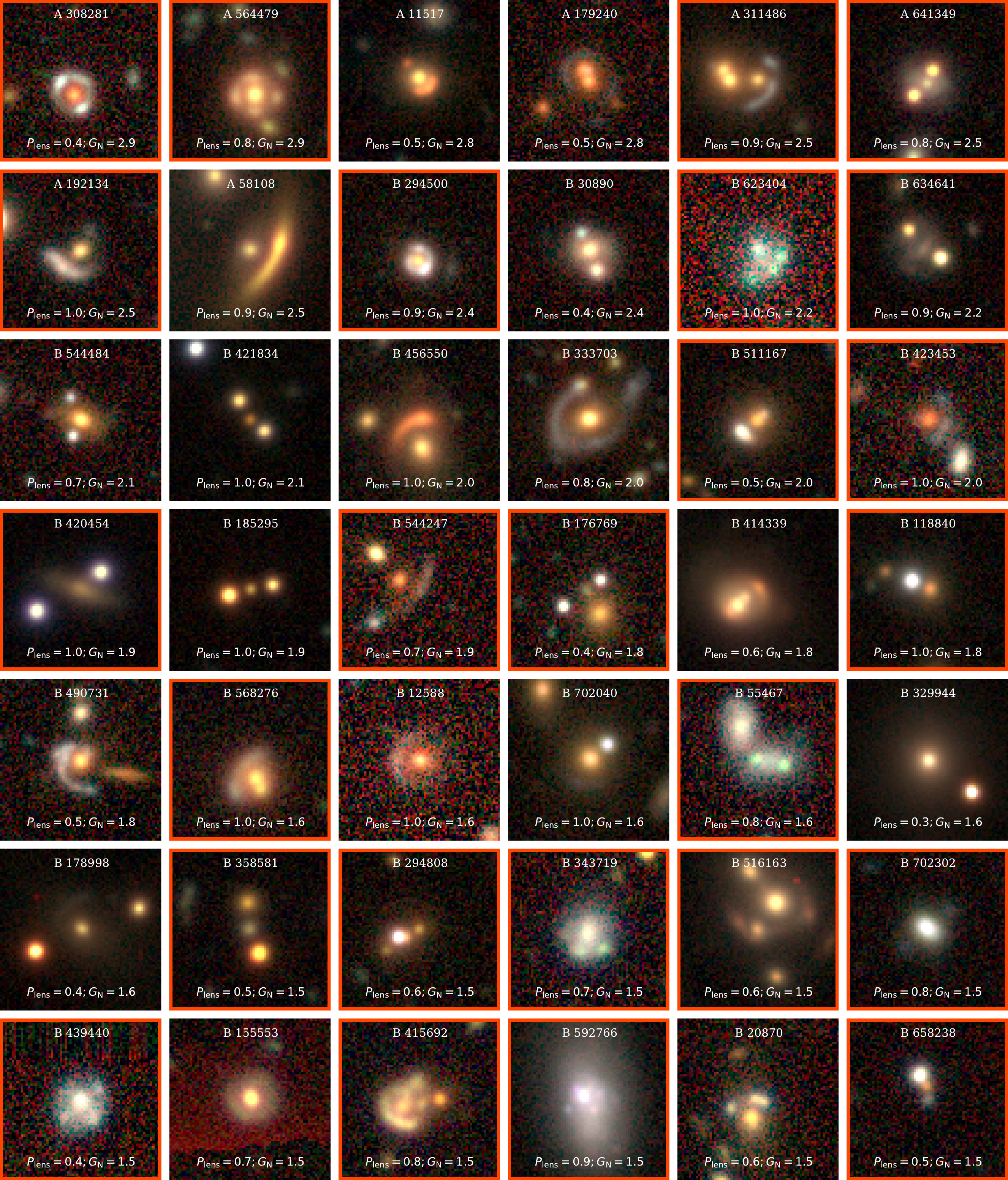}}
	\caption{
		The $64 \times 64$ pixels ($\approx 10\farcs8$ on a side) $riy$-color images of our grade A and B lens candidates are presented. 
		To improve visual contrast, we apply a square-root stretch to the flux values in each HSC cutout. 
		The visual inspection grade and target identification number are displayed at the top of each panel, while the lens probability from our automated classifier is shown at the bottom. 
		Newly identified sources in this study, as well as candidates from \cite{2023A&A...678A.103A}, are highlighted with red rectangles.
	}
	\label{fig:lens_candidates}
\end{figure*}

\begin{figure*}[htb!]
	\centering
	\resizebox{\hsize}{!}{\includegraphics{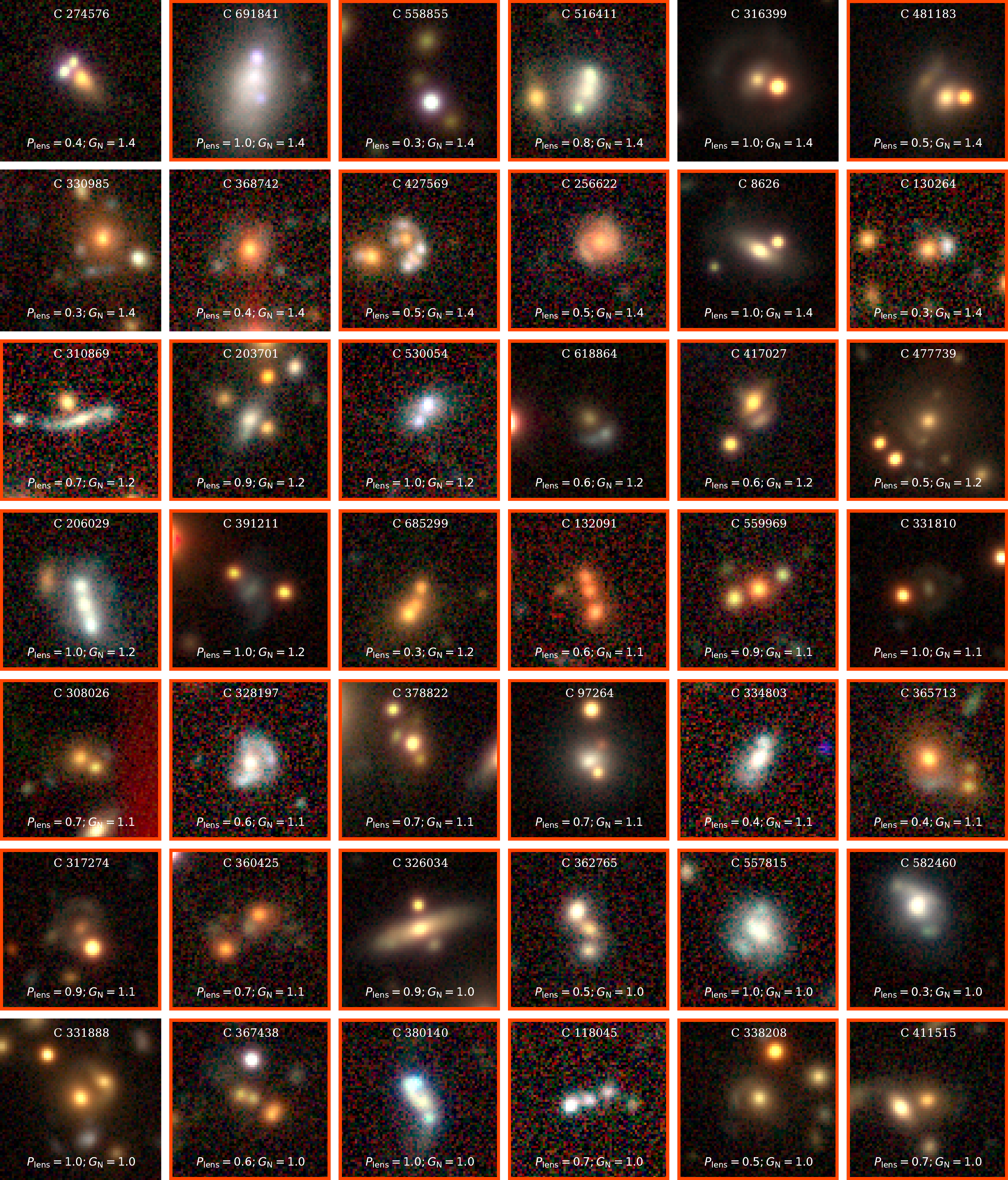}}
	\caption{
		Similar to Figure~\ref{fig:lens_candidates}, but for a subset of our grade C candidates.
	}
	\label{fig:lens_candidates_c}
\end{figure*}

The color images of our lens candidates are presented in Figure~\ref{fig:lens_candidates} and \ref{fig:lens_candidates_c}. 
These candidates were selected using the methodologies discussed in the main text. 
Comprehensive photometric data and the inferred physical parameters for each system can be accessed via \href{https://doi.org/10.5281/zenodo.14718063}{Zenodo}.

\end{appendix}

\end{document}